\documentclass[prd, superscriptaddress, tightenlines, longbibliography, nofootinbib, eqsecnum, amsfonts, amsmath, amssymb, floatfix, notitlepage,twocolumn]{revtex4-1}
\pdfoutput=1

\usepackage[utf8]{inputenc}
\usepackage{mathrsfs}
\usepackage{euscript}
\usepackage{epsfig}
\usepackage{graphics}
\usepackage{graphicx}
\usepackage{amsmath}
\usepackage{amssymb}
\usepackage{bm}
\usepackage[usenames,dvipsnames,svgnames,table]{xcolor}
\usepackage{xspace}
\usepackage{wasysym}
\usepackage{times}
\usepackage{appendix}
\usepackage{lipsum}
\usepackage[nolist,nohyperlinks]{acronym}
\usepackage{float}
\usepackage{simplewick}
\usepackage{tabularx}
\usepackage{booktabs}
\usepackage{natbib, ifthen}
\usepackage{hyperref}

\newcommand{\fsky}{f_{\rm sky}}

\providecommand{\planck}{\textit{Planck}}
\providecommand{\Planck}{\planck}

\newcommand{\mksym}[1]{\ifmmode {\rm #1}\else #1\fi}

\newcommand{\taunl}{\ifmmode {\tau_{\rm{NL}}}\else $\tau_{\rm{NL}}$\fi}

\providecommand{\Planck}{\textit{Planck}}
\providecommand{\planck}{\Planck}

\providecommand{\gea}{\ga}

\providecommand{\agt}{\gea}
\providecommand{\text}[1]{\rm{#1}}

\newcommand{\Msun}{M_\odot}

\newcommand{\begm}{\begin{pmatrix}}
\newcommand{\enm}{\end{pmatrix}}

\newcommand\ba{\begin{eqnarray}}
\newcommand\ea{\end{eqnarray}}
\newcommand\bea{\begin{eqnarray}}
\newcommand\eea{\end{eqnarray}}

\newcommand\be{\begin{equation}}
\newcommand\ee{\end{equation}}

\newcommand{\valpha}{{\boldsymbol{\alpha}}}

\newcommand{\ud}{{\rm d}}

\newcommand{\boldvec}[1]{{\mbox{\boldmath{$#1$}}}}

\providecommand{\vr}{\boldvec{r}}

\newcommand{\vx}{\boldvec{x}}
\newcommand{\vy}{\boldvec{y}}

\defcitealias{Fabbian:2020yjy}{Paper I}
\newcommand{\PaperI}{\citetalias{Fabbian:2020yjy}\xspace}

\newcommand{\isdraft}[1]{#1}

\newcommand{\refchange}[1]{{\isdraft{\color{black} {#1}}}}

\newcommand{\mcut}{M^{\rm cut}}
\newcommand{\websky}{Websky\xspace}
\newcommand{\msun}{M_{\odot}\xspace}

\newcommand{\nlth}{$N_L^{(3/2)}$\xspace}
\newcommand{\nlzero}{$N_L^{(0)}$\xspace}
\newcommand{\nlone}{$N_L^{(1)}$\xspace}

\newcommand{\ximasked}{\xi_{\rm masked}\xspace}
\newcommand{\w}{ {W}}
\renewcommand{\vr}{\boldsymbol{r}}
\newcommand{\fMC}{f_L^{\rm MC}\xspace}
\newcommand{\gMF}{\langle \hat g^{\rm MF}_{LM} \rangle\xspace}
\newcommand{\meandelta}{\bar{\Delta}}

\definecolor{ZurichBlue}{rgb}{.255,.41,.884} 		
\definecolor{ZurichRed}{rgb}{0.9, 0.1, 0} 			
\definecolor{ZurichGreen}{rgb}{.196,.504,.396} 		
\definecolor{ZurichYellow}{rgb}{1,.648,0} 			
\definecolor{dodgerblue}{rgb}{0.12, 0.56, 1.0}
\definecolor{azure}{rgb}{0.0, 0.5, 1.0}
\definecolor{awesome}{rgb}{1.0, 0.13, 0.32}
\definecolor{alizarincrimson}{rgb}{0.82, 0.1, 0.26}
\definecolor{mediumpurple}{rgb}{0.58, 0.44, 0.86}
\definecolor{lasallegreen}{rgb}{0.03, 0.47, 0.19}

\begin{document}
\newcommand{\Sussex}{Department of Physics \& Astronomy, University of Sussex, Brighton BN1 9QH, UK}
\newcommand{\Ferrara}{Dipartimento di Fisica e Scienze della Terra, Universit\`a degli Studi di Ferrara, via Giuseppe Saragat 1, I-44122 Ferrara, Italy}
\newcommand{\INFN}{Istituto Nazionale di Fisica Nucleare (INFN), Sezione di Ferrara, Via Giuseppe Saragat 1, I-44122 Ferrara, Italy}
\newcommand{\Cardiff}{School of Physics and Astronomy, Cardiff University, The Parade, Cardiff, CF24 3AA, UK}
\newcommand{\CCA}{Center for Computational Astrophysics, Flatiron Institute, 162 5th Avenue, 10010, New York, NY, USA}

\title{CMB lensing reconstruction biases from masking extragalactic sources}
\author{Margherita Lembo}\email{lmbmgh@unife.it}
\affiliation{\Sussex}
\affiliation{\Ferrara}
\affiliation{\INFN}

\author{Giulio Fabbian}
\affiliation{\CCA}
\affiliation{\Cardiff}
\affiliation{\Sussex}

\author{Julien Carron}
\affiliation{Universit\'e de Gen\`eve, D\'epartement de Physique Th\'eorique et CAP, 24 Quai Ansermet, CH-1211 Gen\`eve 4, Switzerland}
\affiliation{\Sussex}

\author{Antony Lewis}
\affiliation{\Sussex}
\homepage{http://cosmologist.info}

\begin{abstract}

Observed Cosmic Microwave Background (CMB) temperature and polarization maps can be powerful cosmological probes and used for
CMB lensing reconstruction. However, the CMB maps are inevitably contaminated by foregrounds, some of which are usually masked to perform the analysis. If this mask is correlated to the lensing signal, measurements over the unmasked sky may give biased estimates and hence biased cosmological inferences. For example,
masking extragalactic astrophysical emissions associated with objects located in dark matter halos will systematically
remove parts of the sky that have a mass density higher than average. This can lead to a modified lensed CMB power spectrum over the unmasked area and biased measurements of lensing reconstruction auto- and cross-correlation power spectra with external matter tracers
(from both the direct impact on the lensing power, and via modifications to the lensing-dependent reconstruction power spectrum corrections, \nlzero,
 \nlone and \nlth).
For direct masking of the CMB lensing field, we give an approximate halo-model prediction of the size of the effect, and derive simple analytic models for point sources and threshold masks constructed on a correlated Gaussian foreground field.
We show that biases are significantly reduced by optimal filtering of the CMB maps in the lensing reconstruction, which effectively fills back some of the information in small mask holes.
We test the resulting  lensing power spectrum biases on numerical simulations, masking radio sources, and peaks of thermal Sunyaev-Zeldovich (tSZ) and cosmic infrared background (CIB) emission.
For radio point sources the remaining bias is negligible, but
a temperature lensing reconstruction power spectrum bias remains at the 0.5-5\% level if clusters with a mass larger than $\sim 10^{14}\Msun/h$ are masked, or if peaks in the CIB and tSZ maps are removed with $f_{\rm sky}^{\rm mask} \simeq 0.5$--$7\%$. In any case, these biases can only be measured with a statistical significance $\lesssim 2\sigma$ for future data sets. Moreover, we quantified the impact of the mask biases in the cross-correlation power spectrum between CMB lensing and tSZ and CIB and found them to be larger (up to $\sim 30\%$). We found that masking tSZ-selected galaxy clusters leads to the largest mask biases, potentially detectable with high significance, and should therefore be avoided as much as possible. For the most realistic masks we considered, masking biases can only be measured with marginal significance. We found that the calibration of cluster masses using CMB lensing, in particular for objects at $z\lesssim 0.6$, might be significantly affected by mask biases for near-future observations if the lensing signal recovered inside the mask holes is used without further corrections. Conversely, mass calibration of high redshift objects will still deliver unbiased results.

\end{abstract}


\maketitle

\section{Introduction}
\label{sec:intro}

Cosmic Microwave Background (CMB) observations are inevitably contaminated at some level by foregrounds, from galactic dust to a range of extragalactic signals including the cosmic infrared background (CIB), Sunyaev-Zeldovich effect (SZ), and radio point sources. Extragalactic emissions are correlated to the large-scale structure from which they originate, and hence correlate with other tracers of the matter distribution over a similar redshift range \cite{planck2015-szcatalog,sptpol-szcatalog,advact-szcatalog, Holder:2002wb,Shirasaki:2018wdq,Allison:2015fac,dwek2002, Ade:2013aro,planck2015-cibXtsz,planck2015-sz,Song:2002sg}.
Gravitational lensing is an important effect on the CMB. It both modifies the observed power spectra and produces non-Gaussian signatures in the CMB maps that can be used for lensing reconstruction. The lensing signal correlates with the extragalactic foregrounds and this means that foreground masking has the potential to produce biased inferences from observations of the lensed CMB over the unmasked area.

In principle, the non-blackbody spectrum of the foregrounds (other than kinetic SZ) can be used to clean the foregrounds through multi-frequency observations. Foreground cleaning has been very successfully used, particularly on large scales, but inevitably comes with the cost of increased noise, especially on small scales where the observational noise becomes comparable to the observed signal. Alternatively, the foreground signal can simply be modelled, which is what is often done at the power spectrum level for CMB likelihood analysis. In both cases, it is often necessary to also apply some masking to the brightest sources, including the galactic plane (which is not correlated to large-scale structure and hence does not introduce a direct bias) but also extragalactic sources. For CMB lensing reconstruction, the non-Gaussianity of the foregrounds is important and can produce a direct bias on lensing estimates that is difficult to model~\cite{Osborne:2013nna,vanEngelen:2013rla}. For tSZ foregrounds, the largest non-Gaussianity is associated with the brightest tSZ clusters, and hence can be substantially reduced by cluster masking~\cite{Osborne:2013nna,vanEngelen:2013rla}. Point sources and CIB foreground non-Gaussianity can also be substantially reduced by masking the brightest sources or through component separation.
A variety of other methods have been suggested to reduce lensing biases from small-scale temperature reconstruction~\cite{namikawa2013,Schaan:2018tup,Darwish:2020fwf,Madhavacheril:2018bxi,Fabbian:2019tik}, however, these are usually only applied after the brightest sources have already been masked out. To extract reliable information from small-scale CMB temperature observations it is therefore likely to be necessary to understand the impact of the
source masking, especially if the correlated masking introduces substantial biases.

Recent ACT lensing analyses successfully performed lensing reconstruction on maps where extragalactic sources were subtracted from the map with a dedicated procedure, but not masking the corresponding sky areas~\cite{Darwish:2020fwf,Naess:2020wgi}. Despite the success of the method, it is unclear if it will remain sufficiently accurate for the analysis of future high sensitivity experiments. Source subtraction may itself alter the underlying CMB signal on small scales and become problematic, in particular for extended tSZ-detected clusters, and dedicated assessment of potential systematics introduced by the technique would have to be carried out anyway. 

Masking the observed CMB data with a lensing-correlated mask introduces a number of different effects. Firstly, as shown in \cite[hereafter, Paper I]{Fabbian:2020yjy}, the CMB power spectra on the unmasked area can be significantly altered on small scales (and large scales for B-mode polarization), since there is a net scale-dependent demagnification over the unmasked area. Although estimated to be negligible for \Planck~\cite{PL2018}, the impact is potentially larger for forthcoming high-resolution CMB experiments where more of the information is in the small-scales, precisely where the foregrounds are more important.  This has the potential to bias parameter constraints if not consistently accounted for.
Secondly, if the CMB lensing potential is estimated using standard quadratic estimators~\cite{Hu:2001kj}, the mask complicates its normalization and the noise biases to the lensing power spectrum, even for masks that are uncorrelated to the signal. For uncorrelated masks, these effects can be estimated analytically in some cases (see Appendix~\ref{sec:holes}), or quantified for a fiducial model using independent lensed CMB simulations. After correcting for these effects, if the mask is actually correlated to the signal there will be additional biases because: 1) the areas of high convergence have been preferentially masked directly biasing the actual lensing power over the unmasked area; 2) the normalization, mean-field and Gaussian noise bias (\nlzero) are different because of the differences in the lensed CMB power spectra over the unmasked areas; 3) the \nlone bias due to non-primary lensing contractions~\cite{Kesden:2003cc} is changed due to the different CMB and lensing power; 4) non-Gaussian biases, specifically \nlth related to non-linear large-scale structure growth and post-Born lensing effects~\cite{Bohm:2016gzt,Pratten:2016dsm,Bohm:2018omn,Beck:2018wud,Fabbian:2019tik}, are modified due to the changed non-Gaussianity (e.g. reduced skewness) of the masked field and related changes in the power spectra.

Effects of LSS-correlated masking in the lensing reconstruction might also impact the estimated CMB lensing field itself, and therefore the cosmological constraints involving its statistics beyond the power spectrum \cite{Liu:2016nfs}. Moreover, it might bias the cross-correlation with external matter tracers. These are powerful cosmological and astrophysical probes for a wide range of physical phenomena.
Cross-correlation between CMB lensing and galaxy surveys data helps improving cosmological constraints by breaking parameter degeneracies (e.g., involving galaxy bias) and by measuring nuisance parameters associated with sources of systematic errors (e.g., lensing multiplicative bias, photometric redshift errors) \cite{Vallinotto:2011ge,Schaan:2016ois,Cawthon:2018acr,Schaan:2020qox}. For future galaxy surveys, such as Euclid and LSST, this approach is likely to become the standard analysis to obtain cosmological constraints \cite{Euclid:2021qvm,Sailer:2021yzm}.

The cross-correlation between CMB lensing and the extragalactic emissions can also be a useful source of additional information. The tSZ-CMB lensing cross-correlation ($\kappa\times y$) signal is a unique probe of the physics of the intracluster medium at high-redshift $z\approx 1$, and in relatively low-mass clusters and groups ($10^{13}\lesssim M \lesssim 10^{15} \msun/h$). It is also more sensitive to contributions from structures located in dark matter halos of lower masses than the tSZ auto-power spectrum or the cross-correlation with galaxy lensing measurements \cite{Hill:2013dxa,Battaglia:2014era,Hojjati:2014usa,VanWaerbeke:2013cfa}. Furthermore, it is a powerful cosmological probe due to its strong dependency on $\sigma_8$ and $\Omega_m$ \cite{Osato:2017cva}.

The CIB emission, conversely, is generated by redshifted thermal radiation from UV-heated dusty star-forming galaxies (DSFGs) that have a redshift distribution peaked between $1\lesssim z \lesssim 2$ \cite{Bethermin:2012ta,Maniyar:2020tzw}. The kernel of CMB lensing peaks around the same redshift range and CMB photons are mainly lensed by halos of mass $\approx 10^{11}-10^{13} \msun$ similar to the one of DSFGs \cite{Song:2002sg}. As such, CIB and CMB lensing are highly correlated. Their cross-correlation ($\kappa\times\mathrm{CIB}$) was first measured by \Planck~\cite{Planck:2013qqi} in multiple frequency bands with high statistical significance. Such strong correlation, as high as 80\% at 545GHz, allows the star formation rate density and the mass of the halos hosting CIB to be constrained over a wide range of redshifts, $1\lesssim z\lesssim 7$. Future CMB lensing and CIB measurements of Simons Observatory~\cite[SO hereafter]{Ade:2018sbj} and CCAT-prime~\cite{Aravena:2019tye} are expected to further improve the constraining power on these astrophysical processes \cite{McCarthy:2020qjf}.
The high degree of correlation between CIB and CMB lensing can also be exploited to construct templates of the lensed B-modes for delensing analyses using CIB as an external LSS tracer alone, or in combination with CMB lensing itself \cite{Sherwin:2015baa,Yu:2017djs}.
Potential biases due to LSS-correlated masking on $\kappa\times \mathrm{CIB}$ would not directly translate in a misinterpretation of the delensed B-modes signal (and hence in a bias on $r$) as the correlation between CIB and CMB lensing is usually not assumed from a model but directly measured from the data itself. Nevertheless, it is important to understand if these effects might impact the expected delensing efficiency for future experiments.\\

Building on our work of \PaperI, in this paper we study and quantify the impact of the mask bias for lensing reconstruction. The main aim of this paper is to identify the important sources of correlated mask bias, provided some qualitative understanding of them, and approximately quantify their size. Future work will be required to make accurate and fully self-consistent predictions including the impact of potential foreground-residuals in the masked CMB in addition to the effect of the lensing-correlated mask.

In Sec.~\ref{sec:matter}, we give simple analytic models of the expected effect of foreground masking for various types of mask, taking the masks to apply directly to the lensing convergence field without modelling lensing reconstruction.
 In Sec.~\ref{sec:masks}, we describe the realistic simulations, including extragalactic foreground emissions as well as the effect of non-linear evolution of large-scale structure (LSS), that we used to model the relevant effects and validate our analytic models.
 \refchange{A reader who is only interested in the actual results on lensing reconstruction can skip these two first sections and go directly to Sec.~\ref{sec:recon} where}
 we show the impact of the mask bias on the final estimated lensing potential power spectrum from lensed CMB maps, including the effects of optimal CMB filtering and lensing reconstruction. In Sec.~\ref{sec:cmblens-results}, we also quantify the impact on the cross-correlation power spectrum with the true convergence field and extragalactic foregrounds, and assess the impact on cluster mass calibration if the lensing field recovered inside a cluster mask is used directly without further modelling. Finally, in Sec.~\ref{sec:forecasts}, we estimate the impact of these biases on the near-future CMB measurements.
Modelling the full effect of mask bias on the reconstructed CMB lensing potential field is challenging analytically, but we give some analytic results in Appendix~\ref{sec:holes} for the case of uncorrelated circular mask holes.

For near-future observations, such as SO, the temperature signal still contains a substantial fraction of the available information, so fully exploiting the data will require robust modelling of the temperature signal, which is what we focus on in this work. The contamination induced by foregrounds is less important for CMB lensing reconstruction using polarization \cite{Beck:2020dhe}, since the polarized foreground amplitudes are expected to be substantially lower, but should also be assessed in future work (future observations by CMB-S4~\cite[S4 hereafter]{Abazajian:2019eic} will give lensing reconstructions that are more dominated by polarization).

\section{Lensing power spectra with LSS-correlated masks}\label{sec:matter}
A simple estimate of the effect of an LSS-correlated mask on the CMB lensing power can be obtained by estimating the power spectrum of the true $\kappa$ field over the unmasked area and compare its value with the one computed over the full sky. This operation can only be performed on simulations, since $\kappa$ is not measurable directly but has to be reconstructed from lensed CMB maps.
However direct $\kappa$ masking is more easily modelled analytically, though it would only provide a good estimate of the expected amplitude of the effect if the lensing reconstruction were a fully local noiseless unbiased estimate of the true underlying field. We use it as a simple model to gain some analytic insight into the impact of correlated masking without the complications induced by the lensing reconstruction estimator. We give simple analytic models of the direct masking effects for all the masks described in the next section, and also compare these analytic predictions with measurements based on direct masking of simulations. In the following, we assume sky masks to be uncorrelated with the unlensed CMB and ignore the correlation between CMB lensing and CMB temperature (i.e. $C_L^{T\phi}=0$) induced by the integrated Sachs-Wolfe effect in both the analytical modelling and in the simulation measurements.

\subsection{Halo model}\label{sec:halomodel}
A qualitative explanation of the effect of masking peaks of the density field can be obtained using the halo model of large structure (see e.g. Ref.~\cite{Cooray:2002dia} for a review). In this model, the entire matter of the universe is distributed inside haloes of roughly universal spherically-averaged halo profile $\rho_M(r)$ as a function of the halo mass. In the following we assume $\rho_M(r)$ to be a Navarro-Frenk-White (NFW) profile~\cite{NFW} defined by the mass-concentration relation of \cite{duffy2008}. The matter power spectrum $P(k)$ is built as the sum of two separate parts, the one- and two-halo terms ($1h$ and $2h$ respectively), that correlate points within the same or two different halos respectively. This reads
\begin{eqnarray}
	P(k) &\equiv& P_{1h}(k) + P_{2h}(k)\label{eq:halo-model}\\
	P_{1h}(k) &\propto& \int dM \frac{dn}{dM} |\rho_M(k)|^2\label{eq:1-halo}\\
	P_{2h}(k) &\propto& P^{\rm lin} (k)\left(\int dM \frac{dn}{dM} b(M) \rho_M(k) \right)^2,\label{eq:2-halo}
\end{eqnarray}
where $M$ is the halo mass, ${dn}/{dM}$ is the halo mass function which describes the number density of halos of a given mass, $b(M)$ is the halo bias, $\rho_M(k)$ is the Fourier transform of $\rho_M(r)$ and $P^{\rm lin}(k)$ is the linear matter power spectrum. To compute the integrals of Eqs.~\eqref{eq:1-halo}, \eqref{eq:2-halo} we used the public \texttt{hmvec} code\footnote{\url{https://github.com/simonsobs/hmvec}} and a Sheth and Tormen \cite{sheth-tormen} mass function. To emulate the masking of a nearly mass-limited sample of halos, we truncate the integration over the halo mass function in both terms of Eq.~\eqref{eq:halo-model} to
different mass limits, such as those expected for the SO and S4 tSZ selected clusters (see the next section for more details). This approximation can be refined to include selection function effects of each experiment \citep{Rotti:2020rdl}.
The signature of such a cut is twofold: a large direct suppression of power on small scales from the $1h$ term, and, since the most massive halos form in the highest peaks of the linear density and are highly biased, a suppression on large scales as well.

\begin{figure*} [t!]
\centering
\begin{tabular}{cccc}
\includegraphics[width = 0.497\textwidth]{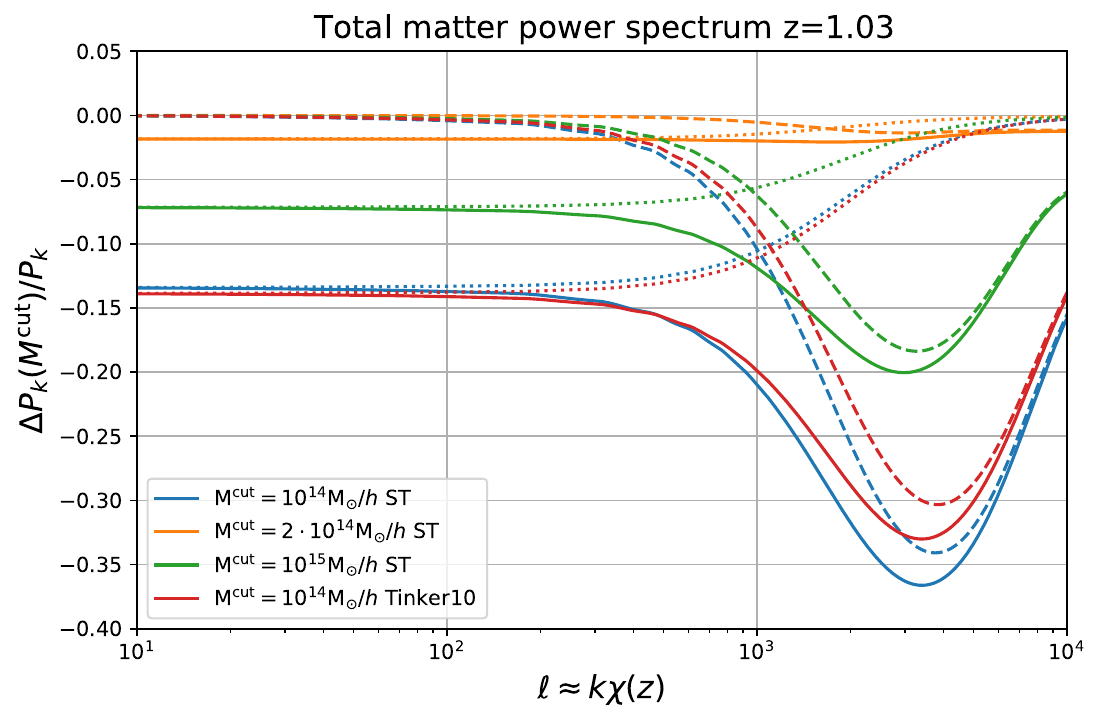}&
\includegraphics[width = 0.503\textwidth]{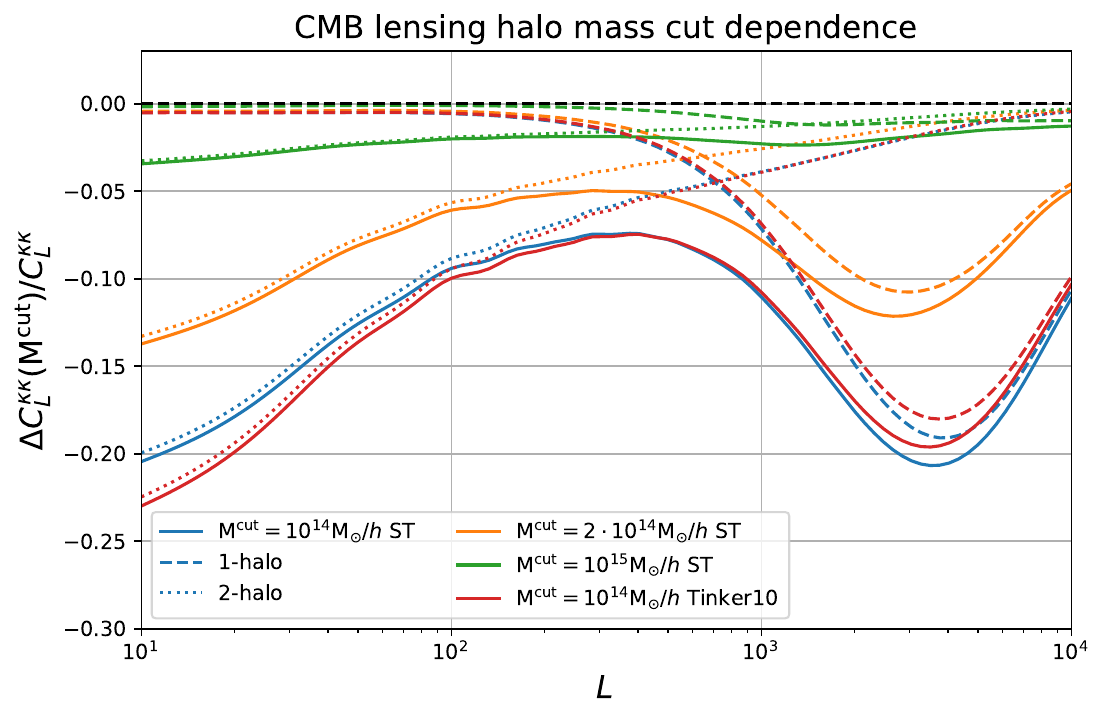}
\end{tabular}
\caption{\emph{Left:} fractional change in the halo-model matter power spectrum due to cutting the most massive haloes at $z\approx 1$, close to the peak of the CMB lensing kernel. The ratio is shown as function of $k\chi \sim \ell$. \emph{Right:} fractional impact of the most massive halos on the corresponding CMB lensing convergence angular power spectrum. In both panels, the dashed and dotted lines show the $1h$ and $2h$  contributions, which sum up to the solid lines. We show results for both Sheth and Tormen \cite{sheth-tormen, st-massfunction01, st-massfunction02} (ST) and Tinker et al. 2010 \cite{tinker2010} mass functions. The choice of the mass function has little impact on the results.}
\label{fig:halomodel}
\end{figure*}

In Fig.~\ref{fig:halomodel}, we show the ratio between the 2D projection of the density power spectra in the Limber approximation after the truncation in the integration, and our reference power spectrum that (very conservatively) uses $10^{18}\msun$ as the upper limit of the integration for two different mass cuts (blue and orange). The solid lines show the ratio of the spectra including both the $1h$ and $2h$ terms at $z=1.03$; the dashed lines show the effect of the mass cut on $1h$ term while the dotted lines show the same effect for the $2h$ term. The suppression of power on small scales is visible, as is a large dip at $\ell\simeq k\chi \sim 3000$, while the $2h$ contribution dominates on large scales.
Since our goal is to assess the impact of masking on the CMB lensing, we therefore used the matter power spectrum with and without the mass cut in the computation of the CMB lensing convergence power spectrum
\begin{equation}
C_L^{\kappa\kappa} =\frac{9H_0^4\Omega^2_{m,0}}{4c^4}\!\int_0^{\chi_s}\!d\chi \left(\frac{\chi_s - \chi}{\chi_s}\right)^2\!P\left(\frac{L+1/2}{\chi},\chi\right),
\label{eq:ckk}
\end{equation}
where $\chi$ is the comoving distance and $\chi_s$ is the comoving distance to the CMB. As noted in \cite{munchmeyer2018}, the $2h$ contribution is not naturally built to match the expectation of linear cosmological perturbation theory. We follow their approach and normalize the $2h$ contribution to match the linear theory prediction when computed over our reference range of integration and apply consistently the same normalization to the cases when the integration range is cut. We regularized the $1h$ term on large scales by exponentially damping the contribution of matter perturbations on scales $k\leq 0.01\mathrm{Mpc}/h$ in the integrand of Eq.~\eqref{eq:ckk} \cite{Cooray:2002dia}. The overall change observed in $C_L^{\kappa\kappa}$ is a consequence of the superposition of the power suppression observed for the matter power spectrum at different redshifts weighted by the lensing kernel.\\*

It may also be possible to extend the halo model approach to make predictions for masking bright infrared sources, or a more direct model of the tSZ intensity. However, we do not pursue this further here, since, as we discuss in Sec.~\ref{sec:recon}, the actual response of the lensing reconstruction to masking is rather different due to the filtering on the CMB maps \cite{Maniyar:2020tzw}.

\subsection{Gaussian foreground peaks model}\label{sec:gaussianmodel}

When the mask is obtained by masking the peaks of a foreground field, we can construct an analytic model by approximating both the foreground field and the lensing convergence as being Gaussian, with some known covariance. More generally, we can consider the effect of masking on the cross-correlation function of any two Gaussian fields (which we could take to both be the lensing convergence, or, for a cross-correlation case, the lensing convergence and some  tracer of the matter density).
In real space, following a similar argument to \PaperI, the main quantity of interest is the masked  correlation function of two fields $A$ and $B$
\begin{align}
\ximasked^{AB}(r) &\equiv \left\langle \w(\vx) A(\vx)\w(\vx')B(\vx')\right\rangle,
\label{eq:maskedxi}
\end{align}
where $\vx' = \vx + \vr$ and $W(\vx)$ is a mask window function.
We assume that some underlying Gaussian statistically-isotropic `foreground' field $f(\vx)$ (i.e. the $\kappa$, $y$ and CIB fields) determines the mask probability locally, so that $\w(\vx)$ only depends on some (in general non-linear) function of $f(\vx)$ at the same point.
For a mask that is constructed by thresholding the foreground to remove the peaks of its emission, the mask window function is given by a simple step function $\w(\vx) = \Theta(\nu\sigma_f-f(\vx))$ where $\nu$ determines the `sigma' value of the cut.
The expectation value in Eq.~\eqref{eq:maskedxi} is then an integral over the correlated Gaussian variables $f(\vx)$, $f(\vx')$, $A(\vx)$ and $B(\vx')$, where the components of the covariance matrix are simply the correlation functions of the fields evaluated at $\vr$, or zero. If we know these correlation functions, the expectation value can therefore be calculated straightforwardly as the two Gaussian integrals over $A(\vx)$ and $B(\vx')$ can be done directly.

For Gaussian fields with any local mask (in the sense defined above), the bias may be written compactly by introducing the Gaussian independent variable $f_{\pm} = f(\vx) \pm f(\vx')$ and integrating analytically over the correlated Gaussian distributions. The bias to the mask-deconvolved correlation function, an estimate of the full-sky correlation function, is then
\begin{align}\label{eq:ABf_bias}
\Delta \xi^{AB}(r) \equiv &\frac{\ximasked^{AB}(r)}{\left\langle W(\vx)W(\vx') \right\rangle} -  \xi^{AB}(r) \\ \nonumber
=&\frac{ \xi^{Af_{+}}(r) \xi^{Bf_{+}}(r)}{\sigma^4_{+}(r)}\frac{\left\langle\left( f^2_{+} -\sigma^2_+\right) W(\vx) W(\vx') \right \rangle} {\left\langle W(\vx)W(\vx')\right\rangle}  \\ \nonumber -& \frac{ \xi^{Af_-}(r) \xi^{B f_-}(r)}{\sigma^4_-(r)}\frac{\left\langle\left( f^2_{-} -\sigma^2_-\right) W(\vx) W(\vx') \right \rangle} {\left\langle W(\vx)W(\vx')\right\rangle} .
\end{align}
In this equation, $\xi^{Xf_{\pm}}(r)$ is $\xi^{Xf}(0) \pm \xi^{Xf}(r)$, and $\sigma^2_{\pm}(r) = 2\sigma^2_f \pm 2\xi_f(r)$.
For a specific given mask, Eq.~\eqref{eq:ABf_bias} provides an empirical estimate of the bias using simulations or data, after computation of spectra and cross-spectra of $W(\vx), (fW)(\vx)$ and $(f^2W)(\vx)$. In the case of a threshold mask
$f_-$ is left unconstrained from the threshold windowing, and $f_+$ is restricted not to exceed $2 \nu \sigma_f$. The averages required result in one-dimensional, very smooth integrals which are easily computed numerically, giving us an alternative fully analytic prediction.

In the limit of large separations, where the foreground fields at the two points are uncorrelated, taking $\sigma_f^2 \gg \xi_f(r)$, $|\xi^{Af}(r)| \ll |\xi^{Af}(0)|$,
$|\xi^{Bf}(r)| \ll |\xi^{Bf}(0)|$, the correction to the correlation function reduces to the simple approximate form
\begin{align}
 \Delta \xi^{AB}(r) \sim \frac{\xi^{Af}(0)\xi^{Bf}(0)}{\sigma_f^2} \frac{\bar{f}^2}{\sigma_f^2} = \bar{A}\bar{B},
\end{align}
where $\bar{X}\equiv \langle W X\rangle/\langle W\rangle$ (with $X\in\{A,B\}$) is the mean value of $X$ evaluated over the unmasked area. If the correlation is instead defined after subtracting the means over the unmasked areas, the limiting value is instead zero as expected.
%

For $r$ sufficiently small that the two points are almost surely both unmasked or inside the same mask hole, with $f(x')\approx f(x)$, we have the other limiting form
\begin{align}
 \Delta \xi^{AB}(r) \sim \frac{\xi^{Af}(0)\xi^{Bf}(0)}{\sigma_f^4}\left(\overline{f^2} - \sigma_f^2\right),
\end{align}
where $\overline{f^2}\equiv \langle f^2 W\rangle/\langle W\rangle$ is the mean value of $f^2$ evaluated over the unmasked area. If the mask systematically removes peaks of $f$, so that $\overline{f^2} <\sigma_f^2$, and $A$ and $B$ have positive correlation to the foreground $f$, this is negative. If the means over the unmasked areas are subtracted before calculating the correlation functions, this becomes
\begin{align}
 \Delta \xi^{AB,0}(r) \sim \frac{\xi^{Af}(0)\xi^{Bf}(0)}{\sigma_f^4}\left(\overline{\sigma}^2 - \sigma_f^2\right),
\end{align}
where $\overline{\sigma}^2 \equiv \overline{f^2} - \bar{f}^2$ is the point variance estimated over the unmasked area.

\subsection{Poisson point sources}
A Poisson model that describe the masking due to radio sources can also be handled analytically assuming a Gaussian distribution for the mean Poisson density. Following \PaperI~, we introduce the Poisson intensity field $\lambda(\vx)$, which defines the probability of observing no source in a given area $\tilde A$ as $\exp\left(-\int_{\tilde A}d^2y \lambda (\vy) \right)$. On each location of a source, a disc of radius $R$ is masked, and the mask $W$ consists of the collection of these discs.

Under the assumption of a Gaussian $\lambda$, we may then write

\begin{align}
	&\Delta \xi^{AB}(r) + \xi^{AB}(r)\nonumber \\ &= \frac{ \left\langle A(\vx) B(\vx') \exp \left( -\int_{\tilde A(\vy)} \lambda(\vy) d^2\vy   \right) \right\rangle }{\left\langle \exp \left( -\int_{\tilde A(\vy)}\lambda(\vy) d^2\vy   \right) \right\rangle}.
\end{align}

In this equation, $\tilde A(\vy)$ is the area $D_R(\vy - \vx) + D_R(\vy - \vx') - D_R(\vy - \vx)D_R(\vy - \vx')$ defined by the overlap of two discs $D_R$ of radii $R$ centred on $\vx'$ and $\vy'$.
Taking $A(\vx)$, $B(\vx')$ and the integral over $\lambda(\vy)$ to be three correlated Gaussian variables, the Gaussian integral can be done analytically in terms of integrals over $\xi^{A\lambda}(r)$ and $\xi^{B\lambda}(r)$.
These integrals can be evaluated directly numerically, or the result can be rewritten compactly in the form
\begin{align}\label{eq:poisson_souces}
	&\Delta \xi^{AB}(r) \\ &=   \left[(\xi^{Af}(r)+\xi^{Af}(0)- \left(\left(\xi^{A\lambda} \cdot D_R\right) \star D_R \right)(r) \right] \nonumber \\ \nonumber
	& \cdot\left[(\xi^{Bf}(r)+\xi^{Bf}(0)- \left(\left(\xi^{B\lambda} \cdot D_R\right) \star D_R \right)(r) \right], \nonumber
\end{align}
where $f(\vx)$ is the Poisson intensity $\lambda(\vx)$ convolved with $D_R$.
Here the $\left(\xi^{A/B\lambda} \cdot D_R\right) \star D_R$ terms are the convolution ($\star$) of the disc $D_R$ with the disc-truncated correlation $D_R(r)\xi^{A/B \lambda}(r)$,
which reduces to $\xi^{A/Bf}(0)$ for $r\ll R$ and vanishes for distances $r > 2R$.
Each full term
therefore changes smoothly
from $\xi^{A/Bf}(0)$ at $r\sim 0$ to $\xi^{A/Bf}(0) + \xi^{A/Bf}(r)$ at $r \ge 2R$.

Note that $\xi^{X f}(0) = -\bar{X}$, so if the means $\bar{A}$ and $\bar{B}$ over the unmasked areas are subtracted this amounts to removing the $\xi^{A/Bf}(0)$ terms in Eq.~\eqref{eq:poisson_souces}.
The entire correction is fourth order in the fields and hence small in either case. If there is a non-zero bispectrum, which is only third order in the fields, the actual real-world effect may be dominated by the non-Gaussian correlations rather than the very small Gaussian prediction.

\section{Simulations }\label{sec:masks}

The analytic models described in the previous section were highly idealized.
To study the impact of a correlated mask on real data, it is necessary to simulate CMB lensing and correlated foreground fields in a realistic way. To do this, we followed \PaperI and used the publicly-available \websky\footnote{\url{ https://mocks.cita.utoronto.ca/index.php/WebSky_Extragalactic_CMB_Mocks}} simulation suite ~\cite{Stein:2020its}.
\websky models the evolution of the matter distribution using the mass-Peak Patch method~\cite{Stein:2018lrh} in a volume of  $\sim 600$ (Gpc/$h$)$^3$  with $\sim 10^{12}$ particles over the redshift interval $0 < z < 4.6$. Despite being and approximate N-body method, mass-Peak Patch has been shown to reproduce the clustering properties of halos with good accuracy for both 2-point and 3-point statistics and their covariances across \cite{Blot:2018oxk,Colavincenzo:2018cgf,Lippich:2018wrx}. The simulation only includes dark matter particles, so we neglected baryonic effects in the analytical modelling discussed in the previous section as well. This realization of the matter distribution is used to produce full-sky maps of the CMB lensing convergence $\kappa$ as well as extragalactic foregrounds based on an analytic halo model calibrated on existing observations and hydrodynamical simulations. The public release contains maps of tSZ and CIB at multiple frequencies, catalogues of radio sources based on the model of Refs.~\cite{Sehgal:2009xv,websky-rs}, and catalogues of the dark matter halos identified in the cosmological simulation.

\subsection{Lensed CMB and CMB lensing}\label{sec:MC_sims}
Following \PaperI, we produced two sets of lensed CMB simulations that are later used to study the impact of correlated masking on lensing reconstruction. These sets share the same 100 unlensed CMB realizations, but they are lensed either using the same fixed deflection field constructed from the \websky $\kappa$ simulation (NG set), or using a different Gaussian random realizations of the deflection field for each map (G set), where the deflection field has the same angular power spectrum as the realization of the \websky $\kappa$ map.
 We used the NG set to isolate the bias as it would appear on real data while the G set was used to compute the error bars of our measurements and calibrate the lensing reconstruction normalization and $N^{(i)}$ biases. As such, the error bars displayed in the figures do not include any non-Gaussian contribution to the covariance. In the following, unless stated otherwise, error bars (or bands) displayed in figures are standard deviations of the plotted variable measured over the set of G simulations.
 An additional G-like set was computed to estimate the mean field of the quadratic estimator in order to debias the reconstructed lensing potential power spectrum.

The \websky lensing $\kappa$ map is constructed using the Born approximation and therefore neglects the effect of post-Born lensing, which is expected to significantly modify the overall non-Gaussianity in the CMB lensing $\kappa$ ~\cite{Pratten:2016dsm} and thus the shape of the \nlth bias in the reconstructed lensing field \cite{Fabbian:2019tik,Beck:2018wud}. Investigating the impact of post-Born effects on the mask bias would require us to include post-Born lensing in the lensed CMB, and (Born) lensing of the extragalactic foregrounds to avoid introducing spurious decorrelation of photon deflections along the line of sight (see, e.g., the discussion in Refs.~\cite{Fabbian:2019tik,Boehm:2019hlv}).
As we do not have a similar set of simulations at our disposal, we investigated the role of post-Born effects only for masks built on the $\kappa$ field using the Born and post-Born CMB lensing maps of Ref.~\cite{Fabbian:2017wfp} \refchange{ \footnote{These adopt a Planck 2013 cosmology which is slightly different from the one used in the \websky simulations.}}.  We used those maps to produce two additional sets of simulations where we lensed the same 100 realizations of unlensed CMB maps with deflection fields constructed from the Born and post-Born $\kappa$ maps following the steps adopted for the \websky $\kappa$ map. We refer to these sets of simulations as BNG and the pBNG respectively. To calibrate the normalization of the estimator, we also created a set of Gaussian simulations, as done for the G set, using the power spectrum of the post-Born $\kappa$ map\footnote{As noted in \cite{Beck:2018wud} differences due to post-Born correction in $C_{L}^{\kappa\kappa}$ are negligible for the purpose of this work.}.

\subsection{Foreground masks}\label{sec:sims-masks}

\begin{figure}[!]
\centering
\includegraphics[width = 0.95 \columnwidth]{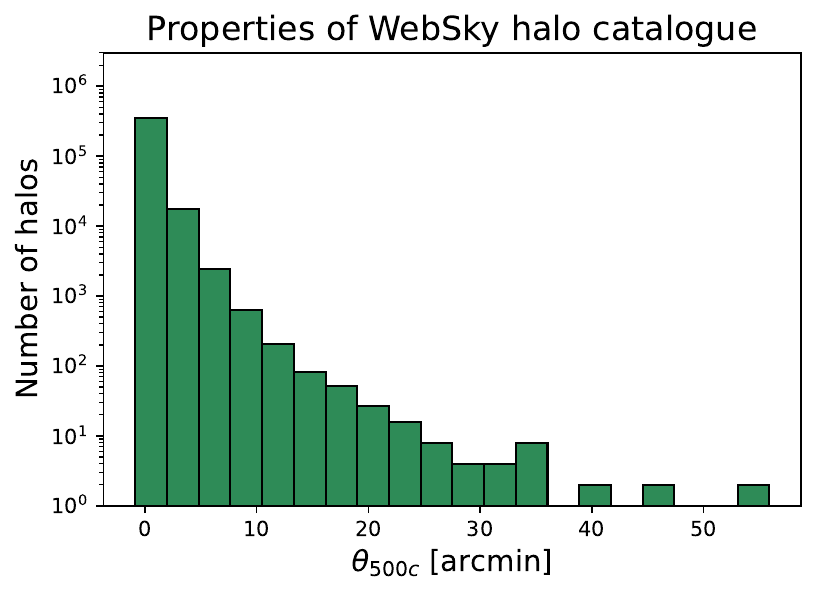}
\caption{Histogram showing the angular size, $\theta_{500,c}$, distribution of halos in the \websky catalogue for selected halos with $M_{500,c} > 10^{14} M_\odot/h$. This is consistent with the mass limit of detectable tSZ clusters in an S4-like survey (see \cite{Ade:2018sbj} for more details).}
\label{fig:hist_catalogue}
\end{figure}

Using the \websky simulations we constructed three qualitatively-distinct kinds of lensing-correlated masks.
\begin{itemize}
\item $W_{\rm halo}$ masks, which roughly mimic masking of objects detected by a mass-limited cluster sample, for example objects detected in a tSZ cluster survey with a given experimental noise level. We created different masks selecting all the halos in the \websky halo catalogue with a mass above a certain mass threshold,  $M^{\rm cut}$ defined in terms of $M_{500,c}$ \footnote{In the following, we usually define halo masses in terms of  spherical overdensity masses in terms of $M_{500,c}$ and $M_{200,m}$. The first is the mass contained within the radius $R_{500,c}$ inside of which the mean interior density is 500 times the critical density. $M_{200,m}$, conversely, is the mass contained within the radius $R_{200,m}$ inside of which the mean interior density is 200 times the mean matter density of the universe.}. For all the selected halos, we masked a disc centred on the halo position with a radius that is a multiple $n$ of the $\theta_{500,c}$ halo angular size. In the following, we will adopt $n = 2$ as our default setup, and will show results for $M^{{\rm cut}}= (1.0, \, 1.8, \, 3.0) \cdot 10^{14}\,M_{\odot}/h$. For reference, Fig.~\ref{fig:hist_catalogue} shows the distributions of angular size $\theta_{500,c}$ in the \websky halo catalogue;

\item Foreground intensity threshold masks $W_{\kappa}, \,W_{\text{\rm CIB}}$ and $W_{y}$. These masks are created by thresholding the \websky $\kappa$, CIB at 217 GHz and the tSZ Compton-$y$ parameter maps respectively such that all the pixels above a specific value are removed from the analysis. We used different thresholds that effectively mask different sky fractions, $f^{\rm mask}_{\rm sky} = 0.6\%,\, 2.3\%,\, 6.7\%$. Before the thresholding step, we smoothed the foregrounds maps with a Gaussian beams of full width at half maximum (FWHM, $\theta_{1/2}$) of $5.1^\prime$ or $1.7^\prime$.
This gives masks with more regular and connected holes, as expected from an experiment with a comparable beam size. By construction, these masks allow us to explore the biases for foregrounds with different degrees of correlation with CMB lensing, where $W_\kappa$ is a limiting-case where the masked field is 100\% correlated, while the CIB map is correlated at the $\gtrsim 70\%$ level at $\ell < 1000$, and the tSZ map at the $30\%$--$50\%$ level. The $W_{\rm halo}$ mask is strongly correlated to the $W_y$ mask, since the main peaks of the $y$-parameter map are associated with massive clusters.

\item $W_{\rm rs}$ masks remove resolved radio point sources (and are the same as those described in \PaperI). For this purpose, we selected all the sources with a measured flux above the $5\sigma$ detection limit for \Planck, SO and S4, and cut out a circular region of the sky around the source. The radius of these holes has been chosen to be $2\theta_{1/2}$ of each frequency channel.
Each of these masks is built as product of three masks obtained by selecting sources in the three frequency bands ranging from 90GHz to 225GHz most relevant for small-scales CMB power spectra measurements (for more details, see Table I of \PaperI).
\end{itemize}

\begin{figure}
\centering
\includegraphics[width = \columnwidth]{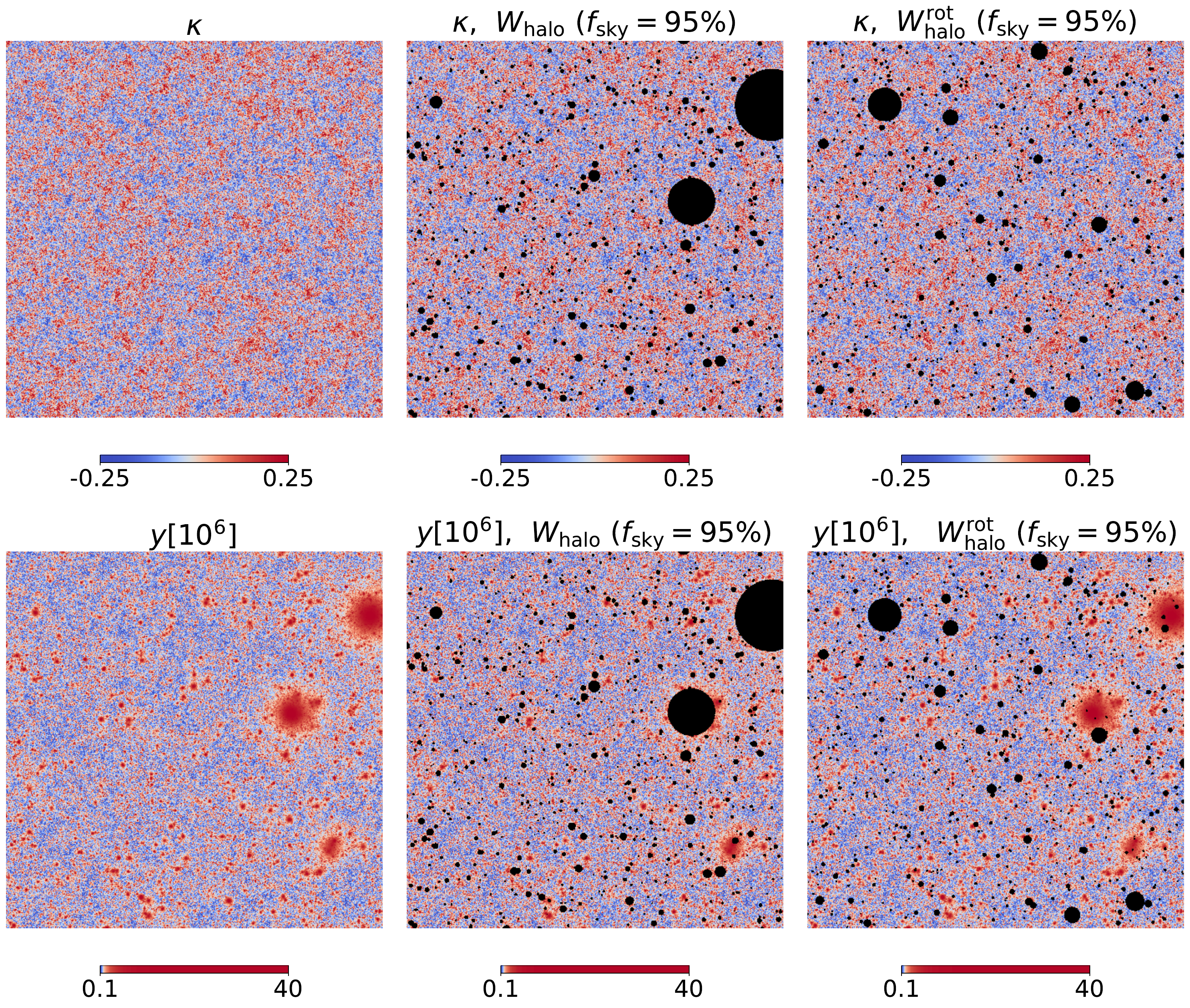}
\caption{20x20 deg$^2$ cutout of the $\kappa$ (top) and tSZ y parameter maps (bottom) of the \websky simulation suite. We highlight the masked pixels in black for one of the $W_\mathrm{halo}$ masks employed in this work. The left column displays the unmasked $\kappa$ and $y$ fields.
The middle column shows the same fields masked with a mask that removes halos with $M>\mcut = 1.8\cdot 10^{14}  M_\odot/h$.  As it is evident from the bottom row plot, the masked areas typically correspond to dense areas of strong tSZ emission.  The right column shows $\kappa$ and $y$ masked with a mask uncorrelated with the underlying lensing field constructed randomizing the position of the halos used for the $W_\mathrm{halo}$ mask.}
\label{fig:masks_all}
\end{figure}

To isolate the impact of mask correlations, \refchange{we apply to each mask a single random rotation to build} 
a new mask, $W^{\rm rot}$, which is effectively uncorrelated with CMB lensing but retains all the other non-trivial mode-coupling effects due to cut sky and hole shapes (we neglect the small area of residual correlation around the poles of the random rotation axis).
In Fig.~\ref{fig:masks_all}, we show a cutout of the full sky $\kappa$ and tSZ $y$-parameter maps from the \websky simulation together with an example of the masks used in our analysis.

\subsection{Biases from direct $\kappa$ masking}\label{sec:matter_1}
\begin{figure*} [t!]
\centering
\begin{tabular}{cccc}
\includegraphics[width=0.35\textwidth]{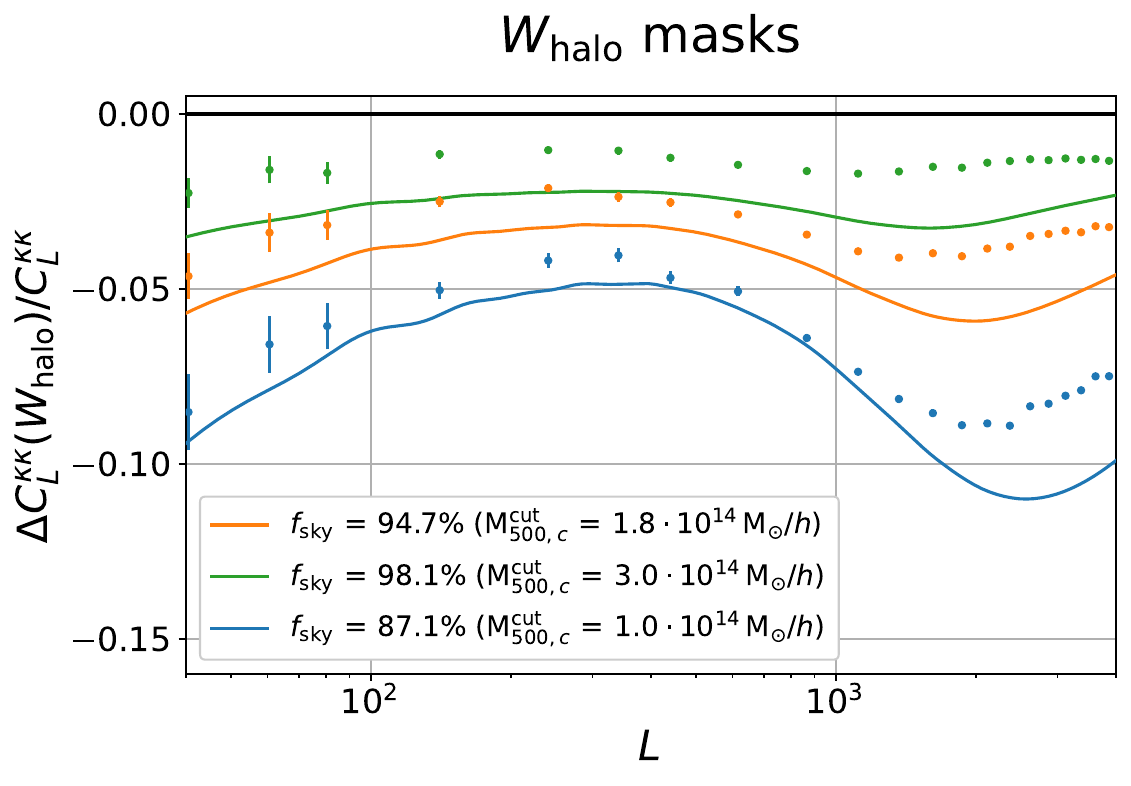} &
\includegraphics[width=0.365\textwidth]{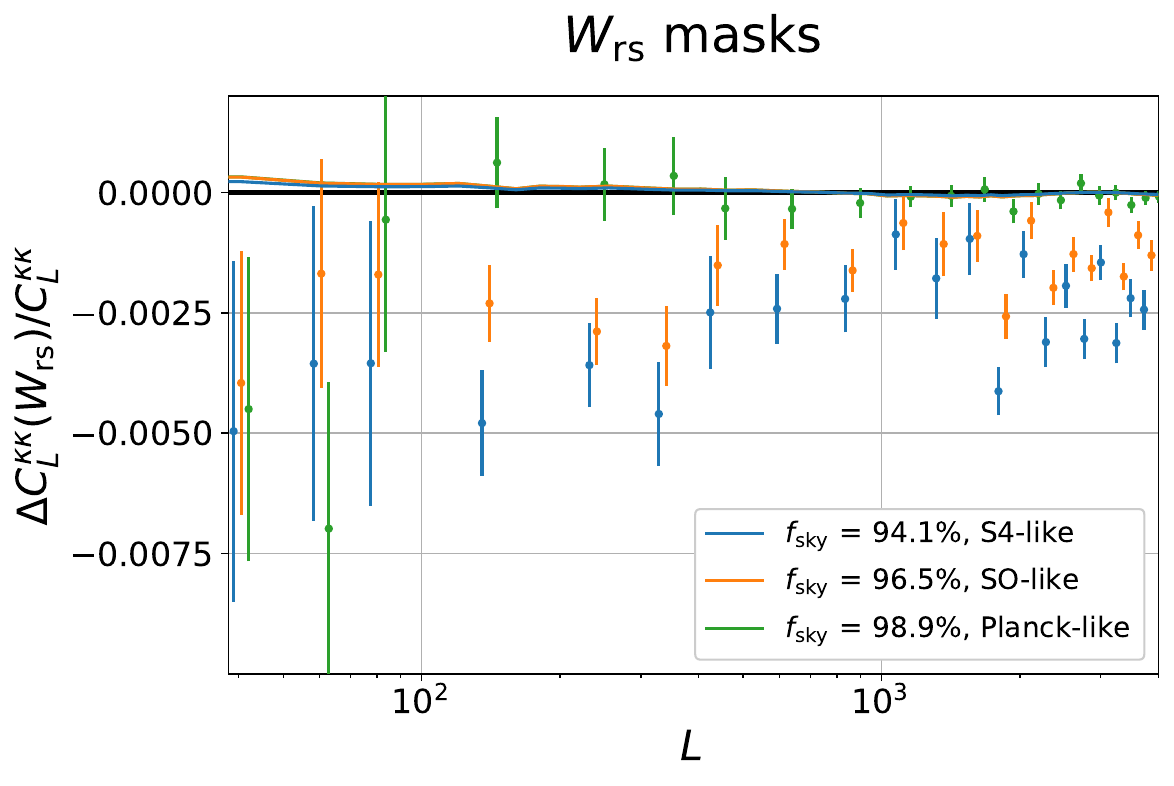}\\
\end{tabular}
\begin{tabular}{cccc}
\includegraphics[width=0.33\textwidth]{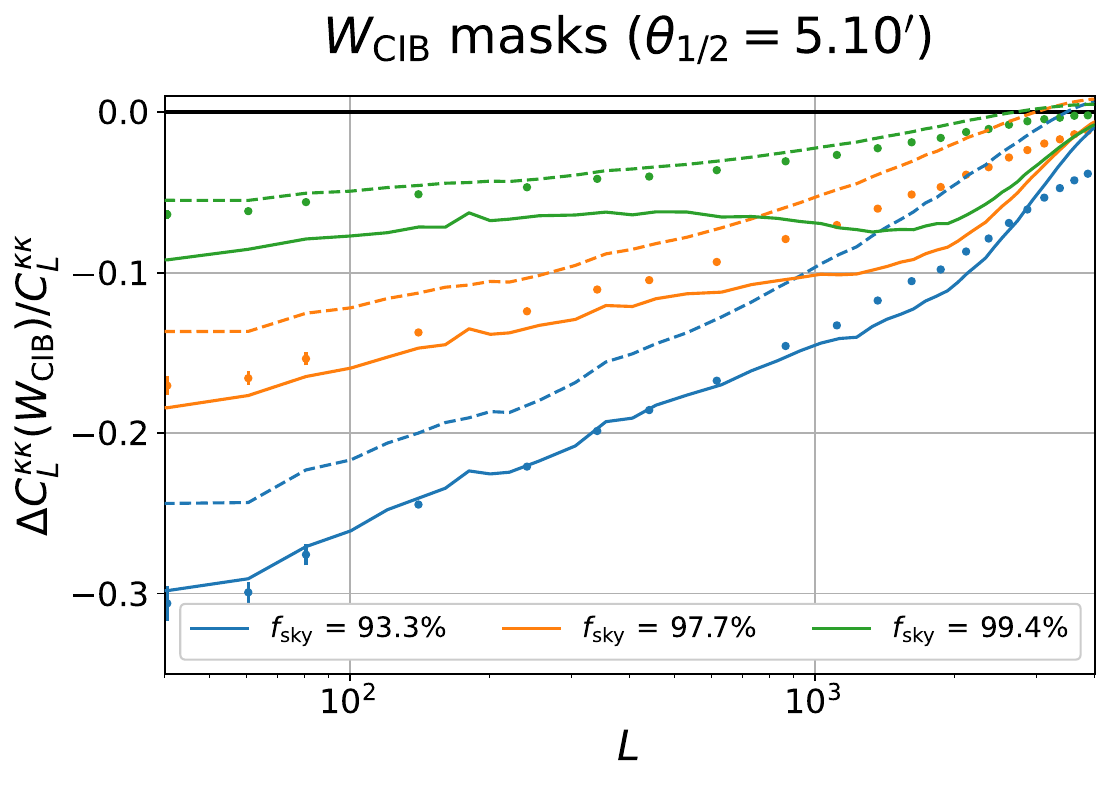} &
\includegraphics[width=0.34\textwidth]{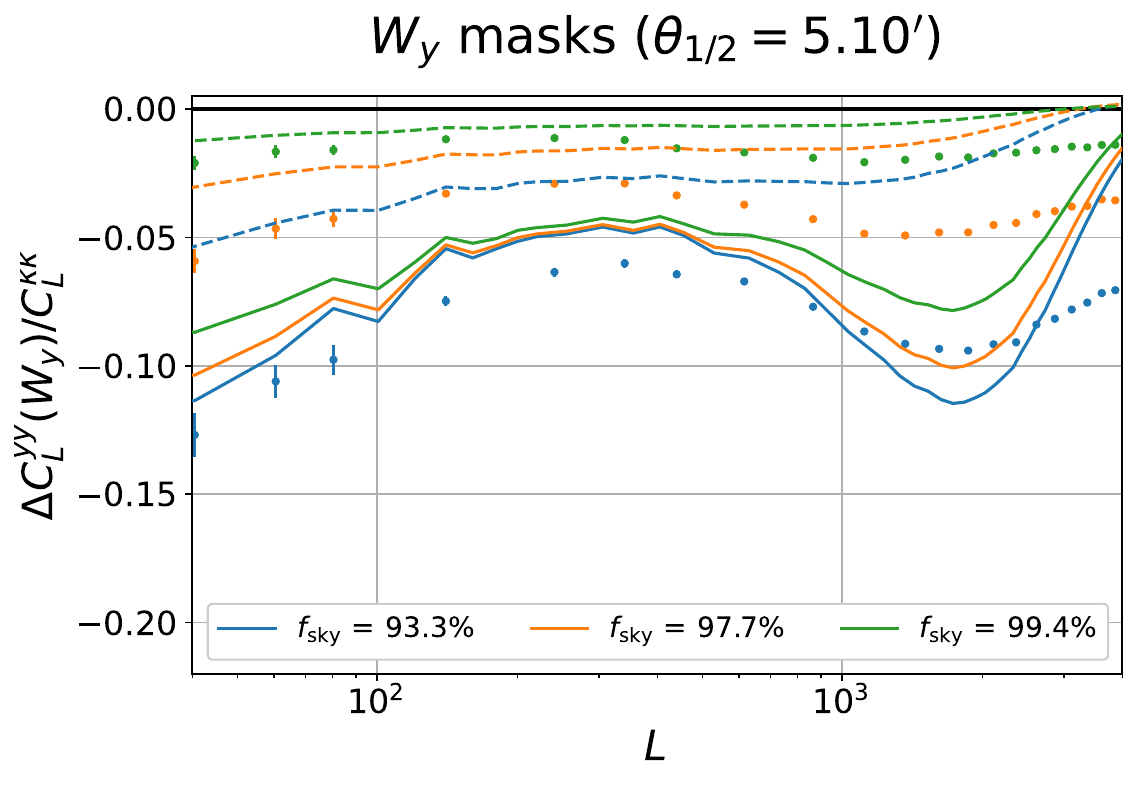} &
\includegraphics[width=0.33\textwidth]{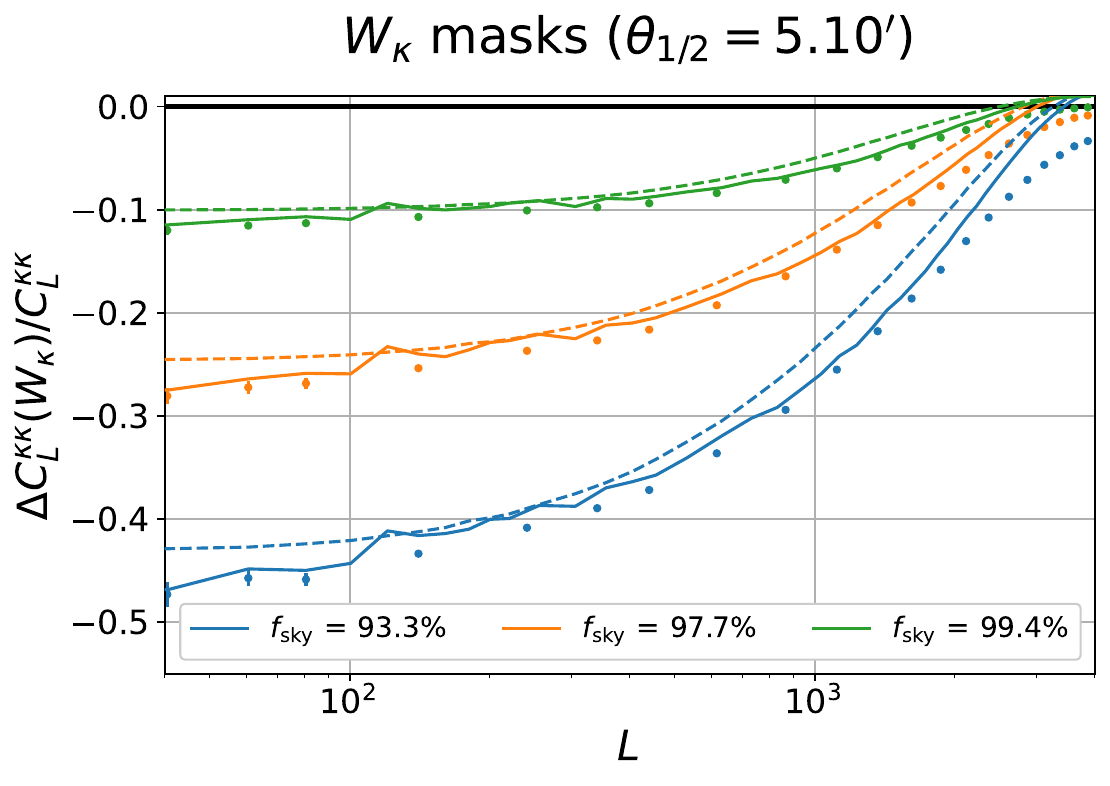}\\
\end{tabular}
\begin{tabular}{cccc}
\includegraphics[width=0.33\textwidth]{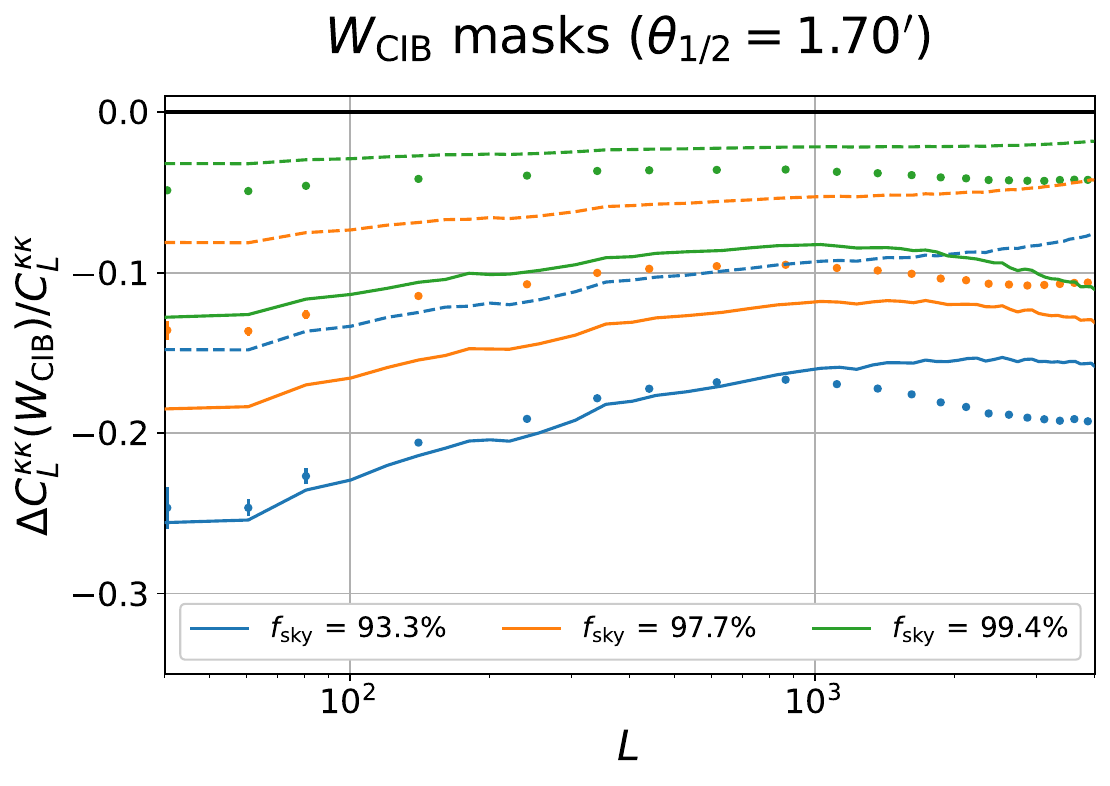}&
\includegraphics[width=0.34\textwidth]{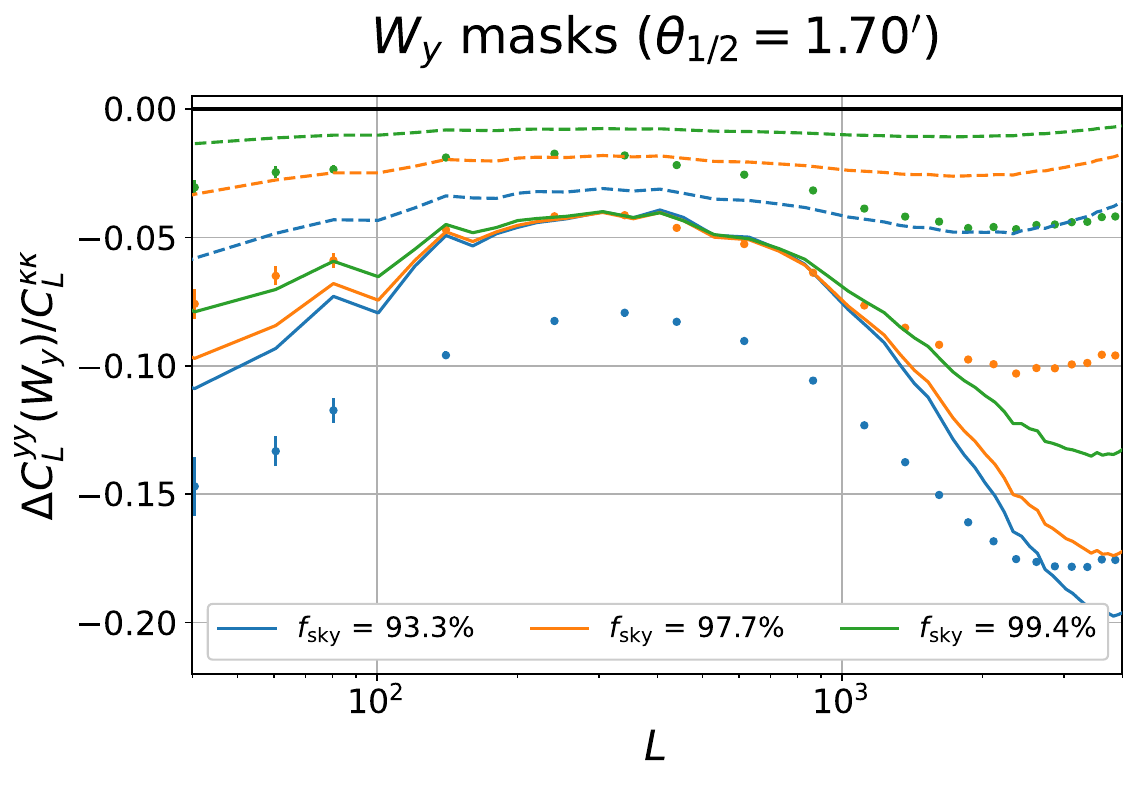} &
\includegraphics[width=0.33\textwidth]{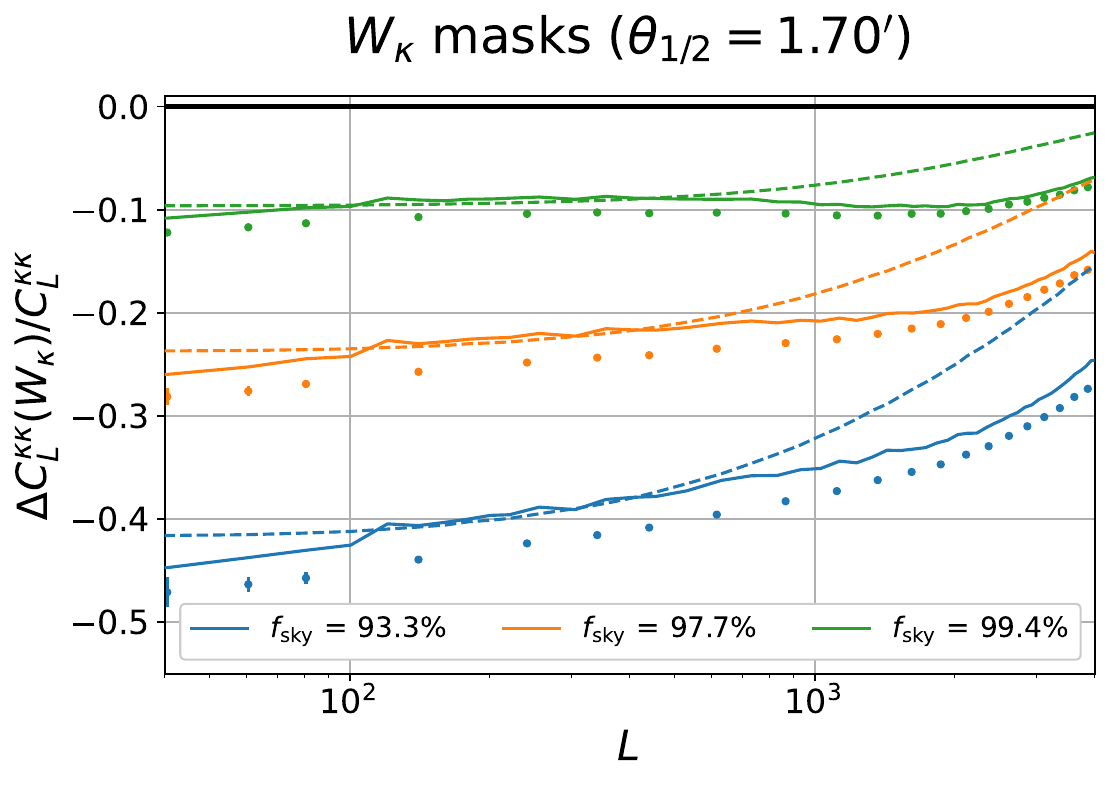}\\
\end{tabular}
\caption{
Effect of LSS-correlated masks on the CMB lensing convergence power spectrum $C_L^{\kappa \kappa}$ as a fractional difference between the $C_L^{\kappa \kappa}$ computed on the masked sky and its value computed on the full sky. Each panel shows results for different masks. Different colours show the result for masks retaining different sky fractions after masking. The  $W_\mathrm{halo}$ plot legend shows the mass limits of the halo samples used for the mask, and the $W_{\rm rs}$ plot legend shows source count detection limit used to construct the radio-source mask for the various listed experiments. Simulation measurements are shown as data points. For $W_\mathrm{halo}$ and $W_{\rm rs}$ masks, the solid lines show the analytic predictions of Sec.~\ref{sec:halomodel} and of Eq.~\eqref{eq:poisson_souces} respectively. For the $W_{\kappa}, W_{y}$ and $W_\mathrm{CIB}$ threshold masks,  the semi-analytic theory predictions are shown in solid and the pure Gaussian model in dashed. For the former we evaluated the expectation values in Eq.~\eqref{eq:ABf_bias} empirically from our simulations while the latter uses a fully analytic model. The middle and bottom rows show results obtained for different smoothings of the foreground fields.}
\label{fig:masking_kappa_noapo}
\end{figure*}

To measure the effect on simulations, we applied the foreground masks described in Sec.~\ref{sec:sims-masks}, $W_X$, to the \websky $\kappa$ map, and estimated the power spectrum over the unmasked area by deconvolving the effect of the mask using the MASTER \cite{Hivon:2001jp} algorithm as implemented in the publicly available \texttt{NaMaster}\footnote{\url{https://github.com/LSSTDESC/NaMaster}} package \cite{namaster}.
We then compared this mask-deconvolved power spectrum with the angular power spectrum of the \websky $\kappa$ computed on the full sky. The results are shown in Fig.~\ref{fig:masking_kappa_noapo}. For each spectra we adopted a variable binning in L: $\Delta_L = 1$ until $L_{\rm max}=10$, 4 bins of $\Delta_L = 20$ ($L_{\rm max}<100$), 4 bins of $\Delta_L = 100$ ($L_{\rm max}<500$), 14 bins of $\Delta_L = 250$ ($L_{\rm max}<4000$) and $\Delta_L = 1000$ for  $L_{\rm max}>4000$.
We estimated the error bars on the measured bias using a jackknife approach. For this purpose we divided the sky map into $N_\mathrm{patches}=48$
subregions of equal areas corresponding to pixels of an Healpix pixelization with nside=2. For the $i$-th subregion $W^{(i)}_{\rm cut}$ and for each LSS-correlated mask $W_X$, we estimated the fractional difference between the $C_L^{\kappa \kappa}$ computed on the sky masked with $W_{X}\cdot W^{(i)}_{\rm cut}$, and its value computed on the sky masked with $W^{(i)}_{\rm cut}$ only. The error bar is then given by the covariance of the fractional differences averaged over all the subregions rescaled by $N_\mathrm{patches} -1$ (see, e.g., Appendix B of Ref. \cite{Makiya:2018pda}).

Since all our masks select areas of the sky where mass over-densities are present, the recovered power spectrum has less power compared to its full sky value.
\refchange{We do not observe this loss of power when we apply the $W^{\rm rot}_{X}$ masks that are uncorrelated with the lensing signal.}

The right panel of Fig.~\ref{fig:halomodel} shows that the halo model describes reasonably well the effect observed when masking the \websky $\kappa$ map with $W_{\rm halo}$ as also shown in the top-right panel of Fig.~\ref{fig:masking_kappa_noapo}; in particular the analytical curves match well the shape of the biases observed in the simulations.
Changing  the halo mass function only gives fairly minor changes to the analytic predictions.
We note that since the halo mass functions and related bias model are usually formulated (or calibrated on simulations) in terms of $M_{200,m}$, while we perform cluster masking of $\kappa$ in terms of $M_{500,c}$, we can compare the curves in the two plots noting that $M_{500,c}\approx 2M_{200,m}$\footnote{We used the \texttt{colossus} code (\url{https://bitbucket.org/bdiemer/colossus/src/master/}) to perform an accurate conversion between the two mass definitions at a specific redshift in the integration.}.
For the $W_{\rm halo}$ mask, reducing $\mcut$ increases the number of objects that are masked and thus also the fraction of sky area that is masked.  This leads to a progressively larger power deficit. Such suppression is more relevant at very large scales and at $L \sim  2000$ in broad agreement with the analytical predictions of Sec.~\ref{sec:halomodel}.
The shape of the biases induced by $W_{\rm halo}$ and $W_y$ are similar. This is due to the fact that, even though they are build in different ways, they  are highly correlated as they mask similar objects and locations as discussed in Sec.~\ref{sec:masks}.

Proceeding clockwise in Fig.~\ref{fig:masking_kappa_noapo}, we show the results for $W_{\rm rs}$ masks. All the relative differences between the $\kappa \kappa$-spectrum computed on the masked sky and the $\kappa \kappa$-spectrum computed on the full sky are quite small but still not compatible with zero, considering the error bars. The theoretical curves, obtained using Eq.~\eqref{eq:poisson_souces}, are not in agreement with the simulation results. This means that the Gaussian foreground peaks model is not completely able to reproduce the behaviour of these radio source masks, consistent with non-Gaussian effects dominating the purely Gaussian predictions.

For the threshold masks, second and third rows of Fig.~\ref{fig:masking_kappa_noapo}, the dashed lines are the theory curves obtained using the pure Gaussian model described in Sec.~\ref{sec:gaussianmodel}, while the solid lines represent the semi-empirical form of Eq.~\eqref{eq:ABf_bias}, where the spectra and cross-spectra of $W(\vx), (fW)(\vx)$ and $(f^2W)(\vx)$ have been computed directly from our set of foreground masks and the \websky $\kappa$, $y$ and CIB fields. Overall, the semi-empirical Gaussian model using the actual masks and foreground fields seems to match better the data points, in particular for $W_{\kappa}$, even though, particularly for $W_{\rm CIB}$ and, even more, $W_y$ masks, there is a clear disagreement at small scales. This is not surprising as $\kappa$ is the most Gaussian of the three foreground fields considered here. The CIB contains a significant shot-noise contribution from individual bright infrared galaxies, and the $y$ from highly collapsed galaxy clusters that induce significant non-Gaussianities in the maps.
We investigate this disagreement for the threshold masks in Appendix~\ref{sec:threshold_masks}.

\section{Lensing reconstruction on masked fields}
\label{sec:recon}

In Sec.~\ref{sec:matter_1} we have directly masked the CMB lensing convergence to roughly quantify the bias induced by the foreground masks described in Sec.~\ref{sec:sims-masks}. However, the true $\kappa$ field is not a direct observable, and has to be reconstructed from the observed CMB maps potentially also in combination with other external matter tracers. Hence, in this section, we quantify the impact of LSS-correlated masks removing the corresponding regions in lensed CMB maps that are then used to perform lensing reconstruction using a quadratic estimator.
For this first analysis we only use the CMB temperature field, since the foreground contamination is less important for polarization and the properties of the polarized sources are currently much less well understood \cite{Gupta:2019kng}.

For the reconstruction of the lensing potential, we mainly followed the same steps as the \Planck~lensing pipeline described in Ref.~\cite{Aghanim:2018oex}. For CMB temperature lensing reconstruction, this is just an optimized and generalized version of the lensing quadratic estimator (QE) of Ref.~\cite{Okamoto:2003zw}. We used the reconstruction pipeline implemented in the public {\tt plancklens} code\footnote{\url{https://github.com/carronj/plancklens}}, and we
refer the reader to Ref.~\cite{Aghanim:2018oex} for more details.
The procedure can be summarized in 4 steps: 1) Optimal filtering of input ``data'' CMB maps;
2) Construction of the lensing quadratic estimator; 3) Mean-field subtraction and normalization of the lensing estimate; 4) Computation of the lensing power spectrum, subtraction of its additive biases, and Monte-Carlo correction of its normalization.

To measure the mask biases in lensing reconstruction, we used the two sets of Monte Carlo simulations of lensed CMB realizations describe in Sec.~\ref{sec:MC_sims}.
We used the NG set to isolate the bias as it would appear on real data, while the G set was used to
compute the mean-field, the RD-$N^{(0)}$ noise bias, and the multiplicative Monte Carlo correction assuming no mask-lensing correlation. We also used the G set to estimate error bars on the auto- and cross-spectra estimators described in the following.
Since the \websky suite includes only a single realization of $\kappa$, our numerical results based on the NG set are limited on large scales by the cosmic variance of this fixed realization of the lensing field.

In the following, we considered two experimental setups representative of an SO-like and an S4-like survey. For SO we assumed an effective white noise level in temperature of $\sim$6.7 $\mu$K-arcmin and $\theta_{1/2}=1.5^\prime$, consistent with the publicly available effective baseline noise configuration after component separation\footnote{Details of the noise model for SO can be found at \url{https://github.com/simonsobsso_noise_models}.
Note that the beam size is not entirely consistent with the $1.7^\prime$ fiducial foreground smoothing that we use, which was chosen to match the smoothing used in \PaperI.}.
For the S4-like survey, we assumed a beam with $\theta_{1/2}=1.0^\prime$, and isotropic uncorrelated $\sim$1 $\mu$K-arcmin noise for temperature\footnote{Details can be found at \url{https://cmb-s4.uchicago.edu/wiki/index.php/Survey_Performance_Expectations}}.
The results shown in the following are computed using SO-like experimental specifications. In addition, we  considered an S4-like experimental setup for a subset of foreground masks: the $W_{\rm rs}$ mask with $f_{\rm sky} = 94.1 \%$, which includes all the sources with a measured flux above the detection limit for S4, the $W_{\mathrm{halo}}$ with $M^{\mathrm{cut}}=10^{14}\Msun/h$, which removes a mass-limited tSZ-selected cluster sample for an S4-like survey \cite{Raghunathan:2021zfi}, and the extreme cases of $W_{\mathrm{CIB}}$ and $W_{y}$ with $f_{\rm sky} =93.3\%$ and $\theta_{1/2}=1.7^\prime$.

\begin{figure*}[!]
\centering
\includegraphics[width =  \textwidth]{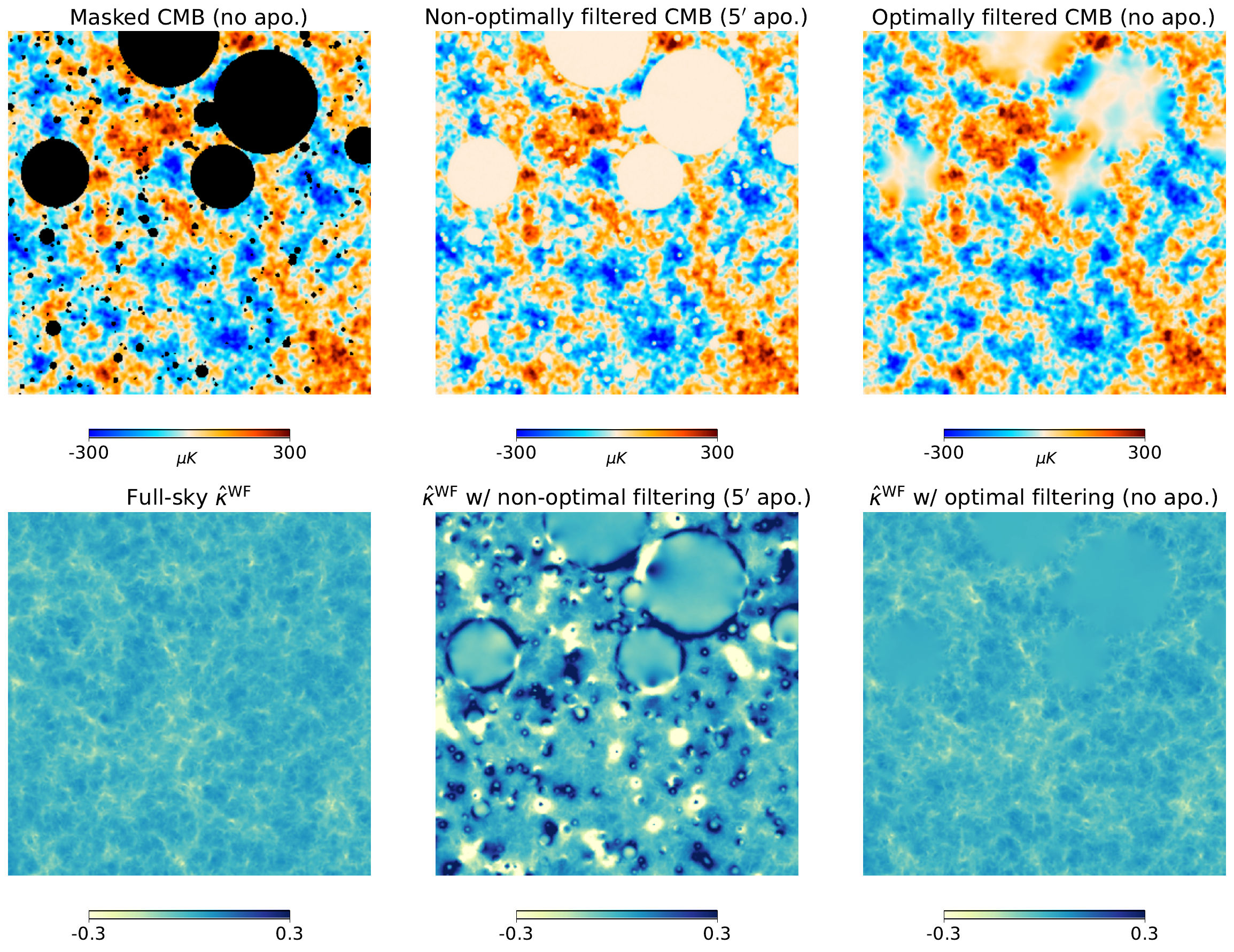}
\caption{Comparison between optimal and non-optimal Wiener filtering. Top panel, from left to right, shows the masked CMB field ($W_{\rm halo}$ with $f_{\rm sky} = 94.7\%$), the filtered CMB field using a non-optimal isotropic method, and the filtered CMB field using optimal filtering as described in Refs.~\cite{Smith:2007rg, Aghanim:2018oex}. Bottom panel, from left to right, shows the Wiener-filtered full-sky reconstruction $\kappa$ map, and the filtered reconstructed $\kappa$ map using non-optimal and optimal filtering. In the non-optimal filtering case, the mask has been apodized using the $C^2$ function (effectively a cosine) implemented in NaMaster with an apodization scale  of $5^\prime$. This helps to slightly reduce the ringing effects, but large hole-induced reconstruction noise edge effects still dominate. In all the $\kappa$ maps a significant fraction of the small-scale structure is the non-Gaussian reconstruction noise.}
\label{fig:optimalfiltering}
\end{figure*}

After adding a realization of the isotropic noise, the input maps are then masked using all the unapodized $W_X$ masks discussed in Sec.~\ref{sec:masks}. The (optimal) filtering step produces Wiener-filtered maps that provide the minimum variance estimate of the full-sky lensed CMB based on the information in the unmasked area. Optimal filtering~\cite{Smith:2007rg,Aghanim:2018oex,Mirmelstein:2019sxi} is particularly valuable with this kind of mask compared to more basic (but faster) inverse-variance-weighted or isotropic filtering. Figure~\ref{fig:optimalfiltering} shows how the optimal filtering operation is able to fill back some information inside small masked regions, effectively recovering information that was masked
(assuming no residual foregrounds outside the masked area). This increases the information available, reduces any complications due to sharp mask cuts, and because the masked area is effectively reduced, can substantially reduce biases when the mask is correlated to the signal.

To build the filtered CMB maps, we used a fiducial lensed CMB temperature spectrum
including multipoles $100 \leq \ell \leq 4000$. Removing multipoles $\ell < 100$ generates little loss of information for lensing reconstruction. Since we approximate the noise as white and isotropic, the noise in the filter is also taken to be isotropic and consistent with the simulated value.

The real-space lensing deflection estimator is then built from a pair of filtered maps discussed above. The gradient part\footnote{The curl component is expected to be zero to a good approximation, and is zero in the \websky lensing field by construction.} of the quadratic lensing deflection estimator, $\hat g_{LM}$, contains information on the lensing potential, which is estimated using \cite{PL2018}
\begin{equation}\label{eq:phi-estimator}
\hat \phi_{LM}\,\equiv \, \frac{1}{\mathcal{R}^{\phi}_L} \Big(\hat g_{LM} - \gMF \Big) \,,
\end{equation}
where $\mathcal{R}^{\phi}_L$ is the non-perturbative response function defined to make the lensing reconstruction unbiased on the full sky~\cite{Hanson:2010rp,Lewis:2011fk,Fabbian:2019tik}, and $\langle \hat g^{\rm MF}_{LM} \rangle$ is the mean field of the estimator.
Note that we are constructing our lensing estimators in the hypothesis of no lensing-mask correlation, as appropriate for quantifying the bias on standard methods.
\refchange{We compute the mean field using the simulations of the G set rather than, for example, defining the mean field by averaging over simulations with lensing fields correlated to the fixed mask.}
Since the mean field depends on the mask, for lensing-correlated masks the mean field is actually correlated to the true lensing potential $\phi$.

\begin{figure}[!]
\includegraphics[width = 0.95 \columnwidth]{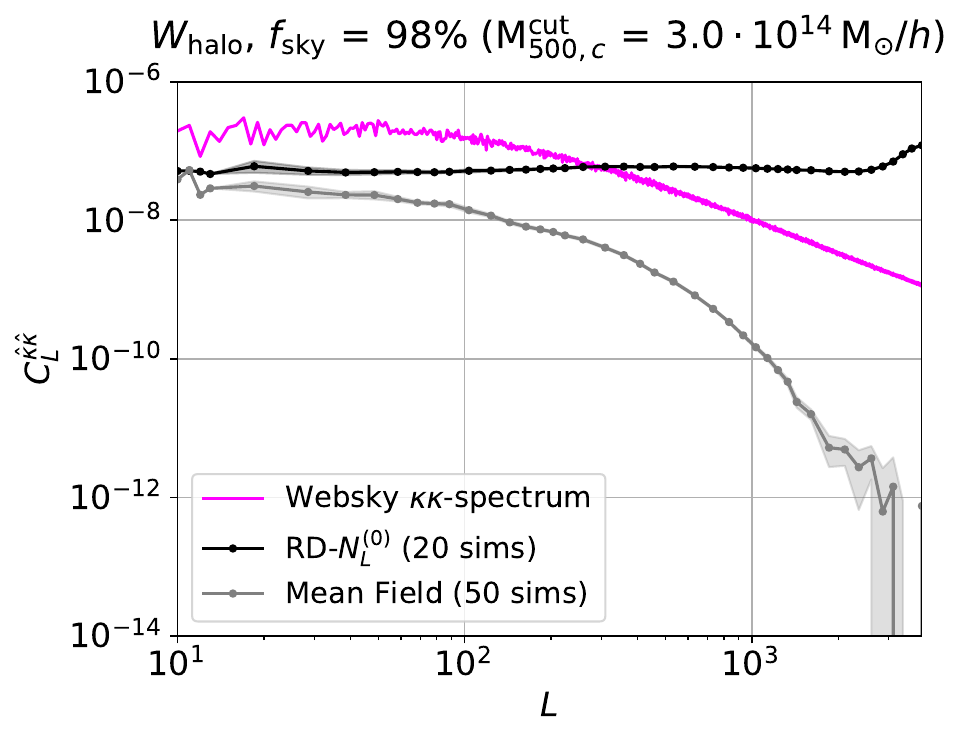}
\caption{The RD-$N^{(0)}_L$ bias (black, as described in Eq.~\eqref{eq:RD-N0}) and mean-field power spectrum  corrected for the response (grey) for the case of the fixed halo mask described in the plot title.
We show (in purple) the \websky full-sky $C_{L}^{\kappa \kappa}$ for comparison.
}
\label{fig:N0-meanfield_3.0halomask}
\end{figure}

In our analysis, we make two estimates of the (uncorrelated) mean field for each mask using two different independent sets of independent 25 G simulations, which we denote $\gMF_1$ and $\gMF_2$ respectively (using 50 G simulations in total, 25 for each mean field).
 The lensing power spectrum is estimated by cross-correlating two lensing map estimates $\hat \phi_1$ and $\hat \phi_2$ following Eq.~\eqref{eq:phi-estimator}, where we subtracted $\gMF_1$ and $\gMF_2$ respectively to avoid reconstruction noise in the mean field cross-correlation.
The estimate of the lensing power spectrum of the single realization is then obtained as
\begin{equation}\label{eq:lensing-power-spectra-esimator}
\hat C_L^{\hat \phi_1 \hat \phi_2} \,\equiv \, \frac{1}{(2L +1) f_{\rm sky}} \sum_{M=-L}^L \hat \phi_{1,LM}^* \hat \phi_{2,LM} \,,
\end{equation}
where $f_{\rm sky} = \sum_p {W_X}_p/N_{\rm pix}$ is the unmasked sky fraction.

From the above estimator, we then subtract the realization-dependent estimate of the Gaussian (disconnected) lensing bias, RD-$N^{(0)}_L$. Subtracting this term, together with the mean-field subtraction, has the effect of removing the disconnected signal expected from Gaussian fluctuations even in the absence of lensing \cite{Hanson:2010rp,Namikawa:2012pe}. The RD-$N^{(0)}_L$ bias is defined as
\begin{equation}\label{eq:RD-N0}
\text{RD-}N^{(0), d }_L \,\equiv\, \langle 4 \hat C^{di}_L - 2\hat C^{ij}_L \rangle \,,
\end{equation}
where angle brackets denote an average over pairs of distinct $i, j$  G simulations.
Here $\hat C^{di}_L$ is the estimator in Eq.~\eqref{eq:lensing-power-spectra-esimator} with $\hat \phi_1$ reconstructed from the $d$-th CMB realization of the NG set and  $\hat \phi_2$ using the $i$-th simulation of the G set; $\hat C^{ij}_L$ is instead the estimator of Eq.~\eqref{eq:lensing-power-spectra-esimator} with $\hat \phi_1$ and  $\hat \phi_2$ reconstructed from the $i$-th and the $j$-th simulations of the G set, respectively.
For each mask (both LSS-correlated and uncorrelated), RD-$N^{(0), d}$ is calculated for each NG `data' simulation, $d$, using 20 different pairs of independent G simulations.

Figure~\ref{fig:N0-meanfield_3.0halomask} shows a typical mean-field power spectrum corrected for the non-perturbative response and an example of the RD-$N^{(0), d}$.

For a given lensing reconstruction field, we can also define the cross-spectrum estimator with the true field
\begin{equation}\label{eq:lensing-cross-spectra-esimator}
C_L^{\hat \phi  \phi} \,\equiv \, \frac{\fMC}{(2L +1) f_{\rm sky}} \sum_{M=-L}^L \hat \phi_{LM}^*  \phi_{LM} \,,
\end{equation}
where $\hat \phi$ is the lensing potential estimate of Eq.~\eqref{eq:phi-estimator}, $\phi$ is the true lensing potential field, and $\fMC$ is a Monte-Carlo (MC) correction defined to make the estimator unbiased for Gaussian lensing potential uncorrelated with the mask. Specifically, we define
\begin{equation}\label{eq:fMC_correction}
\frac{1}{\fMC} \,\equiv \, \left\langle \frac{[C^{\phi\phi}_L]^{-1}}{(2L +1) f_{\rm sky}} \sum_{M=-L}^L \hat \phi_{LM}^*  \phi_{LM} \right\rangle_{50\, {\rm G \,sims}} .
\end{equation}
In practice, this ratio is calculated for the binned spectra rather than individual $L$, using the binning scheme described in Sec.~\ref{sec:matter_1}.
Note that for large (e.g., galactic) masks $\fsky$ is a reasonable approximate normalization so that $\fMC\approx 1$. However, when the mask contains a large number of small holes due to, e.g., point source masking, the $\fsky$ correction becomes a much worse approximation as effectively much less mask area is lost after optimal filtering and reconstruction. We discuss this aspect in more detail in Appendix \ref{sec:holes} and \ref{sec:GAUSS}. Here, the inclusion of $\fsky$ does not affect the result, since it simply amounts to a redefinition of $\fMC$.

We also define the lensing power spectrum estimator
\begin{equation}\label{eq:auto_after-recon}
 C^{\hat \phi_1 \hat \phi_2,\,{\rm RD}}_{L} \,=\, \big(\fMC\big)^2\,\Big( \hat C^{\hat \phi_1 \hat \phi_2}_L \,-\, \text{RD-}N^{(0)}_L \Big) \,,
\end{equation}
where $\fMC$ is defined as above.
However, the $(\fMC)^2$ normalization calibrated on the cross-spectrum may not be the correct normalization factor for the auto-spectrum.
Moreover, even in presence of Gaussian lensing fields and LSS-uncorrelated masks, the estimator in Eq.~\eqref{eq:auto_after-recon} does not provide an unbiased estimate of the CMB lensing power spectrum as it still retains the $N^{(1)}_L$ noise bias induced by signal-dependent contractions~\cite{Kesden:2003cc}. For non-Gaussian lensing field, the estimator also retains the noise term involving the 3-point function of the lensing field (\nlth).
Here we regard $N^{(1)}_L$ and \nlth as part of the signal contained in $ C^{\hat \phi_1 \hat \phi_2,\,{\rm RD}}_{L}$, and only consider differences between the estimated power spectrum (that includes these additional noise bias terms) when using LSS-correlated and uncorrelated masks.

We run the entire end-to-end estimation pipeline 20 times for each mask (both correlated and uncorrelated), taking the `data' each time to be one of the NG lensed CMB simulations. Results are then averaged over these simulations to reduce Monte Carlo noise from variations of the unlensed CMB. However, the cosmic variance of the lensing field is not reduced since all realizations in the NG set share the same single \websky $\kappa$ simulation that is currently available.
To make a comparison with the results obtained in Sec.~\ref{sec:matter}, we plot results for the reconstructed convergence field $\hat \kappa$ instead of $\hat \phi$ \footnote{Note that $\hat \kappa_{LM} \equiv \frac{1}{2} L(L+1) \, \hat \phi_{LM}$}.

\begin{figure*} [t!]
\centering
\begin{tabular}{cccc}
\includegraphics[width=0.32\textwidth]{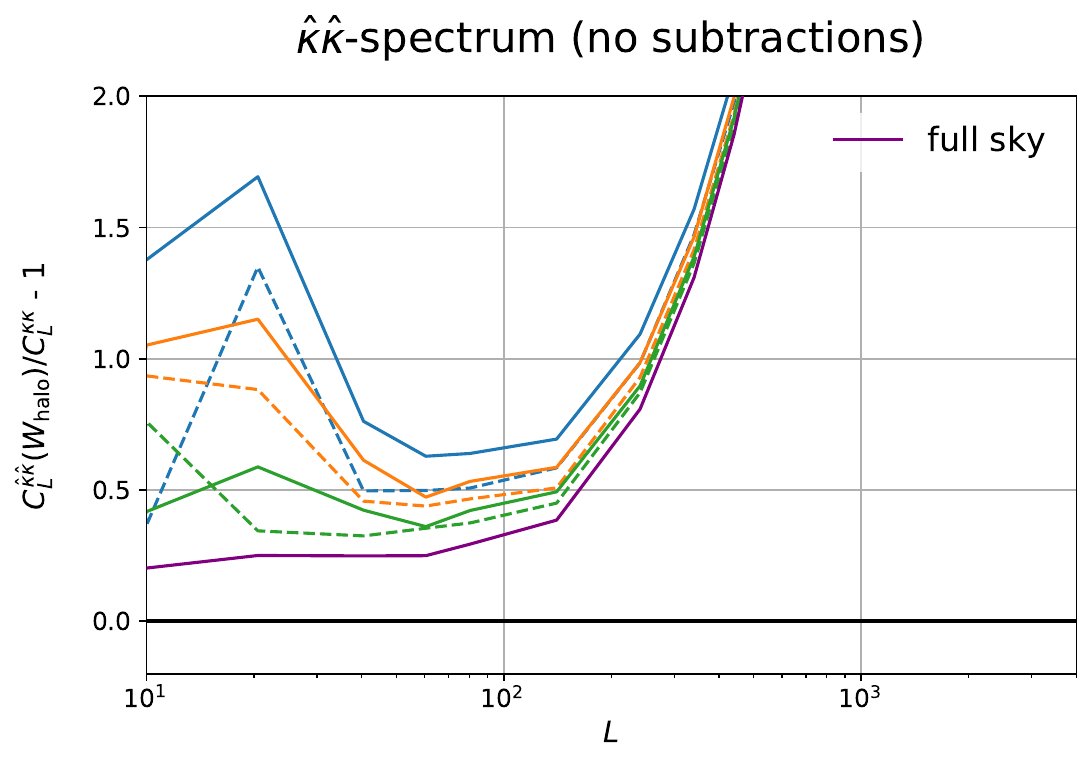} &
\includegraphics[width=0.32\textwidth]{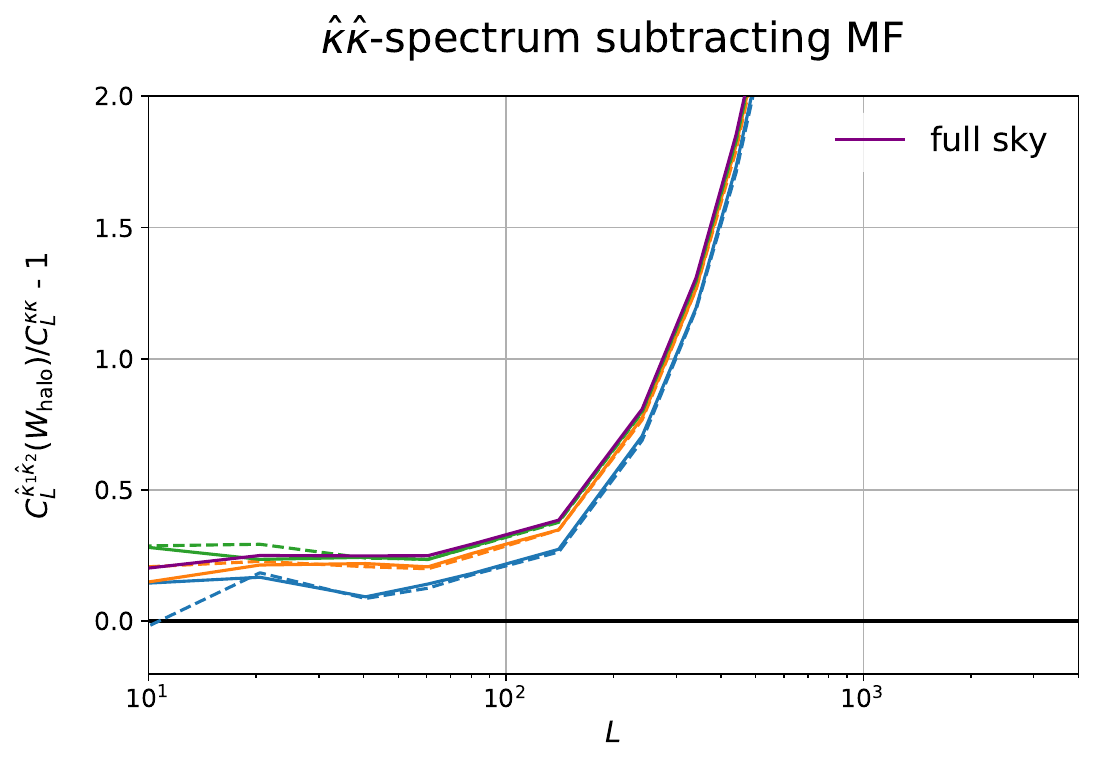} &
\includegraphics[width=0.32\textwidth]{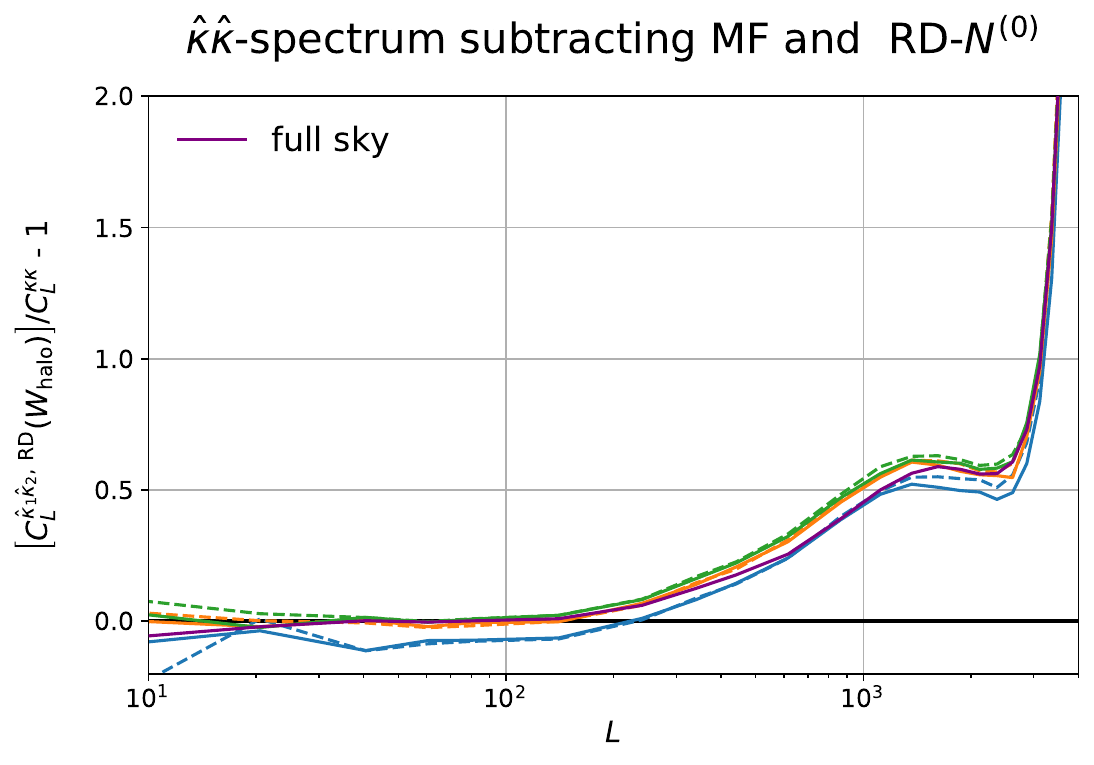}\\
\end{tabular}
\begin{tabular}{cccc}
\includegraphics[width=0.321\textwidth]{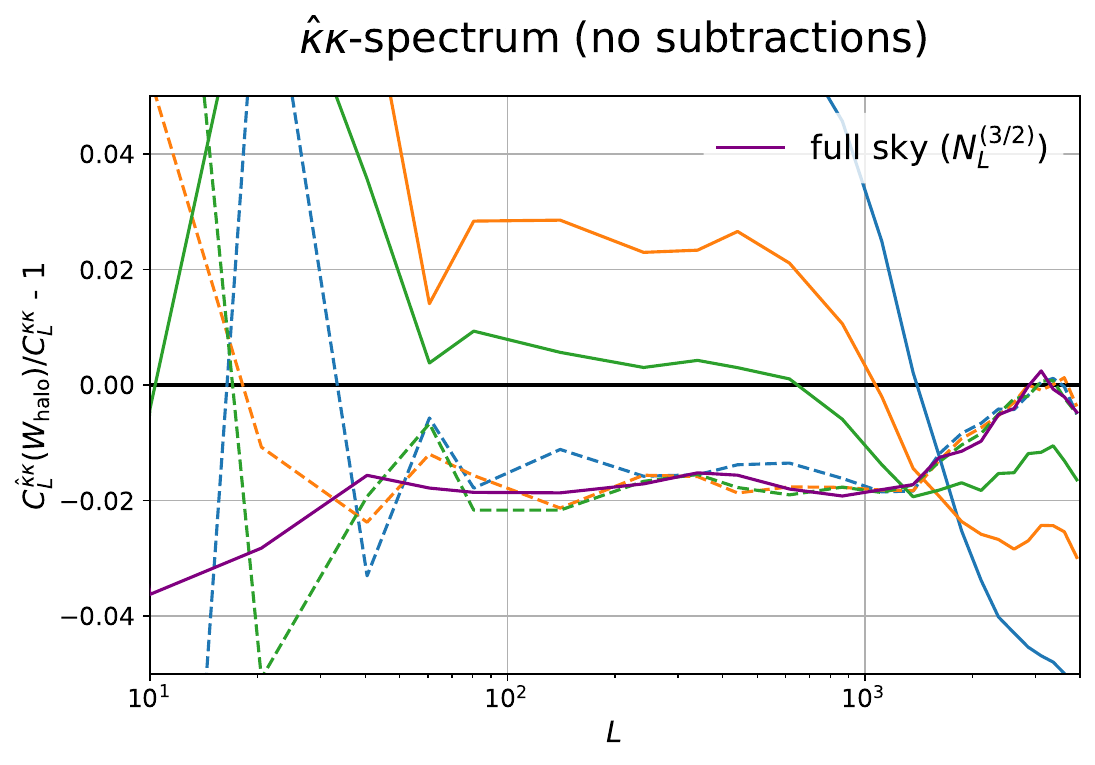}&
\includegraphics[width=0.61\textwidth]{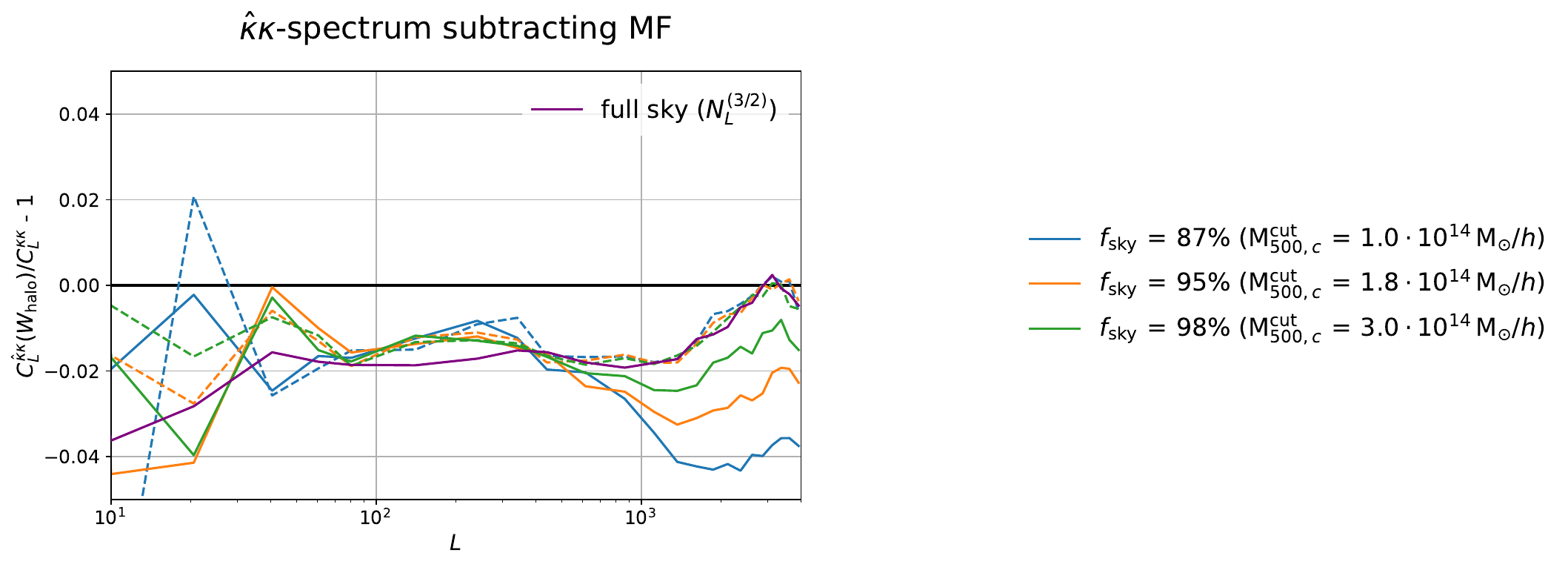} $\quad \,\,\,\,\,\,$\\
\end{tabular}
\caption{\emph{Top:} the left panel shows the raw reconstructed lensing auto-spectra, while the middle and right panel show the reconstructed auto-spectra after subtracting the mean field alone and the mean-field and the RD-$N^{(0)}_L$ noise bias respectively. \emph{Bottom:} the reconstructed cross-spectra with and without the mean-field subtraction are shown in the left and right side respectively. All these curves include the Monte Carlo normalization correction, $\fMC$.
The results are obtained using the $W_{\rm halo}$ masks for different values of the mass detection threshold, $\mcut$, whose corresponding $f_{\rm sky}$ is reported in the legend.
The purple lines correspond to the full-sky analysis, where, for the cross-correlation, the final difference is consistent with \nlth\ reconstruction bias, and for the auto-spectrum is dominated by $N^{(1)}$. Dashed-lines show the results obtained using the LSS-uncorrelated masks, $W^{\rm rot}_{\rm halo}$.
}
\label{fig:lens-rec_auto_Whalo}
\end{figure*}

The effect of the mean-field and RD-$N^{(0)}_L$ bias on the reconstructed $\hat \kappa \hat \kappa$-spectra is shown in the first row of Fig.~\ref{fig:lens-rec_auto_Whalo} as relative differences between the reconstructed $\hat \kappa \hat \kappa$-spectrum and the true \websky $ \kappa  \kappa$-spectrum for three different $W_{\rm halo}$ masks. From left to right, the reconstructed auto-spectra are plotted without mean-field and RD-$N^{(0)}_L$ corrections, then subtracting the mean field only, and finally subtracting both mean field and the RD-$N^{(0)}_L$. The main correction to the reconstructed $\hat \kappa \hat \kappa$-spectra comes from subtracting the RD-$N^{(0)}_L$ bias. The rise on small scales is related to the unsubtracted \nlone. The second row of Fig.~\ref{fig:lens-rec_auto_Whalo} shows the effect of the mean-field subtraction on the relative differences between the cross spectrum $\hat \kappa \kappa$ and the true \websky $ \kappa  \kappa$-spectrum.
Note that for correlated masks, the mean field is also important for the cross-correlation since the mask, and hence the mean field, is correlated to the signal.
The bias induced by the mask on the $\hat \kappa \kappa$ spectra is strongly reduced when we remove the mean-field term, especially at intermediate scales where the difference is then dominated\footnote{Note that the \nlth\ signal is overestimated because the \websky simulations do not include post-Born lensing, which largely has an opposite sign.} by \nlth. In all the plots of Fig.~\ref{fig:lens-rec_auto_Whalo}, the full-sky cases are shown in purple as reference. Similar results are obtained for all the other masks described in Sec.~\ref{sec:sims-masks}.

\section{Numerical results}\label{sec:cmblens-results}
\subsection{CMB lensing power spectrum}
To isolate the effects due to the correlation between the mask and the lensing field, we computed the difference between the two-point correlation function of the reconstructed CMB lensing obtained with the $W_X$ masks and the one obtained with the rotated uncorrelated masks, $W^{\rm rot}_X$. The rotated results have no correlated mask effects, but retain all the other non-trivial mode-coupling effects due to cut sky and hole shapes, as well as \nlth\ and \nlone\ to the extent that they are not modified by an LSS-correlated mask.

\begin{figure*} [t!]
\centering
\begin{tabular}{cccc}
\includegraphics[width=0.35\textwidth]{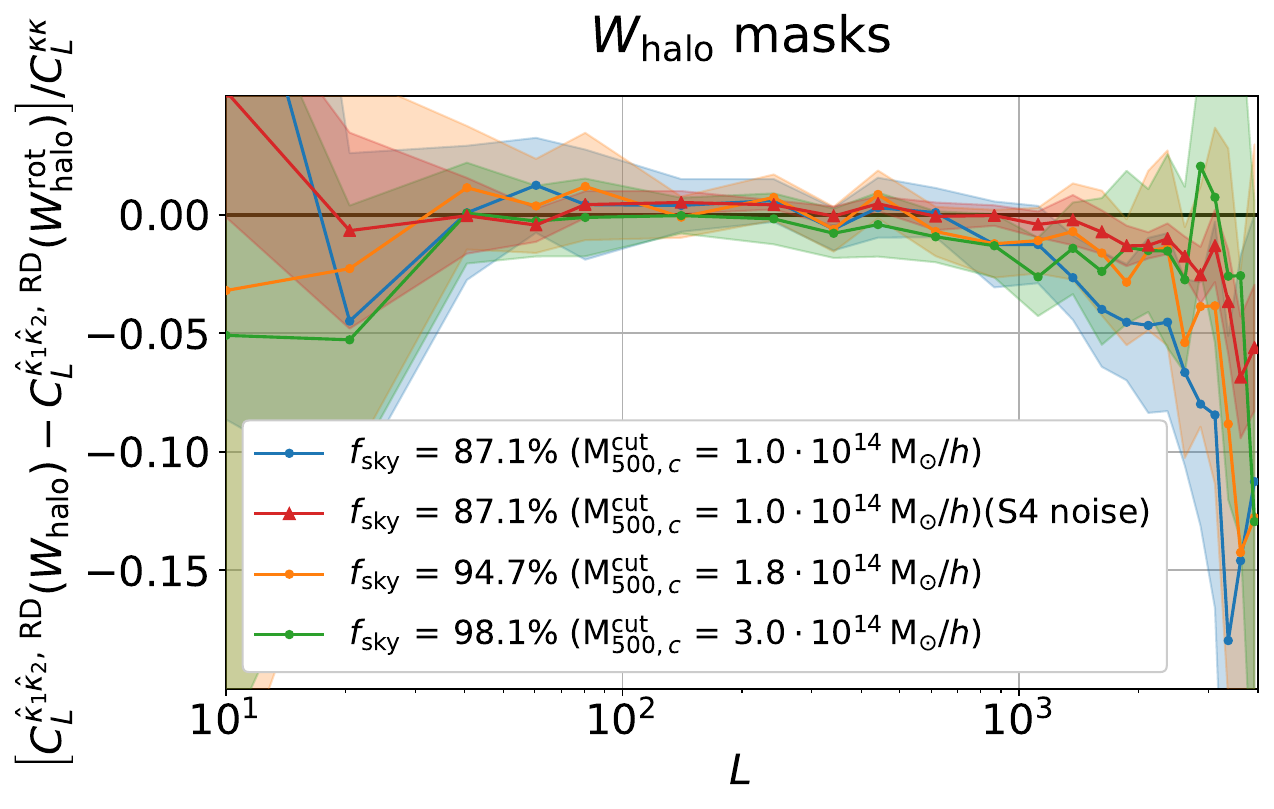}&
\includegraphics[width=0.35\textwidth]{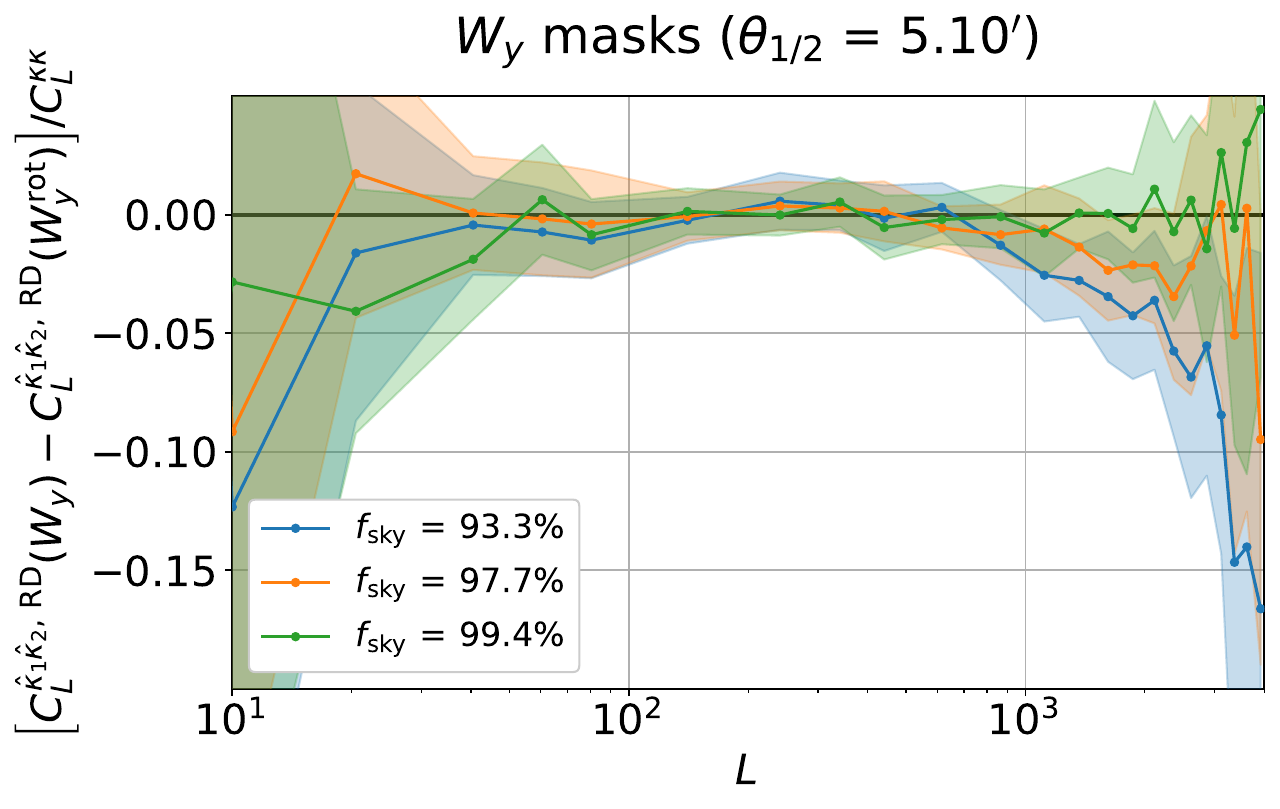}\\
\end{tabular}
\begin{tabular}{cccc}
\includegraphics[width=0.33\textwidth]{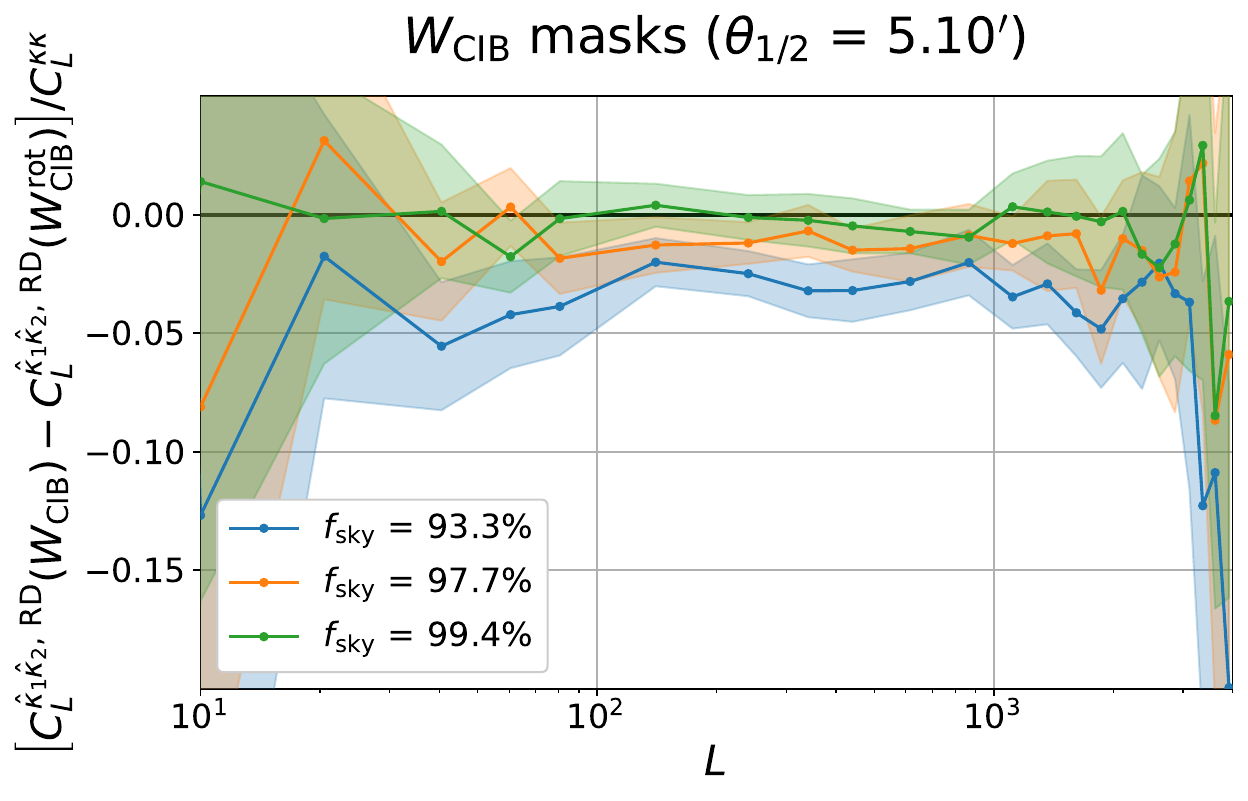} &
\includegraphics[width=0.33\textwidth]{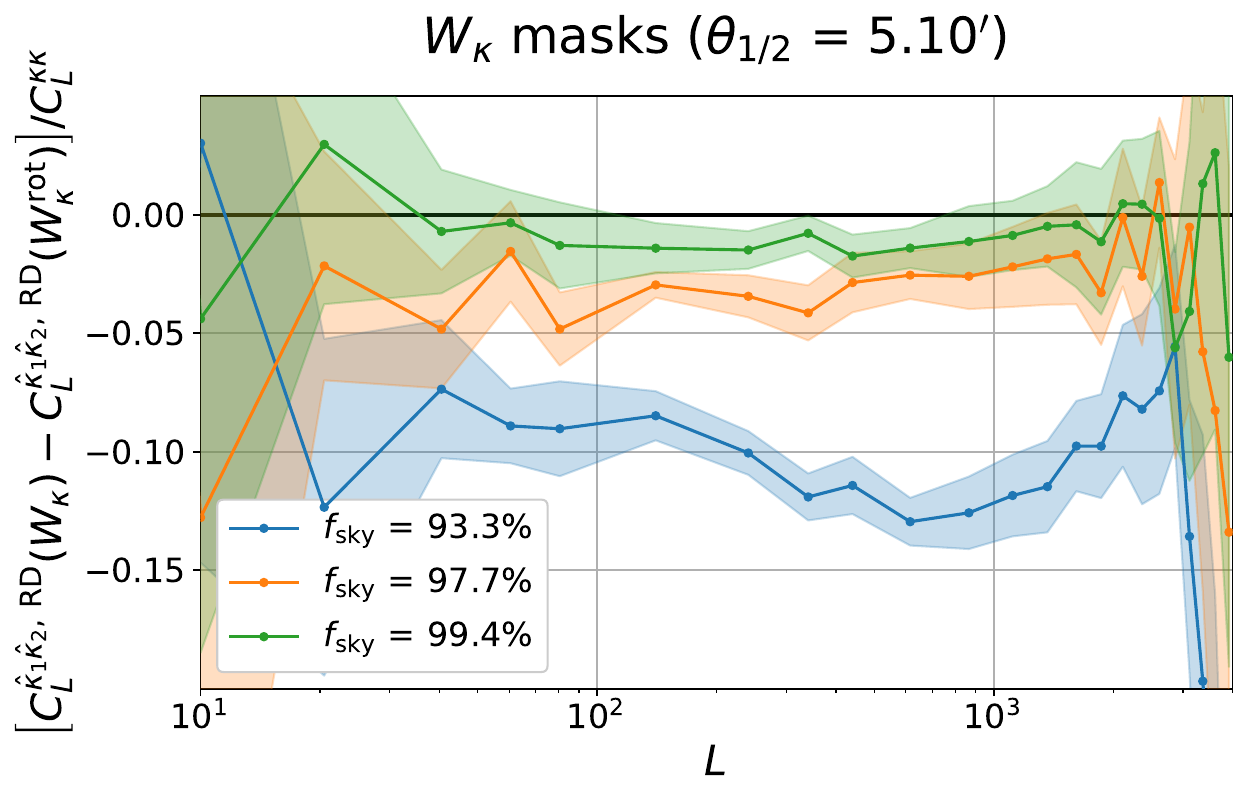} &
\includegraphics[width=0.33\textwidth]{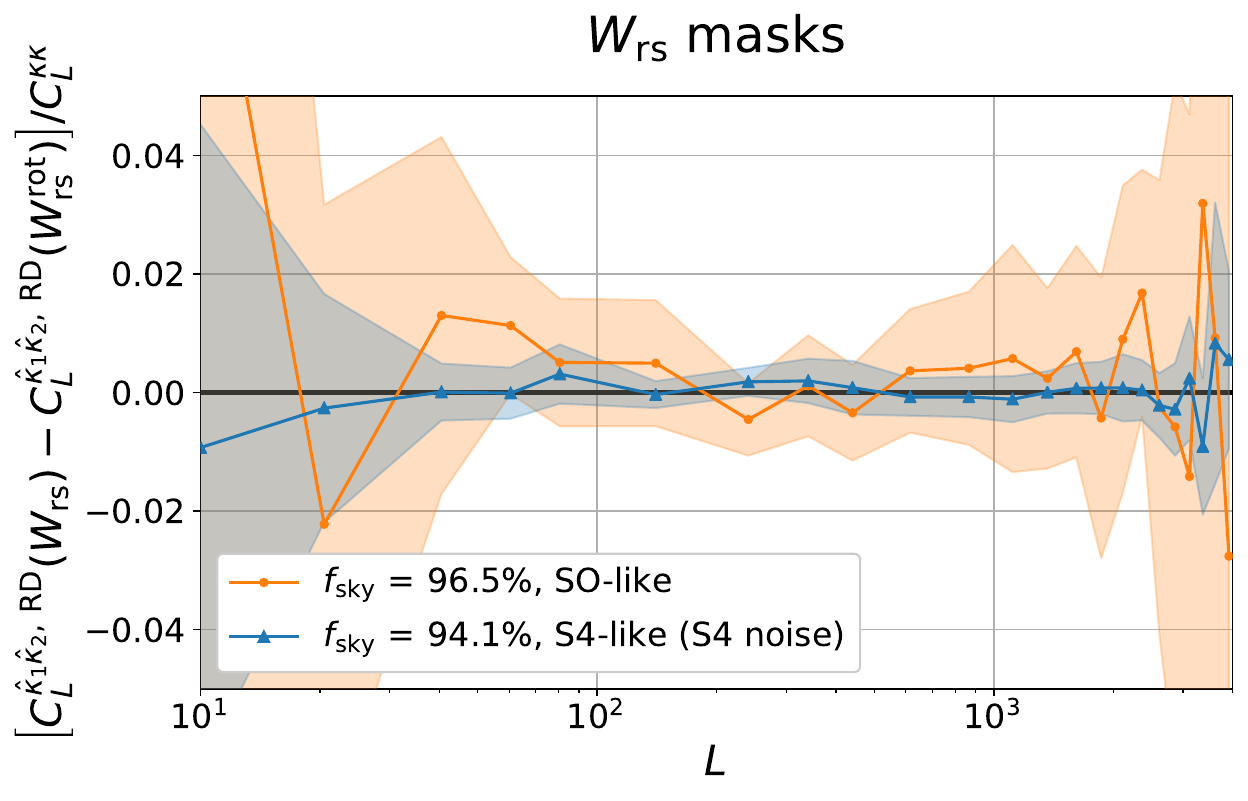}\\
\end{tabular}
\caption{Effect of LSS-correlated masking on the reconstructed CMB lensing convergence auto spectrum obtained with the estimator of Eq.~\eqref{eq:auto_after-recon}. The mask biases are computed as the differences between reconstructions performed with correlated and uncorrelated masks. The plots show the amplitude of the bias relative to the true lensing power spectrum. Different LSS-correlated masks are shown in different panels.  The foreground fields have been smoothed with a $\theta_{1/2}=5.1^\prime$ Gaussian beam prior to thresholding. Unless stated otherwise, we assumed an SO-like CMB temperature noise in the lensing reconstruction.
}
\label{fig:lens-rec_auto}
\end{figure*}

\begin{figure*} [t!]
\centering
\begin{tabular}{cccc}
\includegraphics[width=0.35\textwidth]{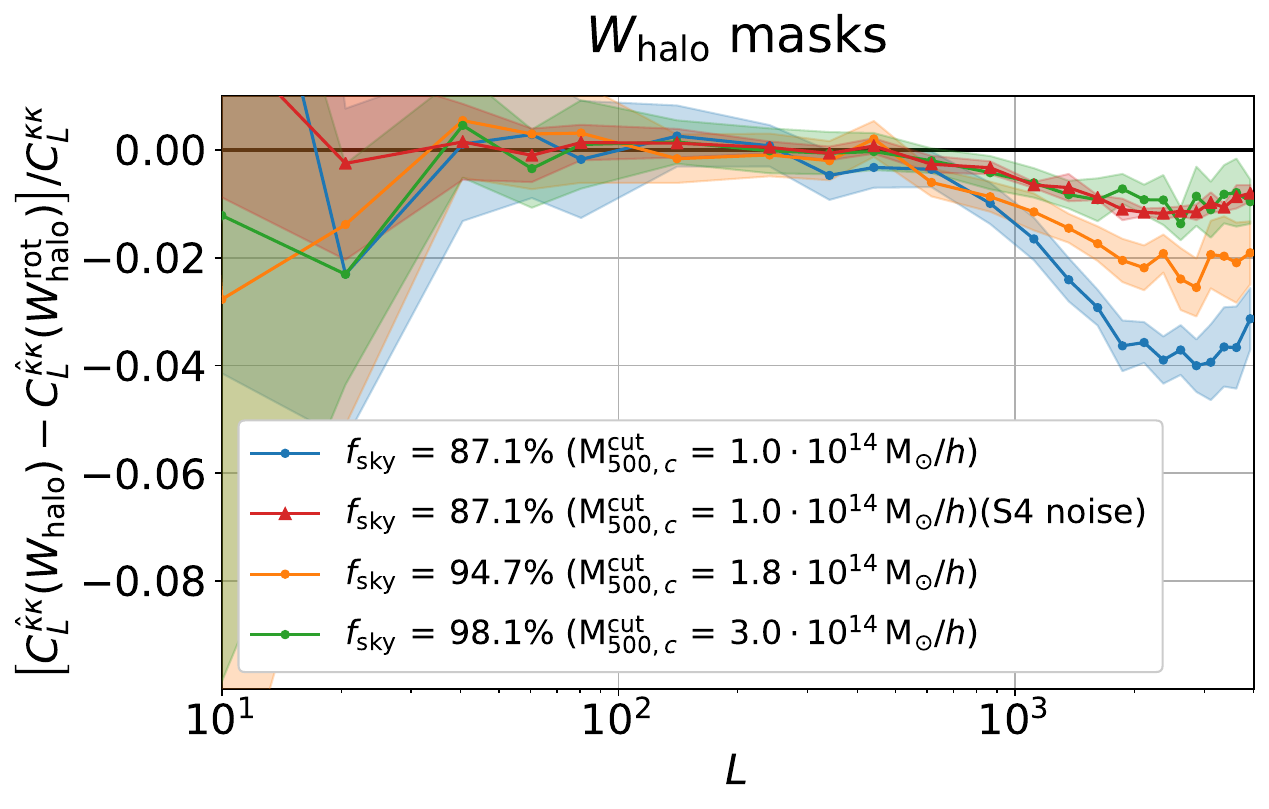}&
\includegraphics[width=0.35\textwidth]{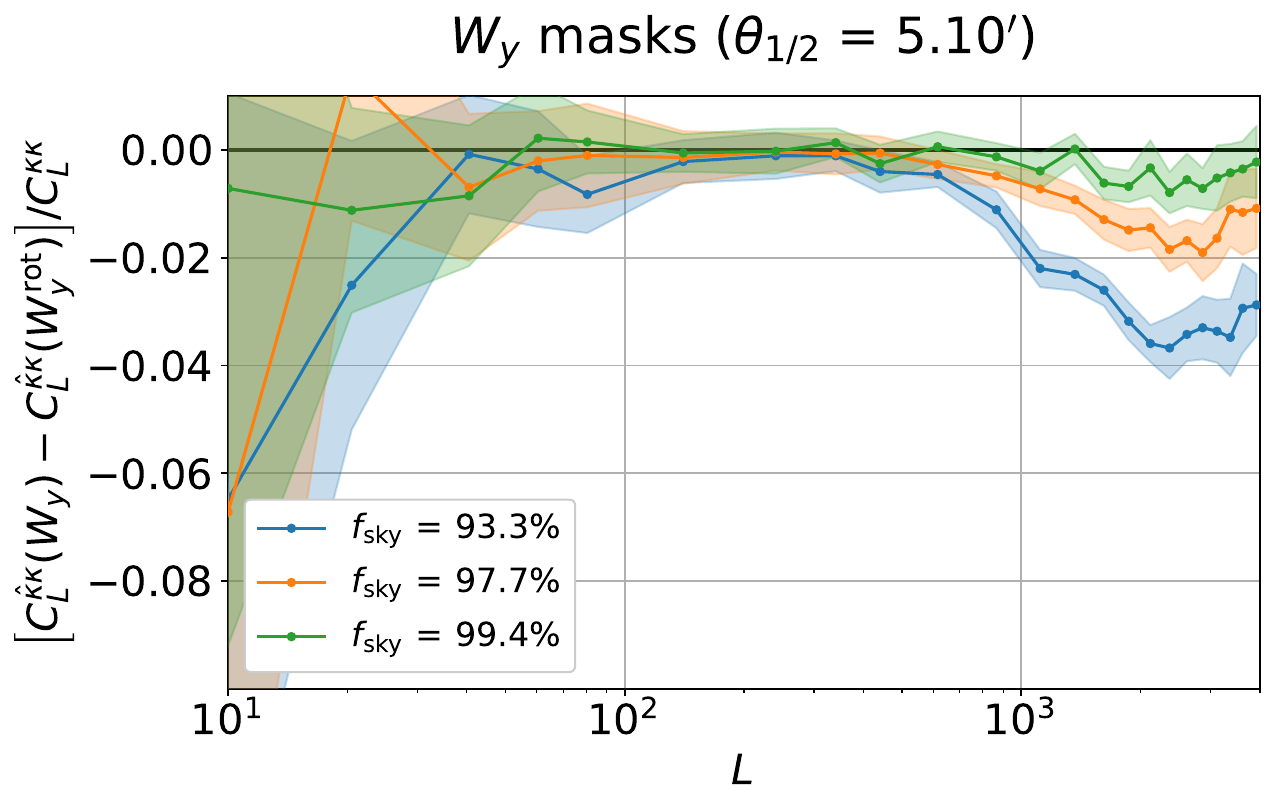}\\
\end{tabular}
\begin{tabular}{cccc}
\includegraphics[width=0.33\textwidth]{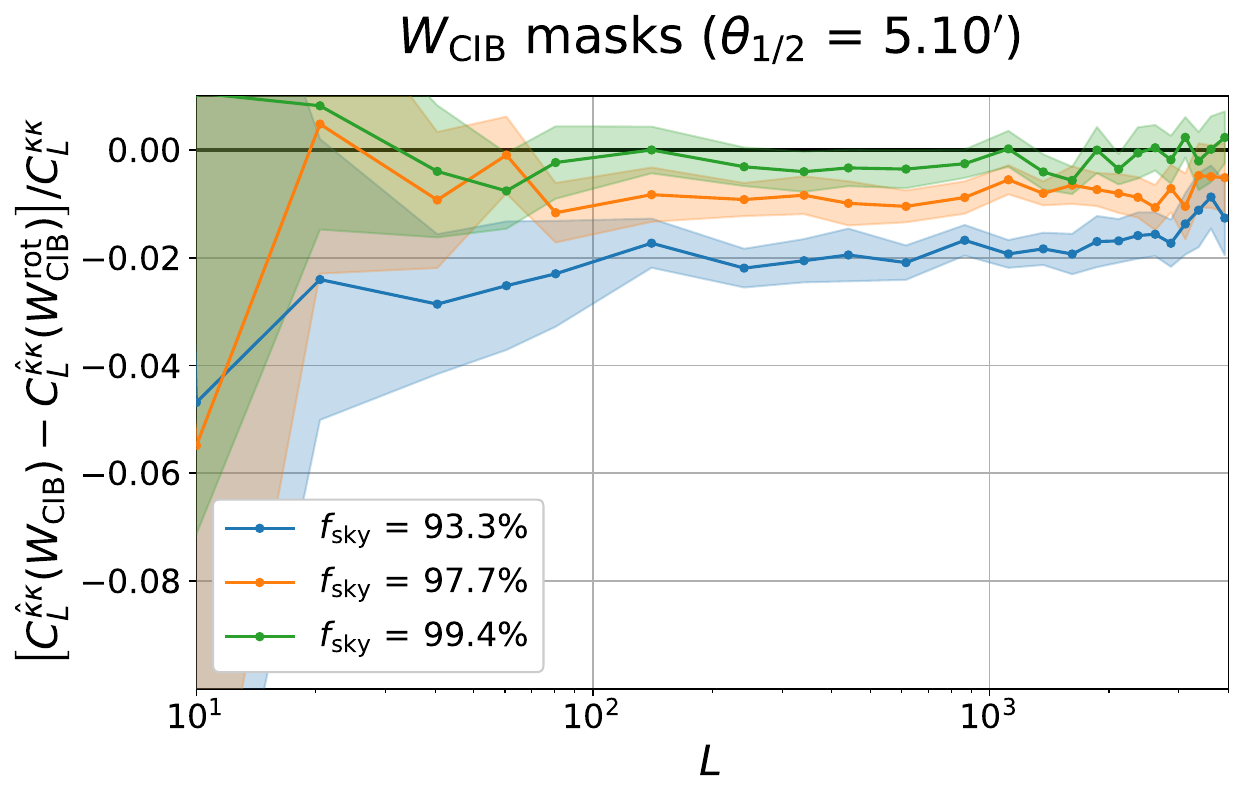} &
\includegraphics[width=0.33\textwidth]{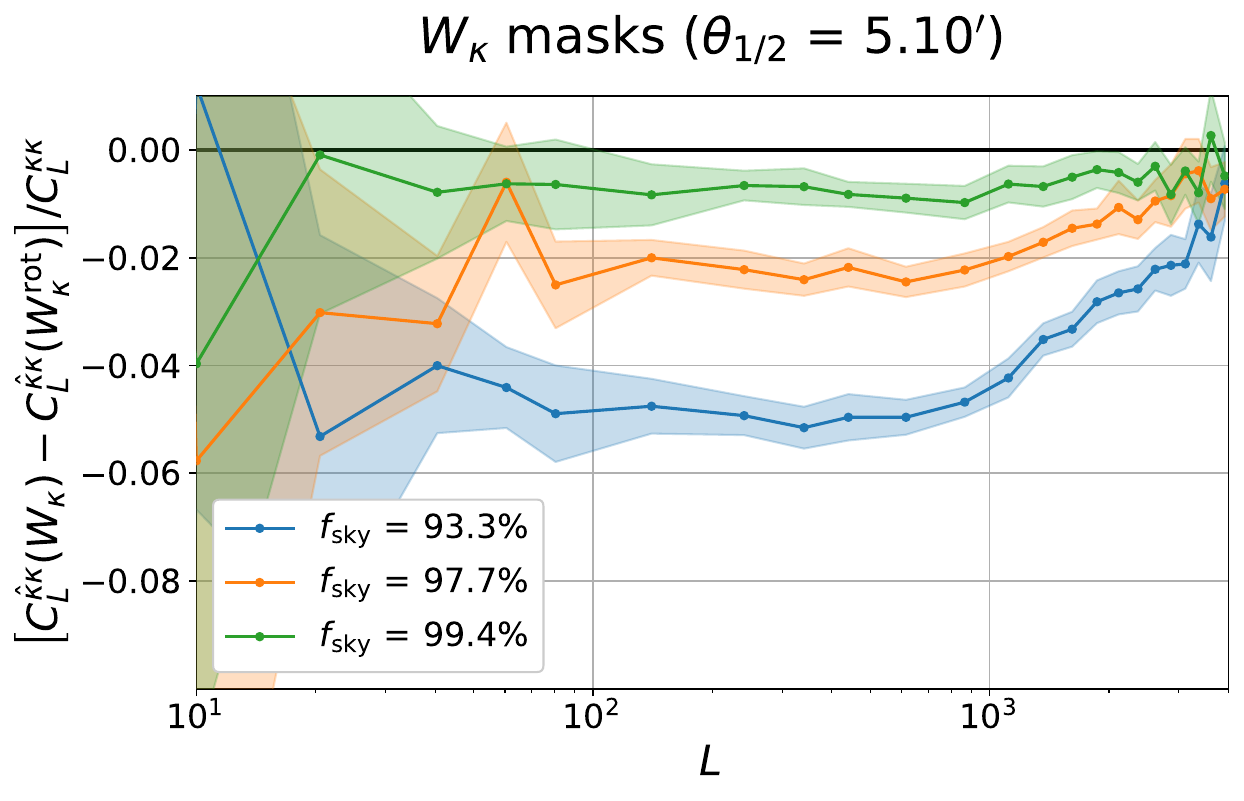} &
\includegraphics[width=0.33\textwidth]{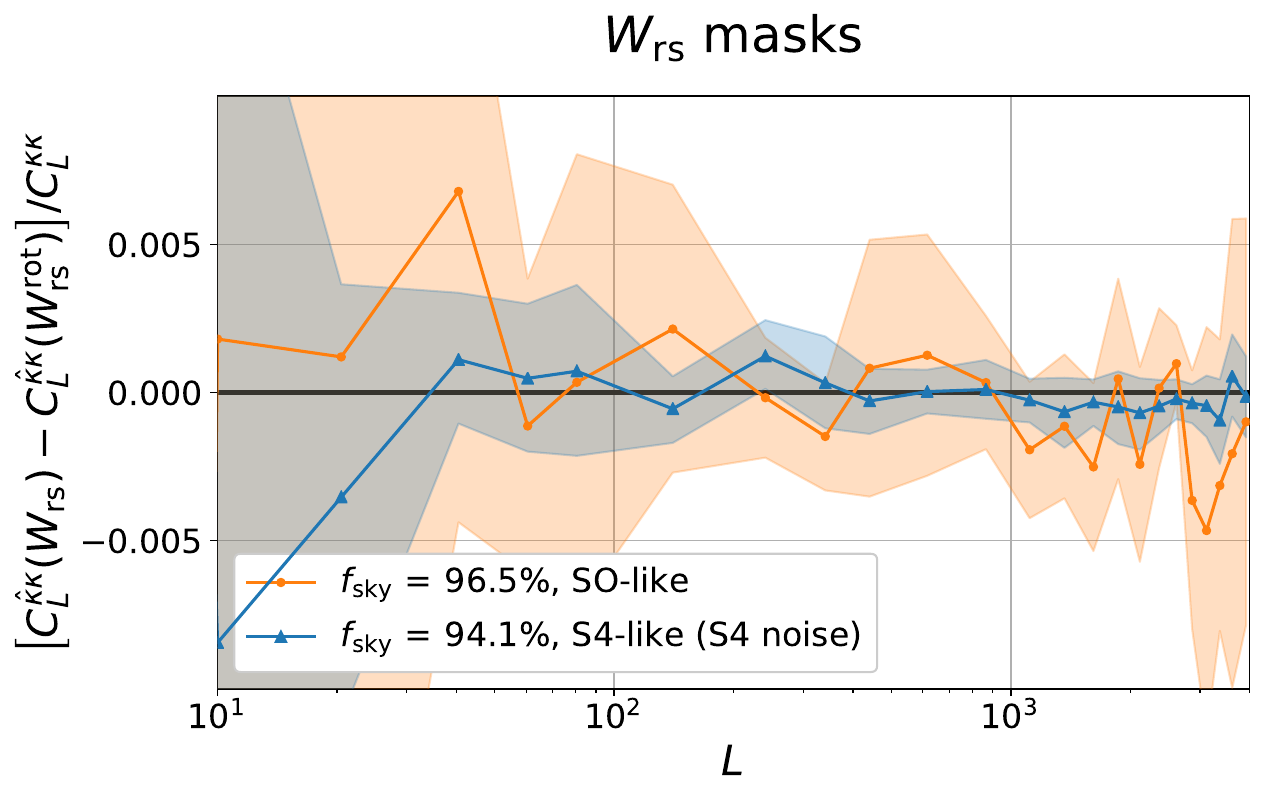}\\
\end{tabular}
\caption{Same as Fig.~\ref{fig:lens-rec_auto} for the cross-spectrum between the reconstructed CMB lensing convergence field and the true lensing field.
}
\label{fig:lens-rec_cross}
\end{figure*}

We show the results of these measurements in Fig.~\ref{fig:lens-rec_auto}, Fig.~\ref{fig:lens-rec_cross} and Fig.~\ref{fig:lens-rec_auto_and_cross_1p7}  for the reconstructed auto and cross spectra respectively, showing the bias for all the masks, $W_{\rm halo}$, $W_{\kappa}$, $W_{\rm CIB}$, $W_y$ and $W_{\rm rs}$. We estimated the error bars of these measurements computing the estimators of Eq.~\eqref{eq:lensing-power-spectra-esimator} and Eq.~\eqref{eq:lensing-cross-spectra-esimator} for both $W_X$ and $W^{\rm rot}_X$ using as ``data'' independent sets of 20 G simulations. The errors are then taken as the standard deviations of the differences between correlated and uncorrelated (randomly rotated) mask results.

As expected, the biases become larger as we increase the masked fraction of the sky. However, the amplitude of the lensing reconstruction biases are significantly reduced with respect to those obtained from directly masking the lensing field presented in Sec.~\ref{sec:matter_1} (see, e.g., Fig.~\ref{fig:lens-rec_auto} and Fig.~\ref{fig:masking_kappa_noapo} for a direct comparison). The optimal filtering used by the lensing reconstruction pipeline substantially reduces the fraction of the lensing information that is removed by the mask, both because the filtering recovers some of the CMB modes inside the mask holes, and because the lensing reconstruction itself is able to recover much of the information about lensing modes on scales larger than the hole size (see Appendix~\ref{sec:holes} for an analytic discussion).

The remaining biases induced by $W_{\rm halo}$ and $W_{y}$ (with $\theta_{1/2} = 5.1^\prime$) masks are mainly relevant on small scales.
The S4-noise case considered for the $W_{\rm halo}$ mask shows a similar trend, with a remaining power spectrum bias of the 2-5\% level for $L\gtrsim 2000$.
For the $W_{\kappa}$ and  $W_{\rm CIB}$ masks the bias is roughly constant across all the scales, and has a magnitude of $\sim 1-10\%$ of the signal depending of the $f_{\rm sky}$.  The bias induced by $W_{\kappa}$ is larger since this is the limiting case where the mask is 100\% correlated with the $\kappa$ field. The bias induced by $W_{\rm rs}$ is negligible for both SO and S4.

\begin{figure}[!]
\centering
\includegraphics[width = \columnwidth]{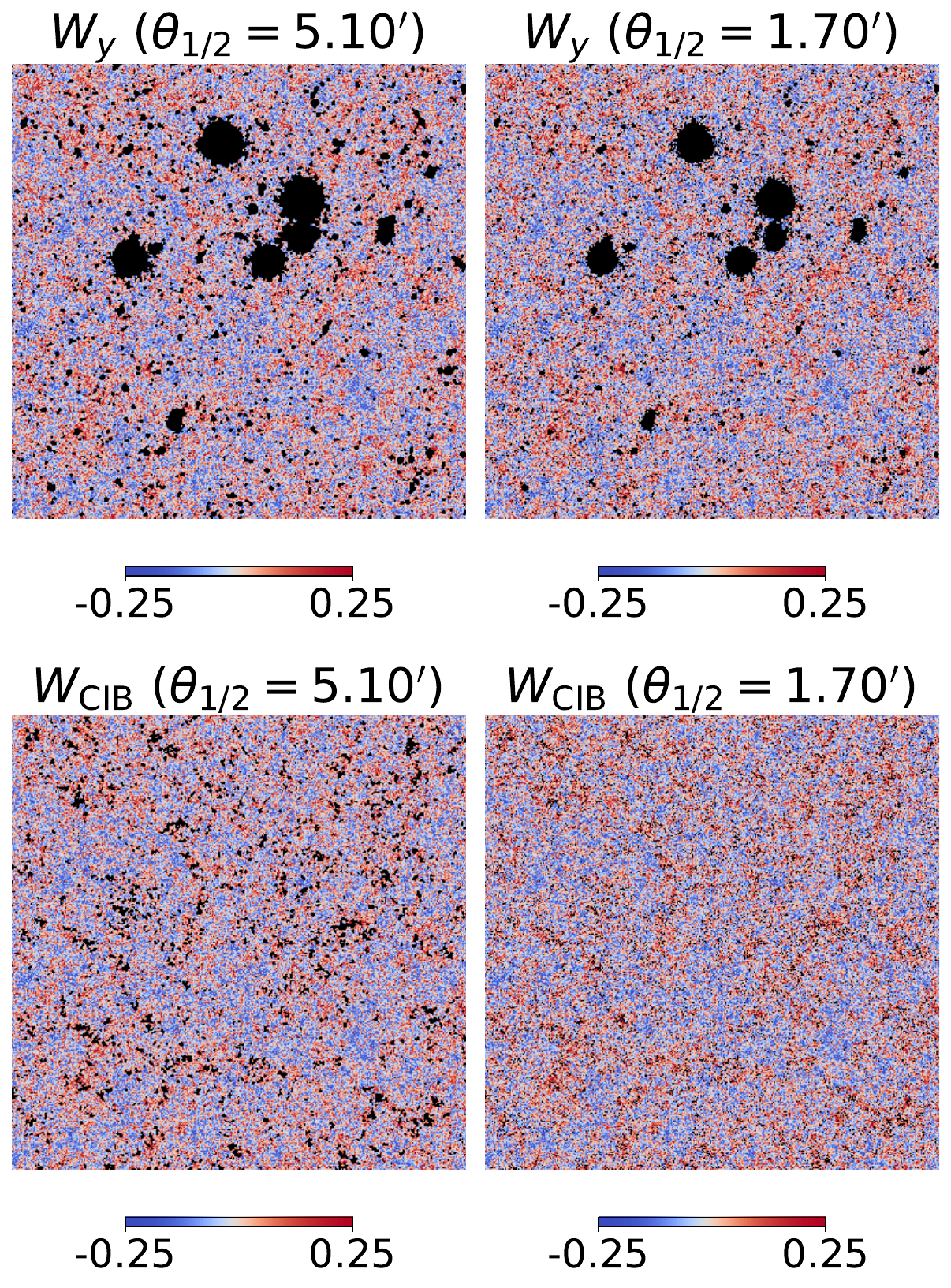}
\caption{Comparison between the masked CMB lensing convergence using $W_y$ and $W_{\rm CIB}$ masks constructed from foreground fields smoothed with different values of $\theta_{1/2}$. While the shape of the masked regions changes significantly, the total masked sky fraction is the same ($f^{\rm mask}_{\rm sky} =6.7\%$).}
\label{fig:smoothing_5p1_vs_1p7}
\end{figure}

In Fig.~\ref{fig:lens-rec_auto} and Fig.~\ref{fig:lens-rec_cross}, the foreground fields used to build the $W_y$, $W_{\rm CIB}$ and $W_\kappa$ masks were smoothed with a Gaussian beam of $\theta_{1/2}=5.1^\prime$ prior to thresholding, similar to the beam of current \Planck~data. This operation leads to masks with large connected holes on the scale of the smoothing. However, future experiments such as SO and S4 will observe the sky at higher angular resolution. To investigate the sensitivity of our result to this smoothing scale, we performed a similar analysis on the masks obtained by smoothing the foreground field with a $\theta_{1/2}=1.7^\prime$ beam prior to thresholding. In Fig.~\ref{fig:smoothing_5p1_vs_1p7}, we show the comparison of the masks obtained with the two different smoothing scales. Figure~\ref{fig:lens-rec_auto_and_cross_1p7} shows the measurements of the mask bias for this set of masks. From the direct-masking results obtained with these masks in Sec.~\ref{sec:matter_1}, we would expect larger biases at the level of the auto spectra, especially at smaller scales.
However, comparing Fig.~\ref{fig:lens-rec_auto_and_cross_1p7} with the results shown in Fig.~\ref{fig:lens-rec_auto} and Fig.~\ref{fig:lens-rec_cross} using the larger smoothing scale, we actually see substantially smaller biases when we mask the same fraction of the sky. Reconstruction with optimal filtering is able to recover more information in the holes when we increase the number of holes but also to reduce their size. There is now only a small residual bias at $L\agt 1000$ for the $W_y$ mask, where the signal is anyway largely dominated by reconstruction noise.
The biases in the S4-noise case are instead completely negligible both for $W_y$ and $W_{\rm CIB}$.
The reconstructed CMB lensing field can also be used to construct templates of the lensed CMB in order to perform delensing of the observed CMB data, in particular the B-mode polarization (see Refs.~\cite{POLARBEAR:2019snn,ACT:2020goa} for recent applications on data). As discussed in Ref.~\cite{Fabbian:2013owa}, an unbiased measurement of the lensing potential at $L\lesssim 1500$ would be sufficient to resolve with sub-percent accuracy the lensing B-mode signal, which is the part of the signal for which the coupling induced by lensing is the most non-local. The mask biases on the reconstructed CMB lensing map and power spectra at these angular scales are small for the most realistic masks we considered, and even more for the S4-like observations we considered, with high-resolution and low-noise, a regime for where delensing would bring larger improvements for cosmological constraints. As such, we  do not expect the mask biases to become a significant problem for CMB internal delensing for future data sets.

\begin{figure*} [t!]
\centering
\begin{tabular}{cccc}
\includegraphics[width=0.35\textwidth]{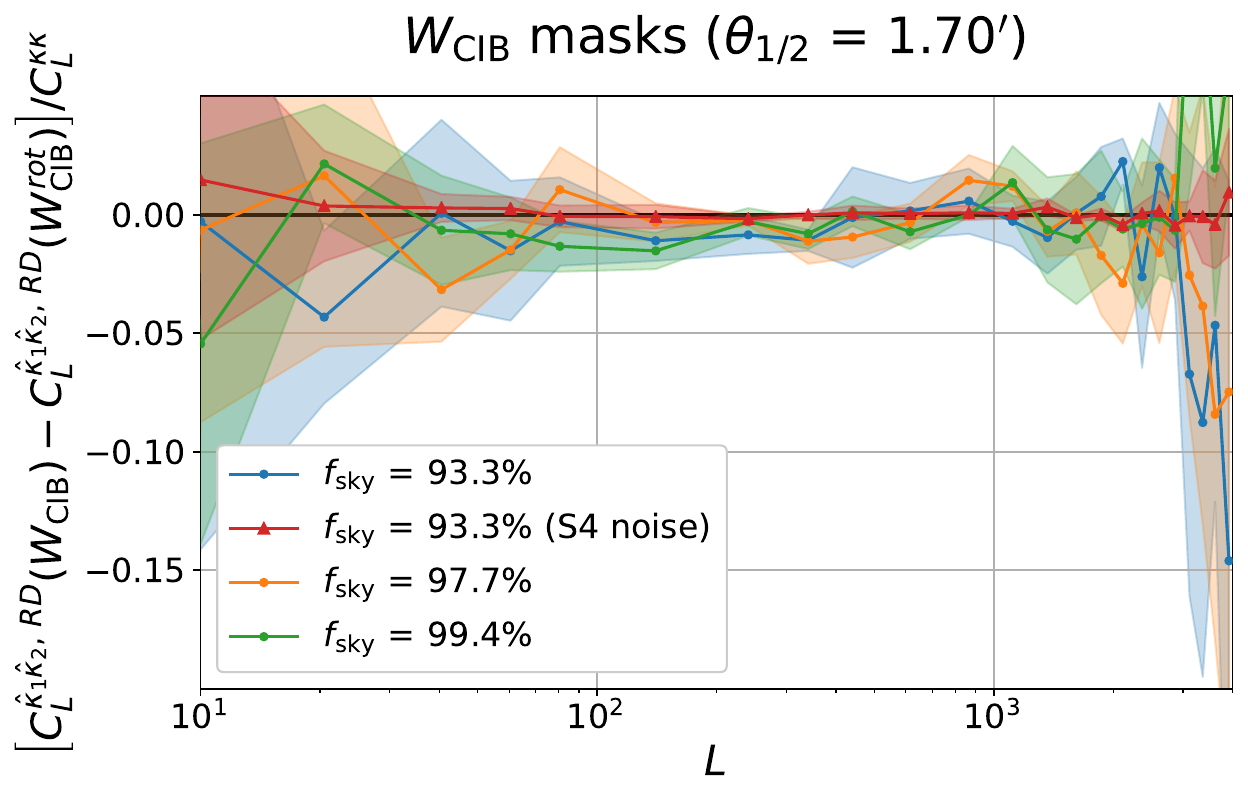} $\qquad$ &
\includegraphics[width=0.35\textwidth]{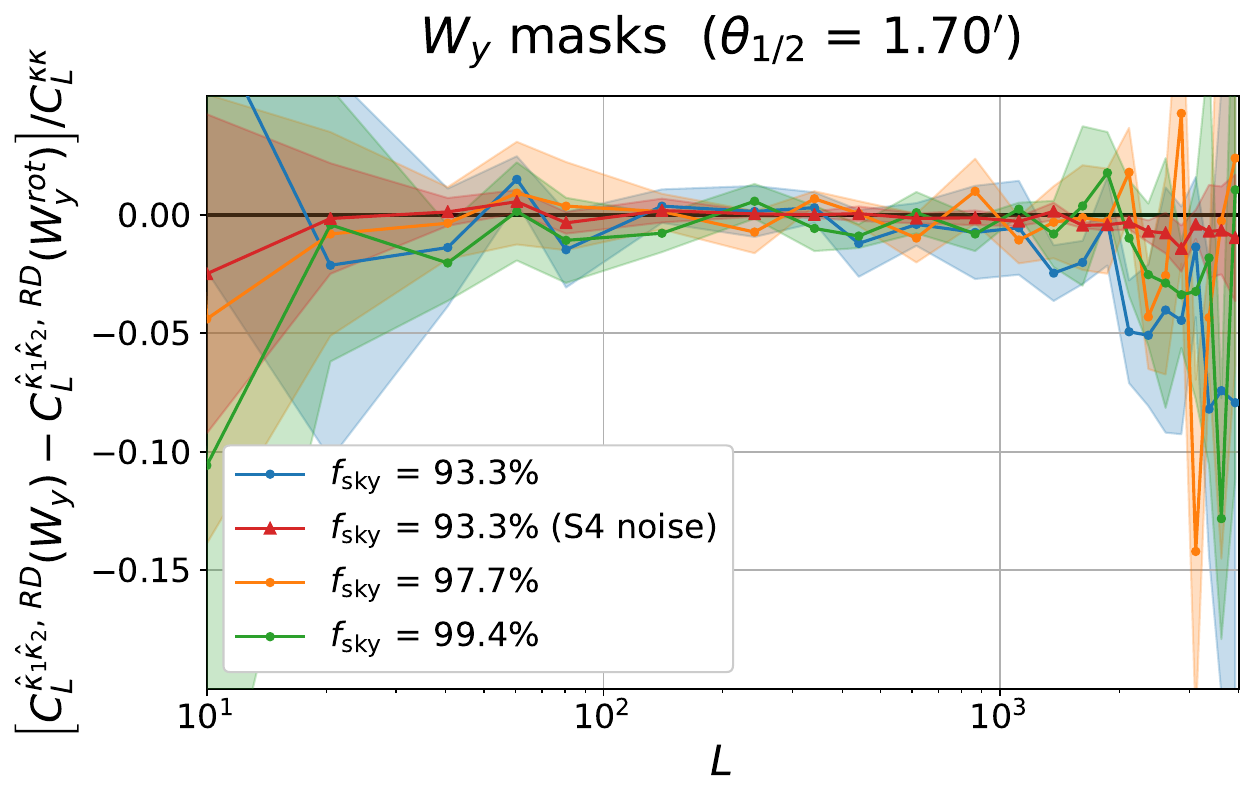}\\
\end{tabular}
\begin{tabular}{cccc}
\includegraphics[width=0.35\textwidth]{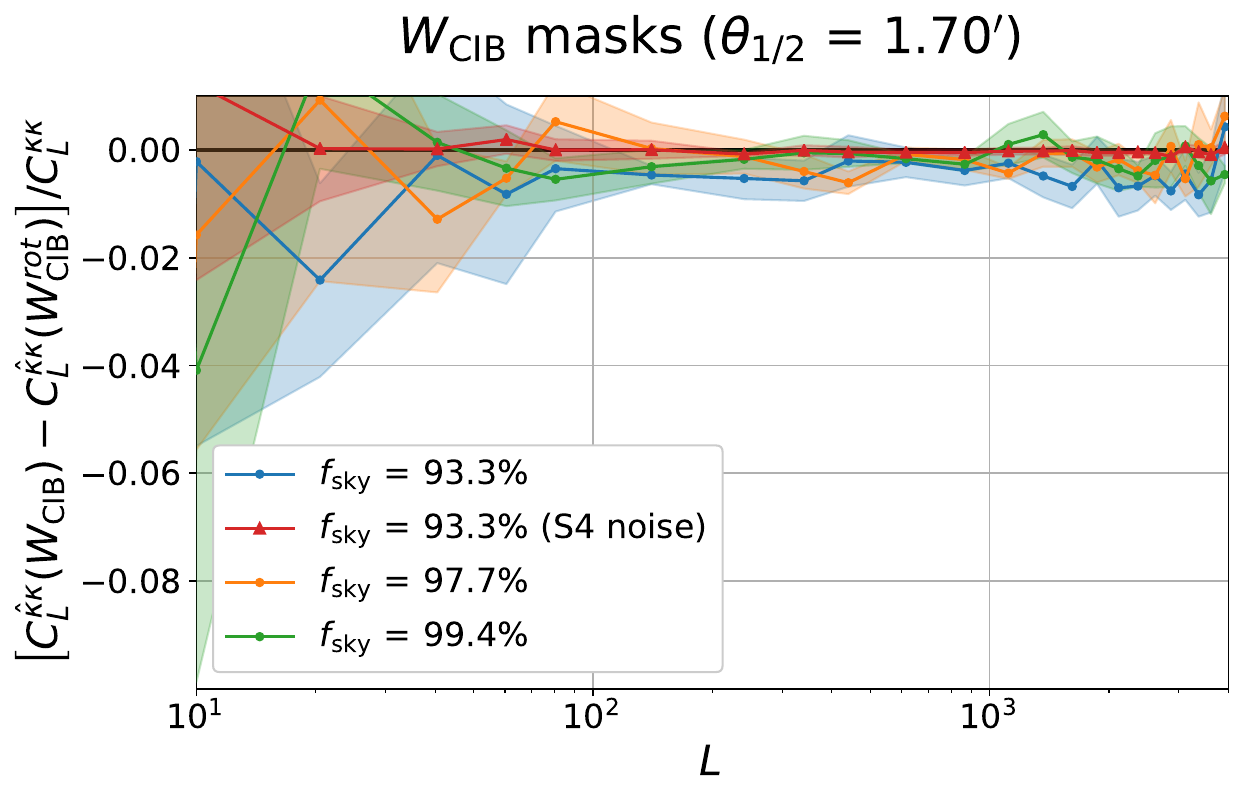} $\qquad$ &
\includegraphics[width=0.35\textwidth]{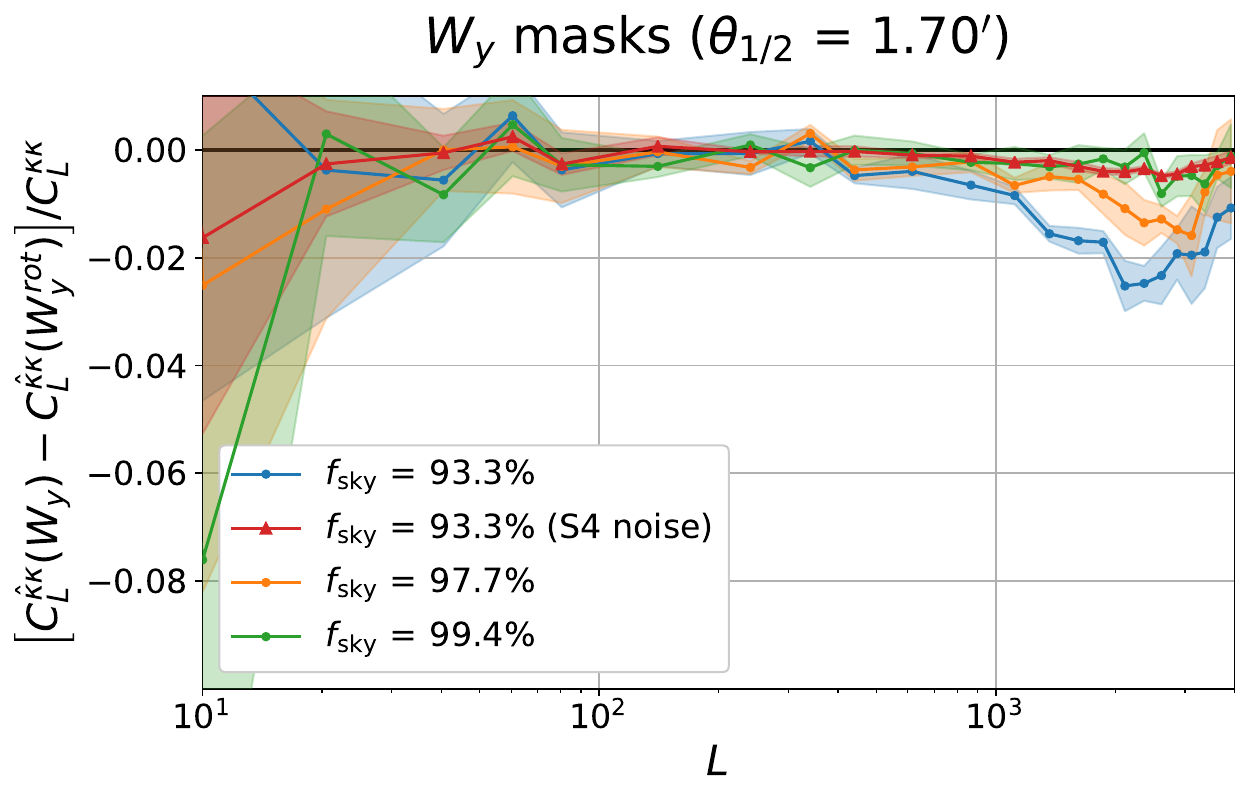}\\
\end{tabular}
\caption{Bias due to the LSS-correlated masking on the reconstructed CMB lensing convergence auto-spectrum (top panels) and cross-spectrum (bottom panel) for a subset of the cases shown in Fig.~\ref{fig:lens-rec_auto}. For these results the masks were constructed smoothing the foreground fields with a $\theta_{1/2}=1.7^\prime$ Gaussian beam prior to thresholding (instead of $\theta_{1/2}=5.1^\prime$ for Fig.~\ref{fig:lens-rec_auto}).}
\label{fig:lens-rec_auto_and_cross_1p7}
\end{figure*}

\begin{figure}[!]
\includegraphics[width = \columnwidth]{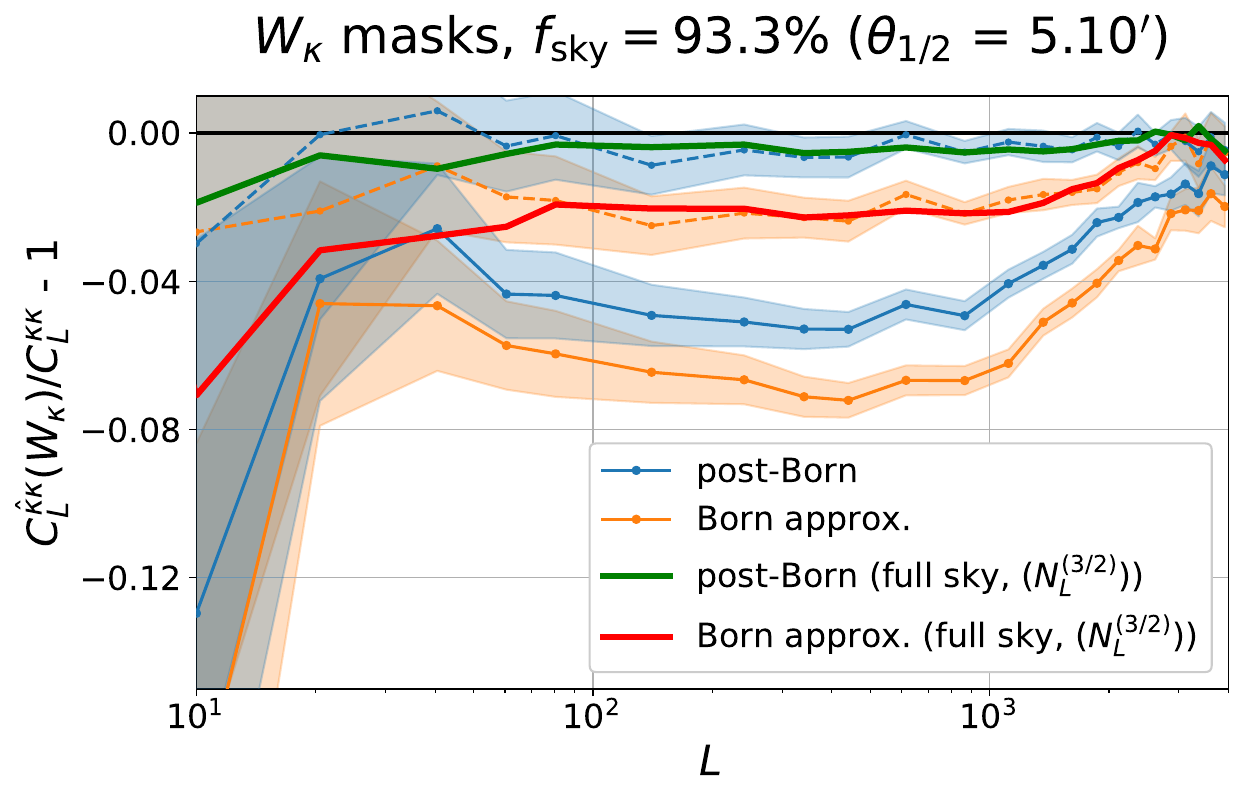}
\caption{Differences between the cross spectra of the reconstructed CMB lensing fields and the true fields relative to the full-sky power spectrum of the true fields. All these curves include mean-field subtraction and $\fMC$ normalization correction.
The results are obtained by performing lensing reconstruction after masking the sky with $W_{\kappa}$ masks constructed from the post-Born and Born $\kappa$ maps of Ref.~\cite{Fabbian:2017wfp} (solid blue and orange respectively). Dashed lines show the results obtained using the LSS-uncorrelated masks $W^{\rm rot}_{\kappa}$. These are consistent with the results of lensing reconstruction on full sky (green and red lines), where the observed discrepancy is only due to \nlth.}
\label{fig:cross_pb}
\end{figure}

\begin{figure}[!]
\includegraphics[width = \columnwidth]{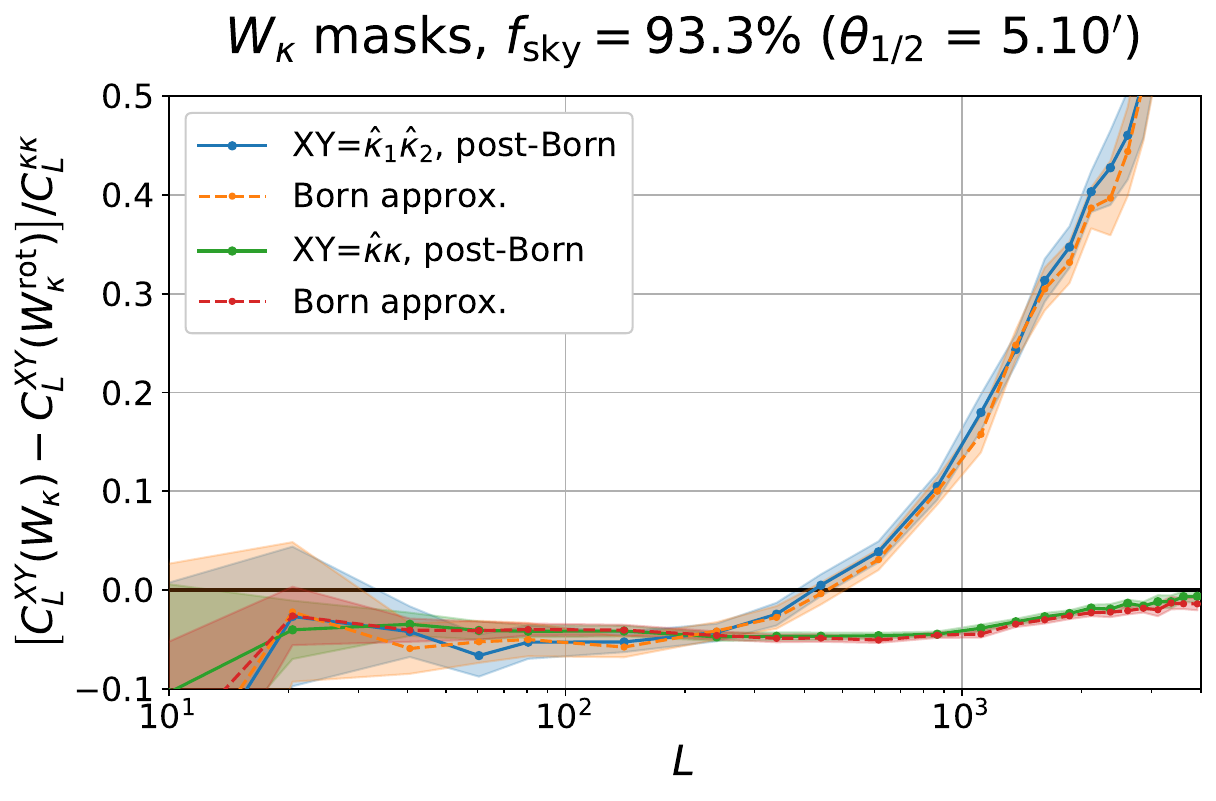}
\caption{
Bias due to the LSS-correlated masking on the reconstructed CMB lensing convergence auto-spectrum (in blue and orange) and cross-spectrum (in green and red) relative to the true full-sky $C_{L}^{\kappa\kappa}$ for $W_{\kappa}$ masks. Results including post-Born lensing are shown in solid lines and those in the Born approximation as dashed lines. All these curves include the $\fMC$ correction and mean-field subtraction but retain the RD-$N^{(0)}_L$ bias.}
\label{fig:cross_pb2}
\end{figure}

To estimate the impact of post-Born lensing on the correlated mask biases, we ran the entire end-to-end mask bias estimation pipeline, as described in Sec.~\ref{sec:recon}, using the pBNG and BNG simulations, computing $\fMC$ corrections and the mean field of the QE on the pBG simulation set. For both the pBNG and BNG simulations we built a new correlated foreground mask, thresholding the corresponding $\kappa$ field such that $f^{\rm mask}_{\rm sky} = 6.7\%$. This limiting case of a $\kappa$ mask is the only case we considered. Post-Born lensing modifies the shape of the \nlth bias as shown in Fig.~\ref{fig:cross_pb}, where the results obtained with rotated masks (dashed lines) are consistent with those obtained on the full-sky (green and red lines, which are consistent with the expected \nlth\ reconstruction bias).
Our results for correlated mask bias are again calculated as differences between spectra on rotated and unrotated masks. As such, the \nlth bias is mainly removed when taking this difference and the impact of post-Born lensing is only important to the extent that the correlated mask changes the post-Born contribution to \nlth compared to the Born case. Since the amplitude of \nlth  is relatively small, the overall impact of post-Born lensing is small (see Fig.~\ref{fig:cross_pb2}). Neglecting post-Born effects is therefore not expected to significantly affect the other results shown for other LSS-correlated masks.

\subsection{Cross-correlation with CMB lensing}\label{sec:xcorr}

In the previous section we showed that using optimal filtering in CMB lensing reconstruction recovers some of the information lost by masking LSS-correlated sky areas, reducing the expected biases in the reconstructed CMB power spectrum. However, this does not guarantee that the CMB lensing field is properly recovered at the sky location masked in the CMB maps used for lensing reconstruction. In this section, we quantitatively assess this problem using statistics involving cross-correlation between CMB lensing and external tracers.

\subsubsection{CIB and tSZ cross-correlation}\label{sec:xcorr}

\begin{figure*}[!htbp]
\includegraphics[width=0.33\textwidth]{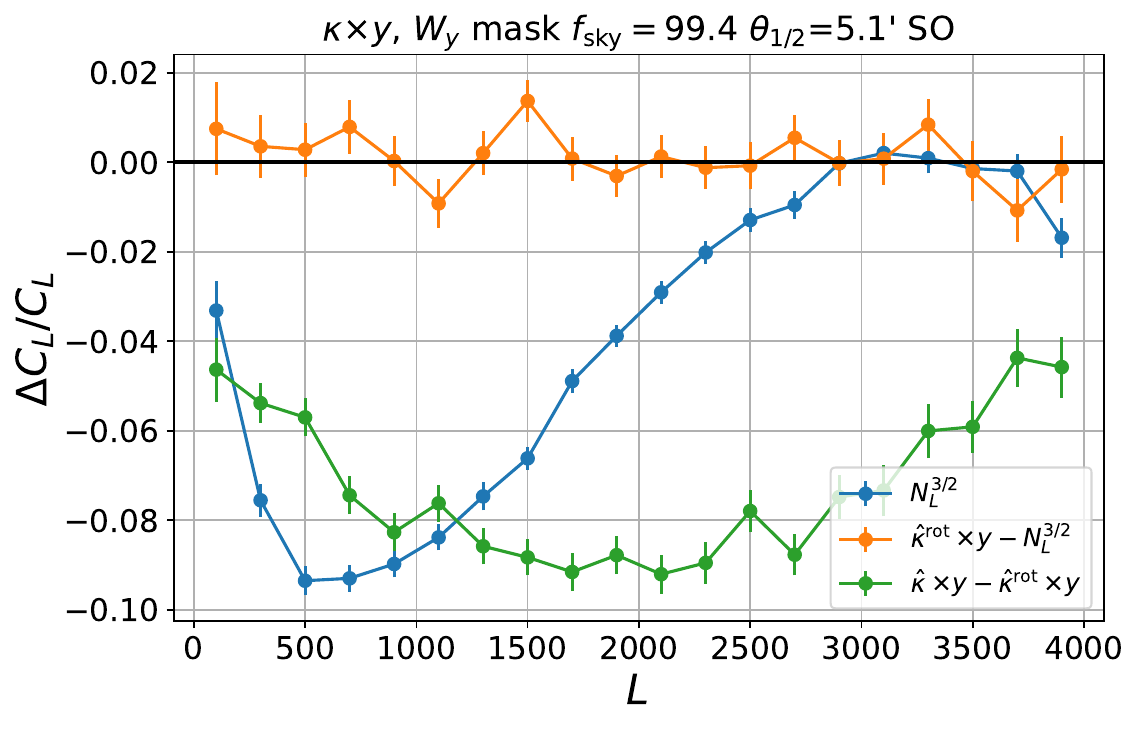}
\includegraphics[width=0.325\textwidth]{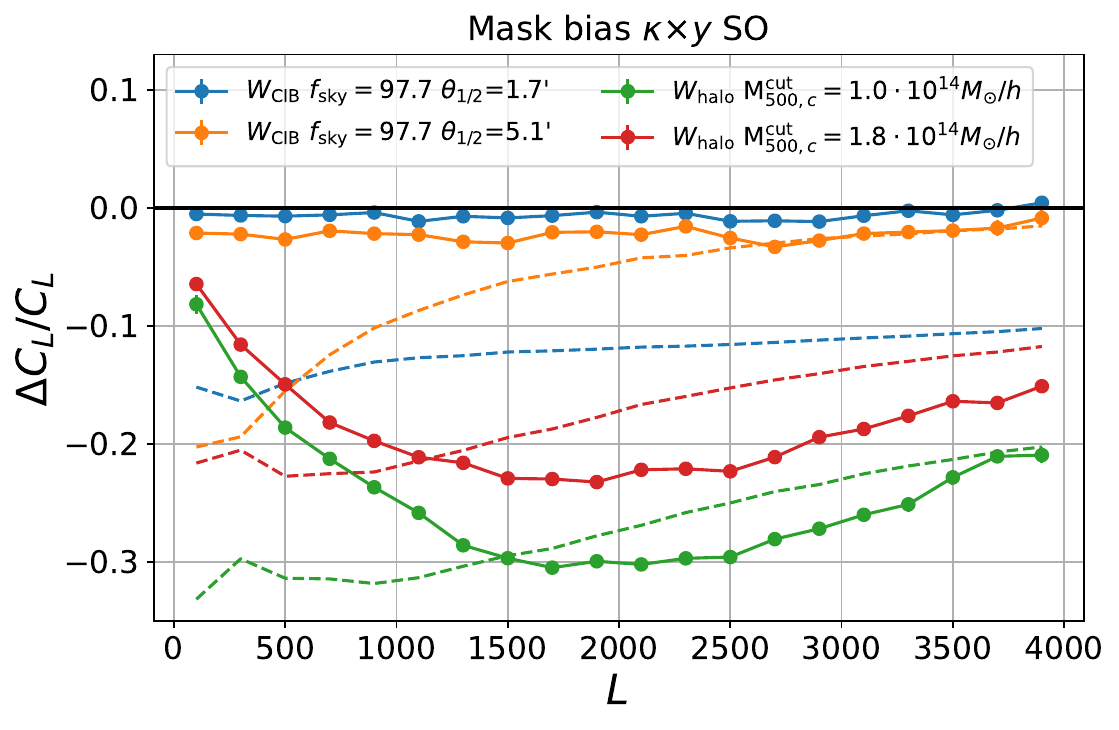}
\includegraphics[width=0.33\textwidth]{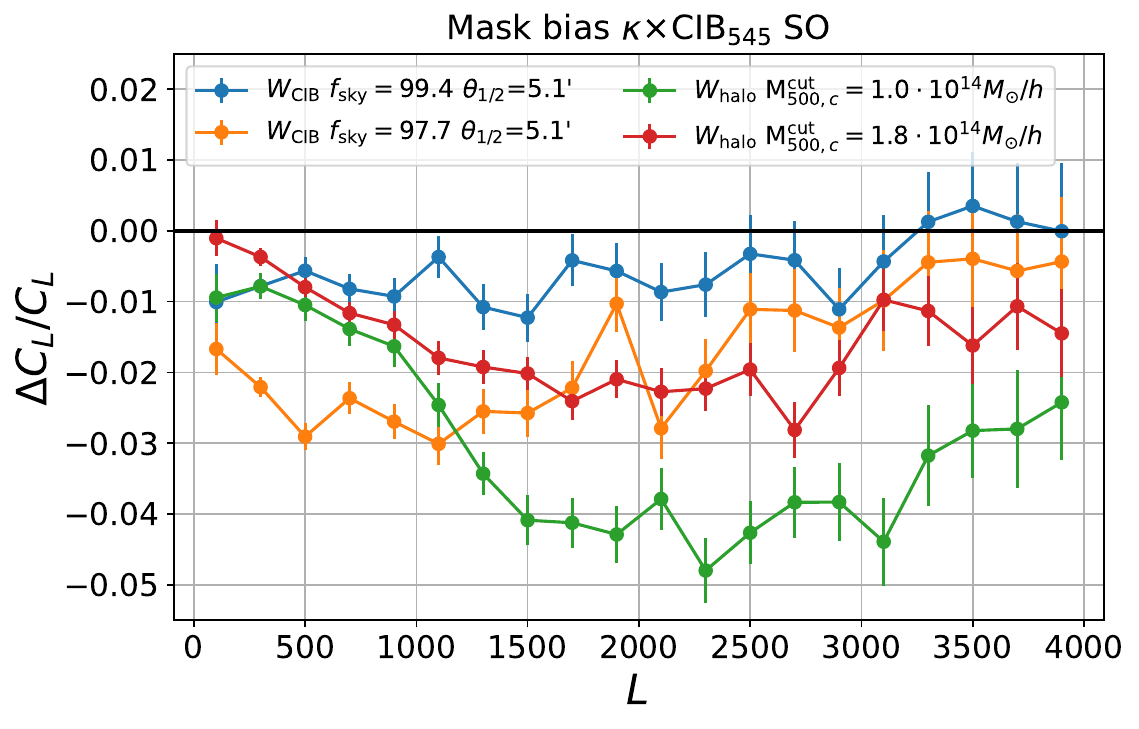}
\caption{In the left panel, we show an example of the bias induced by the LSS-correlated masking on the cross-correlation power spectrum between CMB lensing and tSZ Compton $y$ parameter (green) for SO. In blue, we show the \nlth bias to $C_{L}^{\kappa y}$ due to the non-Gaussianity of the lensing field  itself measured on the full sky. The cross-correlation power spectrum of the reconstructed CMB lensing field with the $y$ map when masking the sky with an LSS-uncorrelated mask is consistent with \nlth (orange) measured on the full-sky. The middle and right panel show the mask bias to the cross-correlation power spectrum between CMB lensing and $y$ and the CIB emission at 545GHz respectively for the most significant masks considered in this work. Such biases are larger than the one obtained for the reconstructed CMB lensing auto-spectrum and cross-spectrum shown in Fig.~\ref{fig:lens-rec_auto} and \ref{fig:lens-rec_cross}. Dashed lines show the expected fractional difference of the power spectrum computed using the true $\kappa$ fields on a masked sky compared to its expected value on the full sky (for display purposes shown divided by a factor of 2). The optimal filtering used for the lensing reconstruction recovers a large fraction of the information lost by masking.}
\label{fig:xcorr-summary}
\end{figure*}

To measure the mask biases on $C^{\kappa y}_{L}$ and $C_{L}^{\kappa \mathrm{CIB}}$ we used a pipeline similar to the one we used to measure the biases on $C^{\kappa\kappa}_{L}$. We computed the cross-correlation power spectrum between the \websky $y$ and CIB map at 545GHz with the lensing map reconstructed from CMB maps masked with our LSS-correlated masks. We then subtracted the same results calculated using the LSS-uncorrelated (rotated) masks.  As for the CMB lensing auto-spectrum, this procedure isolates the effect of the LSS-masking from other reconstruction biases, which in this case are only due to the \nlth bias~\citep{Fabbian:2019tik}. We chose in particular the 545GHz for the CIB emission as it has been successfully employed in delensing studies and offers a good trade-off between signal to noise and dust contamination \cite{Manzotti:2017net,Larsen:2016wpa,Aghanim:2018oex}.
In the left panel of Fig.~\ref{fig:xcorr-summary}, we show that the \nlth bias in cross-correlation obtained for the lensing field reconstructed from an LSS-uncorrelated mask and an SO noise level is consistent with the result obtained from full-sky reconstructions. The full-sky \nlth bias for $C_{L}^{\kappa y}$ is consistent with the one expected for a tracer probing the matter density at a median redshift of $z\approx 0.6$ without post-Born corrections (see, e.g., Fig.~13 of Ref.~\cite{Fabbian:2019tik}).

The other panels of Fig.~\ref{fig:xcorr-summary} show the amplitude of the mask bias relative to the true cross-correlation power spectrum of the \websky maps for a few masks representative of masking infrared sources and galaxy clusters for an SO-like survey. For $C_{L}^{\kappa y}$, masking infrared sources could bias the measured power spectrum by $\sim 2\%$ for an experiment with a \Planck-like angular resolution, or half this for an experiment with an SO-like resolution. Cluster masking has a more dramatic impact, introducing biases of 10--30\% in the measured power spectrum, with a broad peak around the median scales of the masked objects (for our case roughly $6^\prime$). This case is of course an extreme example useful to quantify the error on the recovered correlation inside the masked regions; for practical analyses, using $C_{L}^{\kappa y}$ directly for scientific purposes cluster masking is avoided. However, the optimal filtering recovers a significant amount of information: applying the LSS-correlated mask on the true field would reduce the  $C_{L}^{\kappa y}$ power spectrum by a much larger amount. As an example, in the middle panel of Fig.~\ref{fig:xcorr-summary} the dashed lines show the fractional difference between $C_{L}^{\kappa y}$ computed on a masked sky and the one computed on the full-sky using the true $\kappa$ field. Even for the most extreme cases of cluster masking, optimal filtering recovers about 80\% of the signal for  $L\lesssim 500$ and reduces the bias by a factor of $\sim 2$ at $L\gtrsim 1000$.

For $C_{L}^{\kappa \mathrm{CIB}}$, despite the non-negligible correlation between CIB and tSZ due to infrared emission of galaxies in galaxy cluster environments \cite{Planck:2015emq,Maniyar:2020tzw}, the mask biases are smaller ($\lesssim 5\%$) for the masks considered for this study. For the most realistic cases related to infrared point-sources shown in Fig.~\ref{fig:xcorr-summary}, we found a reduction of power of about 2\% approximately constant for the scales most relevant for delensing  ($L\lesssim 2000$). As these variations are small, we do not expect the delensing efficiencies for CIB-delensing to be strongly impacted by masking biases.


We performed a similar analysis with CMB lensing maps reconstructed from CMB maps with an S4-like noise level. We focussed on a few specific aggressive masks: $W_{y}$ and $W_{\rm CIB}$ threshold masks removing $f_{\rm sky}^{\rm mask}=7.7\%$, computed with a Gaussian smoothing of $\theta_{1/2}=1.7^\prime$, and $W_{\rm halo}$ with $M^{\rm cut}=10^{14}M_\odot/h$. For the latter, shown in Fig.~\ref{fig:xcorr-summary} for SO noise levels, we observed a reduction of the bias on $C_L^{\kappa y}$ and $C_L^{\kappa \mathrm{CIB}}$ by a factor of $\sim 2$ for $L\lesssim 2000$ and by a factor of $\sim 3$ for $2000\lesssim L \lesssim 4000$ thanks to the improved performances of the optimal filtering. Similar trends can be observed for the threshold masks where the reduction of the biases compared to an SO-like noise is even more larger: going from an SO-like to an S4-like noise,  for the $W_{y}$ mask the median bias across all the angular scales changes from 16\% to 6\% for $C_L^{\kappa y}$  and from 1.6\% to 0.2\% for $C_L^{\kappa \mathrm{CIB}}$.

\subsubsection{Cluster mass calibration}
The abundance of galaxy clusters as function of mass and redshift is a highly-sensitive probe of cosmology: it strongly depends on the growth rate as well as on the geometry of the universe. The main challenge for the application of cluster abundance in cosmology is the inference of the true mass of the cluster from observable quantities such as X-ray, tSZ luminosity or optical richness. Gravitational lensing of light sources behind clusters is one of the most promising
techniques to estimate their masses as it is sensitive to the total matter distribution and less affected by complex details of baryonic physics in dense environments. CMB-cluster lensing might be particularly useful for estimating the mass of high-redshift clusters, for which it is difficult to observe background galaxies with sufficient sensitivity, and for providing estimates complementary to galaxy weak-lensing for low redshift clusters as it is sensitive to different systematic effects \cite{Madhavacheril:2017onh}.

Since the signal-to-noise expected for each cluster in CMB lensing maps, even for futuristic surveys, is well below 1 for clusters of $ M_{500,c}\approx 10^{14}\msun/h$ \cite{Raghunathan:2017cle}, cluster masses are usually computed as the average mass of a set of clusters \cite[e.g.,][]{Miyatake:2018lpb,Raghunathan:2017qai}. We therefore quantified the impact on the recovered mean cluster halo mass by stacking the CMB lensing maps reconstructed from different masked fields at the location of the clusters. We selected the objects in the \websky halo catalogue, mimicking a complete mass-limited sample for SO-like noise described in Sec.~\ref{sec:sims-masks}, and estimated the mass of the clusters from the radial profile of the stacked convergence map with the \texttt{cmbhalolensing} public code\footnote{\url{https://github.com/simonsobs/cmbhalolensing}}. For this purpose,  we stacked cut-outs of $25^\prime\times 25^\prime$ stamps around the location of each cluster from our full-sky maps and binned the radial profile of the stack in 25 radial bins. We estimated the covariance of each radial profile by splitting our cluster sample into 192 sub-regions corresponding to the objects located within an Healpix pixelization of nside=4 and adopting a jackknife resampling approach following Ref.~\cite{DES:2018myw}. In the fit we assumed the redshift of the stack to be equal to the mean redshift value of the cluster sample $\bar{z}=0.63$ and used points at a distance $r<10^{\prime}$; we also accounted for the mass dependence of the halo profile concentration parameter $c$ using the scaling relation of Ref.~\cite{duffy2008}. We then quantified the mask bias by comparing the fitted mass value with the one obtained by stacking the full-sky reconstructed $\kappa$ map. To account for the finite sky coverage of future ground-based surveys such as SO and S4, we rescaled the jackknife covariance estimated over the full sky by the sky fraction covered by those experiments before performing the fit. For this purpose we assumed a common sky fraction of $f_{\rm sky}^{\rm obs}=50\%$.

Several methods have been proposed to extract the CMB-cluster lensing signal from CMB temperature and
polarization maps with different level of optimality and sensitivity to extragalactic foreground contaminations \cite{Seljak:1999zn,Dodelson:2004as,Holder:2004rp,Lewis:2005fq,Hu:2007bt,Raghunathan:2019tsz,Madhavacheril:2018bxi}. The standard QE in particular is known to be sub-optimal and biased-low for low CMB noise levels \cite{Maturi:2004zj,Hu:2007bt,Yoo:2008bf}. We quantified this bias by computing the mean cluster mass from the radial profile obtained by stacking the \websky $\kappa$ at the cluster locations and assuming the same covariance of the full-sky reconstructed map. This value is consistent with the mean mass of the sample estimated from the values present in the \websky halo catalog ($\bar{M}_{500,c}=2.64\cdot 10^{14} \msun/h$) while the one obtained with the full-sky reconstructed map is biased low by $\approx 1\sigma$ ($\hat{M}^{\rm full-sky}_{500,c}=(2.29\pm 0.15)\cdot 10^{14} \msun/h$). As a consistency check, we estimated the mean mass from the profile obtained from convergence maps reconstructed from CMB maps masked with $W_X^{\rm rot}$ and found it to be consistent with the results obtained from the full-sky reconstructed maps.\\*

\begin{figure}[!]
\includegraphics[width=\columnwidth]{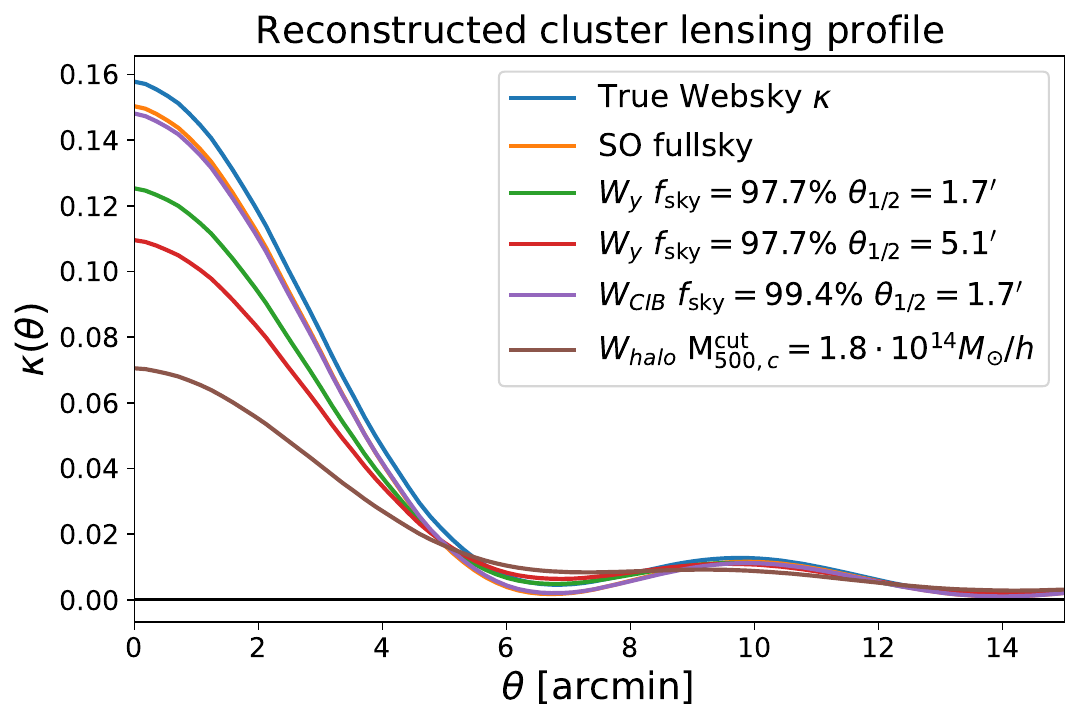}
\caption{The azimuthally averaged radial profile of the convergence maps stacked at the location of SO detected clusters. Different colours show the results obtained on $\kappa$ maps reconstructed from different masked CMB maps and assuming an SO noise level in CMB maps. Masking CMB maps with LSS-correlated masks causes a suppression of the signal at small scales compared to the full-sky reconstruction and modifications of the profile at intermediate scales when masks remove large sky fractions. The full-sky results do not retrieve an unbiased profile as they have been derived with a standard QE reconstruction; as such results obtained stacking directly on the CMB lensing convergence maps of the \websky suite are shown for reference to quantify the importance of the QE bias.}
\label{fig:stacked halo}
\end{figure}

Figure~\ref{fig:stacked halo} shows a comparison of the radial profiles obtained from different LSS-correlated masks. The changes in the halo profile are mainly concentrated towards the halo centre while the largest scales are mainly unaffected. The CIB masks that remove the brightest infrared sources induce changes at the $\approx 1\%$ level in the profile core, but do not significantly affect the mass estimation as the differences with results obtained using the full-sky reconstructed mask are $\lesssim 0.5\sigma$. Masking all the clusters in the sample before lensing reconstruction shows that only about 30\% of the signal at the halo location can be recovered as the recovered mass is $\hat{M}_{500,c}=(0.84\pm 0.17)\cdot 10^{14} \msun/h$. As discussed above, this masking choice is over conservative; however, results obtained with $W_{\rm halo}$ and $W_{y}$ masks are similar if the removed sky fraction is comparable, as these masks are correlated.

We therefore also considered a less extreme case where just the brightest clusters are masked, computing the cluster mass from a convergence map reconstructed from a CMB map masked with a $W_y$ mask removing $f_{\rm sky}^{\rm mask} = 2.3\%$ at a smoothing scale of $1.7^\prime$. This removes a similar sky fraction to an halo mask with the redshift-dependent selection function of the \Planck~cluster catalog ($f_{\rm sky}^{\rm mask}=1.6\%$) \cite{Planck:2015koh} if the clusters are masked within a radius of $2\theta_{500,c}$ from their centres. In this case we observe a bias on the recovered mean cluster mass of about $\approx 2.5\sigma$. To investigate if the mask bias affects clusters in particular redshift bins we repeated the analysis masking only clusters at $z<0.6$, $z>0.6$ and $z>1$. Accounting for the increased statistical uncertainties due to the lower number of sources, we found that the bias is reduced to $1.1\sigma$ for clusters at $z>0.6$, while for objects at $z>1$ the recovered mass is consistent with the one expected from the sample within $\sim 0.5\sigma$. The bias however becomes much more important for low-redshift clusters, where for objects at $z<0.6$ we detected the mask bias at $\sim 3.5\sigma$. This is likely due to the fact that the size of the masked object is about twice as large then the one at $z>0.6$ and therefore optimal filtering is less effective in recovering information.
The mask bias is still measurable at $\sim 1.5\sigma$ significance for less aggressive $W_y$ masks removing $f_{\rm sky}^{\rm mask} = 0.6\%$ at a smoothing scale of $1.7^\prime$. These results suggest that the mask biases will not significantly affect the science case related to cluster mass calibration from CMB lensing mass in the regime where this is the most accurate technique, but more care has to be taken when calibrating the mass of low redshift objects.

We repeated the analysis on CMB lensing maps reconstructed from CMB maps having an S4-like noise level. We focus in particular on a very aggressive case where we masked the sky prior to reconstruction with a $W_{\rm halo}$ mask with $M^{\rm cut}=10^{14}\msun/h$ (consistent with an S4-like cluster sample), and we later stacked the reconstructed $\kappa$ at the location of the same halos that were masked. Despite being unrealistic, such an extreme case is useful to assess the performances of the optimal filtering in recovering small-scale information in the maps. We found that the estimated mass from the stack has a bias at $\sim 1.2\sigma$ level compared to the results obtained stacking a $\kappa$ map reconstructed from full sky observations. Optimal filtering recovers about 90\% of the expected signal at the halo location for S4 noise levels. The same estimate assuming SO-like noise levels and the same mask gives a much larger bias ($\sim 7\sigma$) in the estimated cluster mass with only about 50\% of the expected signal properly recovered. Furthermore, we found that the conclusions on the redshift dependency previously discussed for the SO noise level and the SO cluster sample hold also for the S4 noise and cluster sample, with a mask bias increased to $\sim 3\sigma$, $\sim 1.8\sigma$, $\sim 1\sigma$,  for objects located at $z< 0.6$, $z>0.6$ and $z>1$ respectively. Since the mask bias we considered here for S4 gives a marginal detection significance, despite being an extreme over-conservative case, we do not expect realistic foreground masks to severely impact cluster mass calibration for S4.

Given that the estimator we employed in this section is not only biased but sub-optimal, and given that our noise level is higher than a full minimum-variance lensing reconstruction (we only used temperature data), a bias in the estimated cluster mass over the full sample could be important even if only the brightest fraction of the detected clusters is conservatively masked prior to the reconstruction.
However, a more targeted analysis should be carried out to accurately assess this effect for future data sets including all the complexity of the cluster selection function of each experiment.

\subsubsection{Mask-CMB lensing deflection correlation}
In \PaperI we presented an analytic model of the correlated mask bias on standard CMB pseudo-$C_\ell$ angular power spectrum estimators. This can be evaluated as an effective correction to the lensed CMB correlation function $\tilde\xi(r)$ given by

\begin{equation}
\Delta \tilde{\xi} \approx \partial_r \tilde\xi(r) \meandelta(r),
\label{generaldeconvolved}
\end{equation}
and then converted into a correction on CMB $C_{\ell}$'s. In the last equation $\meandelta(r)$ is the average over the unmasked area of the change in the separation of points due to lensing
\begin{equation}
\meandelta(r)= 2\frac{\langle \alpha_r(\vx) \w(\vx)\w(\vx')\rangle}{\langle \w(\vx)\w(\vx')\rangle},
\label{delta_def}
\end{equation}
where $\vx$, $\vx'=\vx+\vr$ are directions in the sky, $\vr$ their separation vector and $\alpha_r$ the component of the deflection field $\valpha=\nabla\phi$ parallel to $\vr$. $\meandelta(r)$ does not have a general analytical expression but can in principle be calculated empirically. The numerator of Eq.~\eqref{delta_def} can be expressed in terms of the cross-correlation power spectrum between the E-modes of the spin-1 masked deflection field $\alpha \w$, denoted by $E$, and the sky mask $\w$. In the flat sky this reads
\begin{equation}
\langle \alpha_r(\vx) \w(\vx)\w(\vx')\rangle =- \int \frac{\ud l}{2\pi} l C_l^{EW} J_1(lr).
\label{eq:EWpower}
\end{equation}
Thus, if we have an estimate of $C_L^{EW}$ we can calculate the CMB power spectrum bias for any mask. This can be constructed from simulations where we know $\kappa$ (and hence $\valpha$) and the mask $\w$. Alternatively, this can be estimated from data if we have reliable estimates of the lensing deflection field. We used this technique in \PaperI to evaluate the bias induced by a threshold mask built on \Planck~GNILC maps on the CMB temperature power spectrum of \Planck~SMICA maps, and showed that our analytical prediction so computed matched the mask bias observed on data.

\refchange{The estimator of the CMB $C_\ell$ bias using $C_L^{EW}$ requires being able to estimate the deflection field (to calculate $C_L^{EW}$).
If a masked lensing reconstruction is used to estimate this, the estimate of $C_L^{EW}$ may itself be biased due to the bias on the lensing reconstruction.}
To test this, we converted the reconstructed lensing potential $\hat{\phi}$ into a deflection field $\hat{\valpha}$ with a spin-1 spherical harmonics transform\footnote{For this purpose we used the relation between the lensing potential and the  E-modes of the deflection field $\hat{\alpha}^E_{LM}=\sqrt{L(L+1)}\hat{\phi}_{LM}$ and assumed $\hat{\alpha}^B_{LM}=0$.}. We then computed the cross-correlation power spectrum $C^{\hat{E}W}_{L}$ between the reconstructed masked deflection field $\hat{\valpha}W$ and $W$, and compared it with $C^{EW}_{L}$ directly computed using the deflection field extracted from the \websky $\kappa$ map. We focused in particular on the most important realistic cases analysed in \PaperI, i.e. the $W_{\mathrm{CIB}}$ and $W_y$ masks with $f_{\rm sky}=99.4\%$ and for both $\theta_{1/2}=5.1^{\prime}$ and $1.7^{\prime}$, which could all be detectable at more than $5\sigma$ significance for SO noise level. We found the error on $C^{EW}_{L}$ for SO noise levels to be lower than $30\%$ for $L\lesssim 3000$ and lower than $10\%$ for the CIB masking which cannot be avoided by component separation. We evaluated Eq.~\eqref{eq:EWpower} with $C^{\hat{E}W}_{L}$ and found that the analytically-predicted mask bias on $C_{\ell}^{TT}$ is consistent with the one computed using the true $C^{EW}_{L}$ measured from \websky to better than $\sim 0.5\%$ on $L\lesssim 4000$. Even for the most aggressive masks considered here, i.e. $W_{\mathrm{CIB}}$ and $W_y$ that remove $f^{\rm mask}_{\rm sky}=6.7\%$, the error introduced in the prediction of the mask bias using $C^{\hat{E}W}_{L}$ is less than 10\%.  For S4 noise levels and the same aggressive masks, the errors on $C^{EW}_{L}$ are reduced up to a factor of 2 on scales $L\gtrsim 1000$ compared to the SO noise case and therefore we expect the accuracy of the analytical predictions of the biases to further improve at lower experimental noise levels. 

\refchange{The lensing reconstruction mask bias therefore appears to be sufficiently small that the CMB $C_\ell$ bias can be estimated accurately enough to reduce its statistical significance below the detection level.}

\section{Forecasts for future CMB experiments}
\label{sec:forecasts}
\begin{figure*}[t!]
\includegraphics[width = \textwidth]{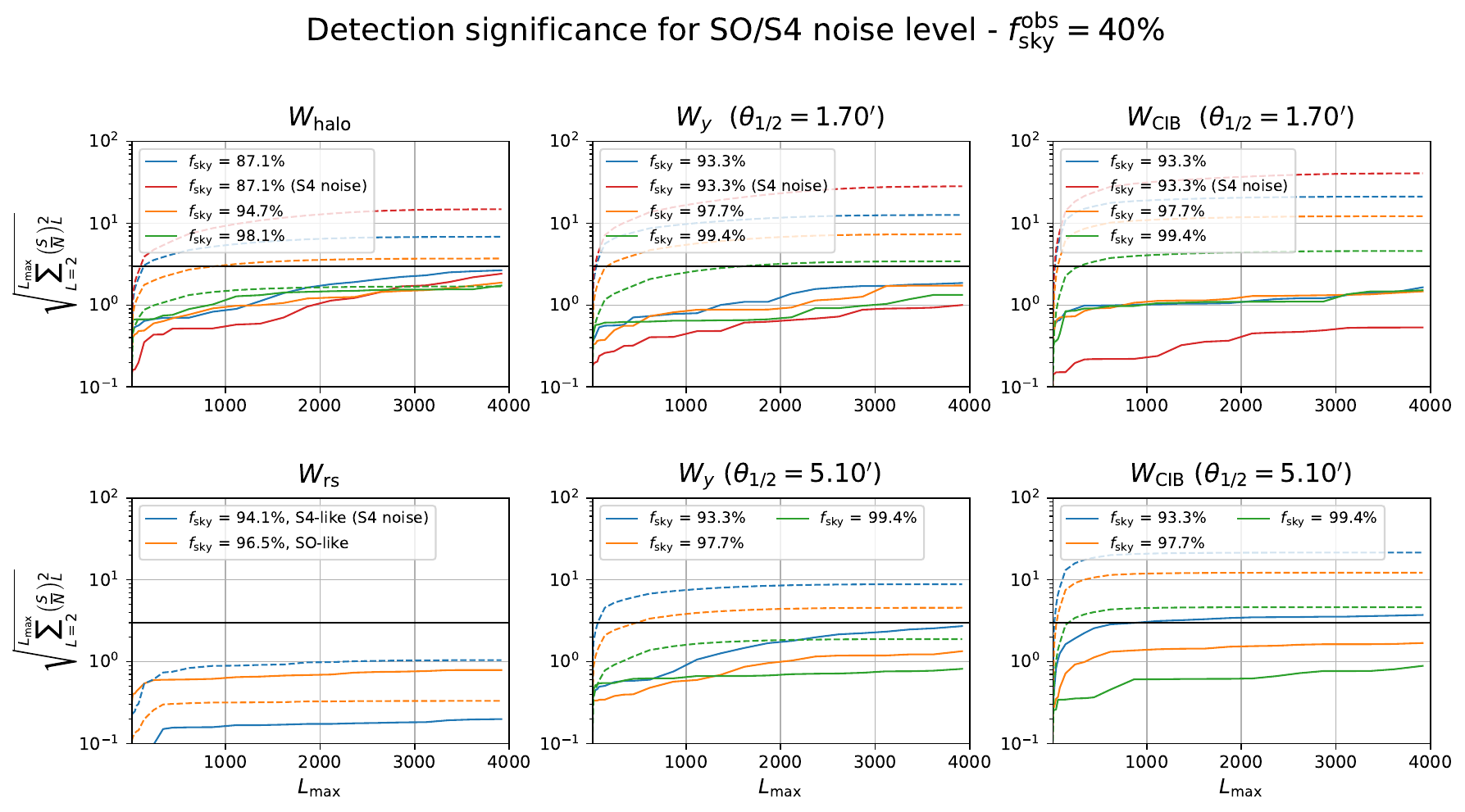}
\caption{Detection significance of the mask bias in the CMB lensing power spectrum for future high-resolution ground-based experiments as a function of maximum multipole $L_{\rm max}$ included in the analysis and for different foreground masks. We conisered an observed sky fraction $f^\mathrm{obs}_\mathrm{sky} = 40\%$. We assumed an SO-like lensing noise for all cases except for those shown in red which assumed an S4-like CMB lensing noise. The dashed lines show the bias obtained directly masking the $\kappa$ field (see Sec.~\ref{sec:matter_1}) while the solid lines show the bias on the reconstructed $\kappa$ field from masked CMB maps (see Sec.~\ref{sec:recon}). 
The horizontal black line highlights a statistical detection significance of $3 \sigma$ for reference.}
\label{fig:snr_SO}
\end{figure*}

In the previous sections, we have shown that LSS-correlated masks can introduce biases on the reconstructed CMB lensing power spectrum that could become non-negligible if not accounted for. We have shown that the biases on the CMB lensing power spectrum are likely to be negligible for radio sources for the immediate future, but may be more important for other masks (a $\sim 2\%$ effect which can increase to $\sim 5\%$ though on scales where the lensing reconstruction noise is important, see Figs.~\ref{fig:lens-rec_auto}, \ref{fig:lens-rec_auto_and_cross_1p7}).

Below, we estimate the detectability of such mask biases for SO and S4 in terms of cumulative signal to noise where we used as the signal the mask biases measured in Sec.~\ref{sec:cmblens-results} and the noise includes the sample and noise variance of each experiment. For this purpose, we assumed a sky coverage of $f^{\rm obs}_{\rm sky} = 40\%$ and the realistic publicly-available noise power spectra for the lensing potential reconstructed using only temperature modes\footnote{For SO we used the so-called baseline noise from \url{https://github.com/simonsobs/so_noise_models}. For S4 we took the noise power spectra available at \url{https://cmb-s4.uchicago.edu/wiki/index.php/Survey_Performance_Expectations}.}.
We fix all cosmological parameters, so the results are an upper limit on the impact on any cosmology constraint.

The results for all the $W_X$ masks considered in this analysis are summarized in Fig.~\ref{fig:snr_SO}, where we show the detection significance of the mask biases as a function of maximum lensing multipole $L_{\rm max}$ included in the analysis. We do not include the results for the $W_\kappa$ mask since $\kappa$ is not directly observable but, for comparison, we also show the expected biases from directly masking the $\kappa$ convergence field as done in Sec.~\ref{sec:matter_1}.
Our results clearly show that, using optimal filtering of the CMB maps for the lensing reconstruction, the biases are much smaller than they would be if $\kappa$ were masked directly (which would give a signal detectable with high significance, well above $5\sigma$ in future measurements).

For $W_{\rm halo}$, masking up to 1-2\% of the sky (consistent with the sky fraction removed by masking SO-like tSZ cluster sample), the mask bias can be measured with a statistical significance $\lesssim 2\sigma$ while for cluster samples more similar to the S4 ones, where the  mass limit of $10^{14}M_{\odot}/h$ and the masked sky fraction is $\sim 10\%$, the mask bias can be detected at $\lesssim 3\sigma$ in the reconstructed lensing field, using both the SO and the S4 noise. Similar results can be observed for the threshold mask $W_y$ when the foreground field has been smoothed with $\theta_{1/2}=5.1^\prime$ Gaussian beam and a similar sky fraction as in the $W_{\rm halo}$ case is masked, given that the two types of masks are correlated.

We note that cluster masking may not be used in  final CMB lensing analyses, and the tSZ contamination can also be reduced by analysing component-separated CMB maps where components with a tSZ-like SED are projected out during component separation, or using dedicated modified quadratic estimators \cite{Madhavacheril:2018bxi,Patil:2019zbm}. To assess the validity of the methods and to quantify potential residual emission in these alternative analyses, however, it is common practice to compare the results with more conservative analyses based on cluster masking as also done in the \Planck~2018 lensing analysis \cite{PL2018}. As such, our results show that care is required when performing these kind of comparisons, as mask biases may lead to misleading inconsistencies in the comparison of the CMB lensing estimates.

For the $W_{\rm CIB}$ (with $\theta_{1/2}=5.1^\prime$) masks, the bias detectability remains $\lesssim 1\sigma$ when the masked sky regions removes $\sim 5\%$ of the sky or less. For more aggressive masks, removing $f_{\rm sky}^{\rm mask}\approx  7\%$, the bias will be detectable at more than $3\sigma$ significance. The increased significance is not unexpected since the CIB is highly correlated with the CMB lensing field, especially at $L \lesssim 1000$, and therefore a mask that removes these peaks naturally enhances the mask bias effect.

The detection significance for both $W_y$ and $W_{\rm CIB}$ masks built from higher-resolution observations (i.e. with $\theta_{1/2}=\refchange{1.7}^\prime$  smoothing) will produce biases that will be measured with lower statistical significance compared to the  $\theta_{1/2}=5.1^\prime$ cases. They always stay $\lesssim 2\sigma$ for SO and $\lesssim 1\sigma$ for the S4 cases we considered, showing the improved performance of the optimal filtering in recovering information in presence of lower noise level and smaller holes.

For radio sources, $W_{\rm rs}$, the detection significance of the bias is below the $1\sigma$ level, whether we are masking all the radio sources up to the detection limit of SO or S4.

We finally stress that the bias for any Poisson source mask is expected to be very small, as for radio sources, since the mask is largely determined by random Poisson sampling of the background distributions rather than tracing lensing-correlated perturbations closely. For example, as long as masks removing infrared sources are only removing Poisson sources (i.e. at very high or low redshift), rather than peaks of the full CIB field, their bias should also be negligible. Although masking peaks in the CIB emission is not common practice in CMB analysis, our $W_{\rm CIB}$ masks can potentially mimic the effect of masking bright infrared point sources, where a fraction of them is from strongly lensed objects that are therefore highly correlated with the matter distribution along the line of sight on much smaller scales than CMB lensing is sensitive to \cite{spt-dsfg2020}. We note that \websky simulations were not constructed to reproduce the source number counts in the infrared, nor include any effect of magnification bias which would affect the number of detected sources in CMB maps for a fixed noise level. However, preliminary analyses showed a good agreement between the expected source number counts at the highest fluxes  from semi-analytical models \cite{Cai:2013wna,Lapi:2012xp} and the source counts expected from the halo-model adopted in the \websky simulation to construct the CIB maps \cite{websky-ir}. Our \websky simulation results should therefore provide a reasonable ballpark estimate of the total effect, since the uncorrelated Poisson part of the distribution only affects the mask bias via the total $\fsky$ masked.

We performed a similar forecast analysis for the mask biases in the cross-correlation between CMB lensing and external tracers presented in Sec.~\ref{sec:xcorr}. In this case, we assumed the publicly available SO and S4 noise power spectra for the $y$ map achievable with an ILC component separation and for the lensing noise; for the CIB we fitted a white noise level and an effective beam from the noise-dominated regime of the beam-deconvolved power spectrum of the GNILC maps at 545GHz. We obtained a noise level of $4.84 \mathrm{Jy/sterad}$ and a Gaussian beam with $\theta_{1/2}=4.65^\prime$. We also considered future CIB measurements of CCAT-prime and assumed the beam and noise power spectra described in Ref.~\cite{Choi:2019rrt}. As expected, biases in cross-correlation are more important than those reported above for the CMB lensing spectrum. We considered mainly threshold masks smoothed with $\theta_{1/2}=1.7^\prime$, as it is the case more relevant for future experiments and $W_{\rm halo}$ masks. For SO, we found that $W_{\rm CIB}$ and $W_y$ threshold masks do not produce any significant bias on $C_{L}^{\kappa \mathrm{CIB}}$ if they remove less than 6\% of the sky both for \Planck~and CCAT-prime noise levels and resolution. Biases due to $W_y$ masks are however more harmful for the analysis of $C_{L}^{\kappa y}$ even if only a minor fraction of the brightest clusters is removed. $W_y$ masks biases will in fact be detectable with a significance of $\sim 3\sigma $ if the mask removes 0.6\% of the observed sky and of $5\sigma$ when masking 2.3\% of the sky (consistent with the sky fraction removed by masking Planck tSZ-detected clusters).

The significance increases to $10\sigma$ and $13\sigma$ for $W_{\rm halo}$ masks removing SO and S4-like clusters respectively. These masks will also leave detectable biases in $C_{L}^{\kappa \mathrm{CIB}}$ at $2.5\sigma$ and $4.5\sigma$, respectively.\\
For an S4-like survey, we saw in Sec.~\ref{sec:xcorr} that the mask biases are reduced compared to an SO-like one; however, the noise of the reconstructed CMB lensing and $y$ maps is also lower and the detection significance might still be comparable to the ones of an SO-like survey. Nevertheless, we found that the overall detection significance of the mask biases of a given mask is reduced for an S4 survey compared to an SO-like one. Even for the extreme cases discussed in Sec.~\ref{sec:xcorr}, the mask biases on $C_{L}^{\kappa \mathrm{CIB}}$ are negligible as they will be measured with a detection significance $\lesssim 1.5\sigma$. For $C_{L}^{\kappa y}$, biases induced by $W_{\rm CIB}$ will not be detectable and the significance of mask biases induced by $W_{y}$ and $W_{\rm halo}$ masks will be reduced by about a factor of 2. This reduction is enough to reduce the detection significance of mask biases of Planck-like detected clusters to marginal significance ($\sim 2.5\sigma$) but more aggressive cluster masking will introduce biases in $C_L^{\kappa y}$ that will be detectable at high significance.

Finally, we note that if CMB polarization is also used in the lensing reconstruction, and the polarization mask is much smaller than the temperature mask as expected, all the biases that we found here for the temperature are expected to be reduced. Also, combined minimum-variance and optimized estimates~\cite{Hirata:2002jy, Hirata:2003ka, Carron:2017mqf} of the lensing potential should have these biases reduced as the relative importance of polarization-based reconstruction channels will increase for future observations.

\section{Conclusions}
\label{sec:conclusions}
In this work, we studied the impact that foreground masks correlated with the large-scale structure distribution could have on the reconstructed CMB lensing potential map and power spectrum.
Future high-resolution ground-based CMB experiments, such as SO and S4, will resolve a much larger populations of extragalactic sources than current experiments, so masking larger areas of resolved tSZ-selected clusters and radio sources may be necessary and its impact carefully quantified. In~\PaperI, we already showed how such masks can potentially give large biases on the CMB temperature and polarization power spectra, even if the masked sky area is small.
Building on these results, in this paper we have shown that:
\begin{itemize}
  \item  Extra-galactic foregrounds masks that are correlated to CMB lensing give substantial biases on simple lensing power spectrum estimates if the lensing convergence field were masked directly (or no information inside the masked sky area can be recovered).
  \item Significantly smaller biases are obtained on the reconstructed CMB lensing field and power spectrum if the CMB fields used for the reconstruction are optimally filtered, effectively
    recovering signal inside small mask holes.
  \item Simple Halo Model or analytic Gaussian models that we derive can provide a qualitative understanding of the effect when the lensing field is masked directly, though non-Gaussianity of the extra-galactic foregrounds is important to model for accurate results.
  \item Radio source masks, and any other mask constructed from Poisson sources, should give a bias that is safely negligible.
 Masking foreground peaks in the tSZ or the infrared emission, or large halos, can instead give larger biases and should be avoided if possible.
  \item Biases on the reconstructed CMB lensing power spectrum from CMB temperature will mostly be measured with only marginal significance ($\lesssim2\sigma$) for forthcoming experiments.
  \item Biases in the cross-correlation between CMB lensing and tSZ or CIB are detected with higher significance for realistic foreground masks  ($\lesssim3-4\sigma$ for SO)  but can become much larger for more aggressive masks.
Thanks to improved performance of the optimal filtering in presence of low noise levels, the importance of such biases is greatly reduced for S4, however, not enough to make them completely negligible.
  \item At SO noise levels, cluster mass calibration with CMB lensing can be significantly affected by mask biases only for low-$z$ objects but will still deliver unbiased results at high-$z$. This holds even if only the brightest clusters are masked.  Cluster mass calibration for S4, conversely will likely not be affected by mask biases when realistic foreground masks are employed.
  \item Since masking biases in the reconstructed CMB lensing are small, the large masking biases expected for CMB power spectra can be predicted and marginalized self-consistently using CMB lensing as an LSS tracers and the analytical model we described in \PaperI.
\end{itemize}
We showed that mask biases are small for lensing reconstruction using optimal CMB filtering, but
alternative methods based on sky map inpainting or source subtraction at the map-making level, combined with an isotropic QE instead of optimal filtering, might also be effective in avoiding sharp mask edge effects and recovering information inside small holes, but would have to be assessed separately. Applying an additional optimal filtering step on the reconstructed lensing field may also be able to recover information in small mask holes that were not filled at the level of the CMB map (c.f. Ref.~\cite{Mirmelstein:2019sxi}).

A detailed analysis including CMB polarization in the lensing reconstruction as well as polarized foregrounds is left for future work, but biases are likely to be smaller given the expected levels of polarization of radio and infrared sources and other extragalactic emissions.
 A study of the impact of these correlated mask biases on the estimation of cosmological parameters is also left for future works.

\section*{Acknowledgments}
We thank Anthony Challinor for useful discussion and Boris Bolliet, Emmanuel Schaan and Blake Sherwin for comments.
AL, GF and JC acknowledge support from the European Research Council under the European Union's Seventh Framework Programme (FP/2007-2013) / ERC Grant Agreement No. [616170], and support by the UK STFC grants ST/P000525/1 (AL) and, ST/T000473/1 (AL and GF). GF acknowledges the support of the European Research Council under the Marie Sk\l{}odowska Curie actions through the Individual Global Fellowship No.~892401 PiCOGAMBAS. JC acknowledges support from a SNSF Eccellenza Professorial Fellowship (No. 186879). ML acknowledges the support of the Fondazione Angelo Della Riccia fellowship.

The sky simulations used in this paper were developed by the \websky Extragalactic CMB Mocks team, with the continuous support of the Canadian Institute for Theoretical Astrophysics (CITA), the Canadian Institute for Advanced Research (CIFAR), and the Natural Sciences and Engineering Council of Canada (NSERC), and were generated on the Niagara supercomputer at the SciNet HPC Consortium~\cite{2019arXiv190713600P}. SciNet is funded by: the Canada Foundation for Innovation under the auspices of Compute Canada; the Government of Ontario; Ontario Research Fund - Research Excellence; and the University of Toronto.

\noindent

\appendix
\section{CMB lensing reconstruction responses for masks with holes}
\newcommand{\Fiso}[0]{F^{\rm iso}}
\label{sec:holes}
\begin{figure}[h!]
\includegraphics[width = \columnwidth]{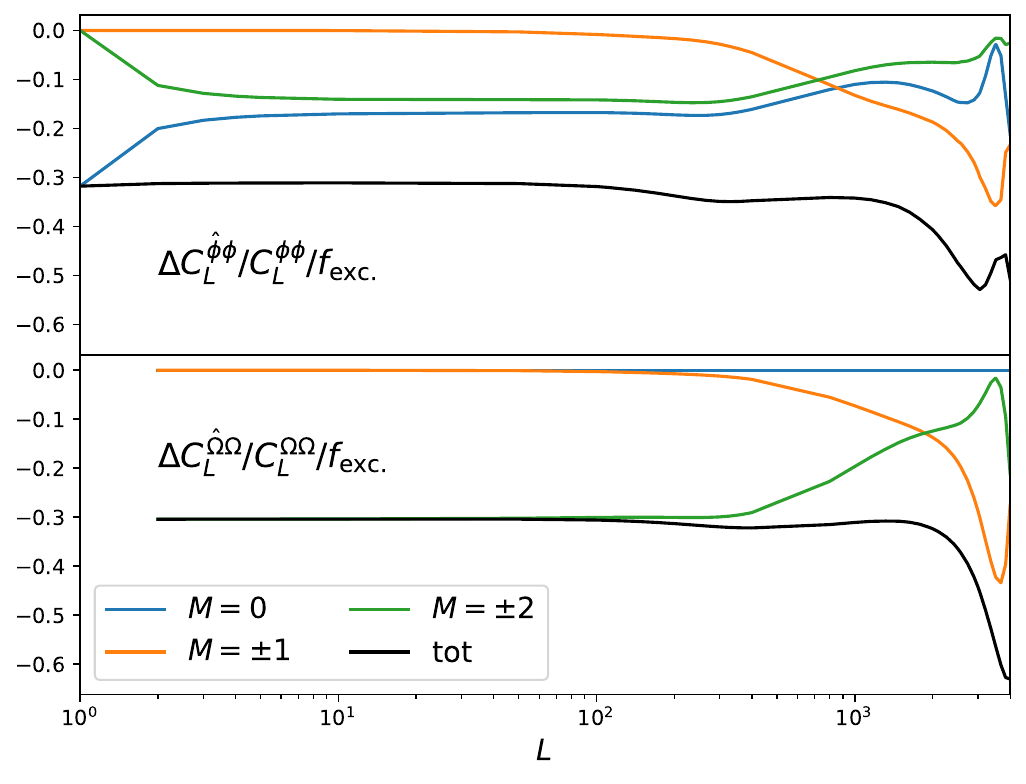}
\caption{Relative correction to the cross-spectrum of the lensing temperature QE to the true signal due to a mask built out of a population of very small holes, for the baseline experimental configuration in this work. In this regime,  for fixed $\ell_{\rm max}$ used in the QE construction and fixed instrumental noise level, the correction scales proportionally to the excised sky area $f_{\rm exc}$ for hole size $\ll 1/\ell_{\rm max}
$; the lines show the correction divided by $f_{\rm exc}$ so this scaling is taken out.
The top and bottom panel show the gradient and curl part of the lensing QE. The total correction is negative, and significantly closer to zero than the -1 that would be expected if information were lost proportional to the masked sky fraction $f_{\rm exc}$ (but not zero, since not all relevant small scales modes can be accurately recovered at this noise level). As demonstrated in the text, in the absence of optimal filtering, this correction is much larger, being equal to -2 for all $L$. The coloured lines show the first few multipolar contributions (for a hole centred the pole), displaying the expected mostly monopolar ($\hat\phi$) and quadrupolar  ($\hat\phi, \hat \Omega$) response of the estimator.
}
\label{fig:PS_GR}
\end{figure}

Masking the CMB maps impacts the lensing reconstruction even for a mask that is independent of large-scale structure. The CMB over the masked area cannot be perfectly recovered by optimal filtering, and after mean-field subtraction a signature remains in the estimators that is inaccurately captured by the standard isotropic lensing responses~\cite{Okamoto:2003zw} devised for a full-sky analysis. In CMB lensing estimation analyses, various corrections including this one are usually encapsulated inside a single Monte-Carlo correction, at the map or power spectrum level~\cite{Wu:2019hek, Darwish:2020fwf, Aghanim:2018oex}.

	This masking effect cannot be perfectly separated on data from correlated-mask signatures and the bispectrum contribution $N^{(3/2)}_L$, but simulations with the same mask can be used to estimate the effect by analysing results from Gaussian realizations. In this Appendix, we aim at some quantitative understanding of the impact of uncorrelated masking of CMB maps with analytical methods, assuming Gaussian unlensed fields, Gaussian lensing potential and perfectly linear response of the QE to lensing.

In order to do this, we calculate the response  $\mathcal R^{MM'}_{LL}$ (now a dense matrix instead of the isotropic response $\mathcal R_L$) that connects the estimator to the underlying deflection in a number of cases relevant to our study. For lensing, there are two effects: 1) the lensing gradient response gets modified and 2) masking introduces leakage between the lensing gradient and lensing curl, where the lensing curl estimator picks up a term proportional to $\phi$ and vice-versa.

We use notation consistent to the \Planck~lensing curved-sky pipeline~\cite{Carron:2019swk}, where a separable QE $\hat q$ of spin $s_o + t_o$ (1, for lensing) can be written in real space (possibly as a sum of terms) like
\begin{equation}\label{eq:qe}
\begin{split}
_{s_o + t_o}\hat q(\hat n) = &\left(\sum_{\ell m} w_\ell^{s_o s_i} \:_{s_i}\bar X_{\ell m}\:_{s_o}Y_{\ell m}(\hat n)\right) \\ \cdot &  \left(\sum_{\ell m} w_\ell^{t_o t_i}\:_{t_i}\bar X_{\ell m} \:_{t_o}Y_{\ell m}(\hat n)\right), 	
\end{split}
\end{equation}
for a set of in-spins $s_i, t_i$, out-spins $s_o, t_o$, and weight functions $w$  (left and right leg weights being distinguished from the context). On a masked sky, the CMB $\bar X$ fed into the QE is anisotropically related to the true CMB via a filtering matrix $F$,
\begin{equation}
	_{s}\bar X_{\ell m} = \:_{ss'}F^{mm'}_{\ell\ell'} \:_{s'}X^{\rm cmb}_{\ell' m'}.
\label{Fdefine}
\end{equation}
In this work we use exclusively optimal filtering, which re-captures CMB signal in the CMB maps by reconstructing the maximum a posteriori full sky CMB map, including at locations inside the holes. To do this we assume a fiducial likelihood model for the CMB, with fiducial spectra $C_\ell^{\rm fid}$, a noise variance map $N$ and a fiducial beam and transfer function $\mathcal B^{\rm fid}$. The filtering matrix is then 
\begin{equation}
	F = C^{\rm fid, -1}\left[C^{\rm fid, -1} + \mathcal B^{\dagger, \rm fid} N^{-1}\mathcal B^{\rm fid}\right]^{-1}\mathcal B^{\dagger, \rm fid} N^{-1}\mathcal B^{\rm sky},
\end{equation}
where $\mathcal B^{\rm sky}$ is the sky beam and transfer function. The matrix is in practice rectangular, since typically reconstruction of only a limited set of sky modes is attempted. Of special interest for us is the ideal case of temperature reconstruction using an azimuthally symmetric boolean mask $W(\cos \theta)$ with homogeneous uncorrelated noise variance $N_{\rm lev}$ and isotropic beam and transfer functions $b_\ell$. Writing  $Y_{\ell m}(\theta,\phi) = e^{i m \phi} \:\lambda_{\ell m}(\cos \theta)$, the inverse noise matrices in~\eqref{Fdefine} then become
\begin{equation}\nonumber
\begin{split}
	&\left[\mathcal B N^{-1} \mathcal B\right]_{\ell \ell'}^{m m'} = \delta_{mm'} \frac{b_\ell b_\ell'}{N_{\rm lev}} A^{m}_{\ell \ell'}, \\
	A^{m}_{\ell \ell'}&\equiv2\pi  \int_{-1}^1 d\cos \theta \:W(\cos \theta) \:  \lambda_{\ell m} (\cos \theta)\:\lambda_{\ell' m}(\cos \theta)
\end{split}
\end{equation}

The integral may be evaluated quickly for all $\ell$ and $\ell'$ using the recursion relations of the associated Legendre polynomials, as described by Ref.~\cite{Wandelt:2000av}.

It is interesting to compare optimal filtering results with the case where the CMB inside the mask holes is simply taken to be zero, so the harmonic modes of the CMB are taken directly on the (possibly apodized-) masked CMB map. In this case the filtering matrix is
\begin{equation}
	F^{(m, \rm apo)}_{\ell \ell'} = \frac{1}{C^{\rm fid}_\ell + N_{\rm lev}/ b_\ell^{\rm fid,2 }}\frac{1}{b_\ell^{\rm fid}}A^m_{\ell \ell'} b^{\rm sky}_{\ell'}
\label{eq:apofilt}
\end{equation}
where $A$ is the exact same matrix as above where now $W(\cos\theta)$ can include some radial apodization function. We first discuss the simple but illuminating case of a small hole in ~\ref{sec:small_holes}. The general case is then dealt with in section~\ref{sec:large_holes}.

\subsection{Small hole}\label{sec:small_holes}
In the case of a very tiny hole, large scale temperature modes inside the hole can be well reconstructed, while only the noise level of the map prevents, to some extent, reconstruction of the small scales modes. Hence, for a fixed experimental configuration, one expects the correction to the lensing cross-spectrum and auto-spectrum to scale with the excised area $f_{\rm exc.}$, and to grow with the $\ell_{\rm max}$ of the CMB fed into the QE (while $\ell_{\rm max}$  is not noise dominated) but to remain substantially smaller than $f_{\rm exc.}$ itself.

In Fig~\ref{fig:PS_GR} we show the cross-spectrum correction for our baseline configuration, where $\ell_{\rm max} = 4000$, $N_{\rm lev}^T =\left( 6.67 \mu\rm{K}\right)^2$ and a  $\theta_{1/2}=1.5^\prime$ beam. The black lines show the total correction
\begin{equation}\label{eq:cross2input}
 \frac{\Delta C_L^{\hat q q^{\rm in}}}{C_L^{q^{\rm in} q^{\rm in}}} = \frac 1 {2L + 1}	\sum_{M=-L}^L \frac{\mathcal R^{qq}_{LM,LM}}{\mathcal R_L^{qq,\rm fid}},
\end{equation}
in units of the excised area. In the top and bottom plots, $\hat q$ stands for the lensing gradient $\hat \phi$ and lensing curl potential estimator $\hat \Omega$ respectively. The temperature QE gets its information from very small scales, and in our configuration we still find a number of relevant modes which are too noisy to be reconstructed. This results in a relative correction of$\sim -0.3 f_{\rm exc.}$, where the prefactor $0.3$ is still somewhat comparable to unity. By comparison, restraining the QE to use $\ell_{\rm max} = 3000$ or $2000$ for the same noise level reduces the large scale correction by a factor of 3 or 40, to $\sim 0.1 f_{\rm exc.}$ or $\sim 0.0075 f_{\rm exc.}$ respectively.

After rotating the sphere so that the mask covers the pole and is azimuthally symmetric, the response matrix is diagonal in $M$, and $M$ corresponds for large-scale lenses to the local multipolar structure. The coloured lines in Fig~\ref{fig:PS_GR} show the first few $M$ multipole contributions. As might be expected, we find a large-scale mostly monopolar and quadrupolar response for $\hat \phi$, corresponding to distortions in the local estimated spin-0 convergence and spin-2 shear (for $L \ge 2$) respectively. In contrast, there is no monopolar curl potential response, since close to the pole this corresponds to an undetectable rigid rotation. Since the lensing deflections $\alpha$ are of the order of a couple of arcminutes, the distortion at the pole should not affect the reconstruction much further away than that. Consistent with this, we find that the correction converges very quickly, requiring $M_{\rm max}$ at most $\sin \alpha\cdot (2L_{\rm max} + 1) \approx 5$.
\subsection{Larger holes}\label{sec:large_holes}
We now obtain the analytic expression for the response and discuss a more general implementations.

For a generic separable QE in Eq.~\eqref{eq:qe}, the response matrix always involves the product of two Gaunt integrals each with one lensing $L, M$ and two CMB $\ell m$'s. The first integral contains the observed deflection and two data legs multipoles, together with the QE weights. The second integral the CMB sky lensing and CMB, together with the response weights. The filtering matrix couples the CMB $\ell$'s and $m$'s together. Concretely, let $_{ab}\mathcal R$ denotes the linear response of the spin $a$ estimator to the sky spin-$b$ source,
\begin{equation}
	_a\hat q_{LM} =\:_{ab}\mathcal R^{MM'}_{LL'} \:_b{q}_{L'M'}.
\end{equation}
Also let $R^{b, s}$ be the response weights of the spin $s$ CMB to the sky spin-$b$ source (for lensing, $R^{\pm 1, s} = \pm \sqrt{(\ell \pm s)(\ell \mp s + 1))}$~\cite{Challinor:2002cd, Carron:2019swk}), and
\begin{equation}
	C_{\ell}^{st} \delta_{mm'}\delta_{\ell \ell'} = \left \langle{\:_{s}X_{\ell m}^{\rm cmb}\: _{t}X_{\ell m}^{\dagger, \rm cmb}} \right\rangle
\end{equation}
The response matrix takes then the following form
\begin{multline*}
   \:_{ab}\mathcal R^{MM'}_{LL'}= (-1)^a
    \begin{pmatrix} w^{s_o s_i} & w^{t_o t_i} & 1  \\
                    \ell_1 & \ell_2 & L \\
                    m_1 & m_2 & -M \\
                    s_o & t_o & -a
    \end{pmatrix}  \:_{s_i s}F^{m_1m_1'}_{\ell_1 \ell_{1'}}\: _{t_i t}F^{m_2 m_2'}_{\ell_2 \ell_{2'}}
    \\ \cdot   \left\{
    \begin{pmatrix}
    1 & R^{b, t}C^{t-s} & 1 \\
    \ell'_{1} & \ell'_{2} & L' \\
    m'_1 & m'_2 & -M' \\
    s & b-s & -b
    \end{pmatrix}^\dagger
    +
    \begin{pmatrix}
    R^{b, s}C^{s-t} & 1 & 1 \\
    \ell_{1'} & \ell_{2'} & L' \\
    m'_1 & m'_2 & -M' \\
    b-t & t & -b
    \end{pmatrix}^\dagger
    \right\}
\end{multline*}
where the matrices stand for the weighted Gaunt integrals
\begin{equation}\label{eq:carronmatrices}
\begin{pmatrix} c^1 & c^2 & c^3 \\ \ell_1 & \ell_2 & \ell_3 \\ m_1 & m_2 & m_3 \\ s_1 & s_2 &s_3 \end{pmatrix} \equiv
 \int d^2n\:\prod_{i = 1}^3\left(c^i_{\ell_i}\:_{s_i}Y_{\ell_i m_i}(\hat n)\right)
 \end{equation}
where always $s_1 + s_2 + s_3 = 0$.
In terms of the more usual Wigner 3j symbols, this is
\begin{equation}\nonumber
\frac{1}{\sqrt{4\pi}}\left(\prod_{i = 1}^3 c^i_{\ell_i} \sqrt{2\ell_i + 1} \right)
\begin{pmatrix}
  \ell_1 & \ell_2 & \ell_3\\
  m_1 & m_2 & m_3
\end{pmatrix}
\begin{pmatrix}
  \ell_1 & \ell_2 & \ell_3\\
  -s_1 & -s_2 & -s_3
\end{pmatrix}.
\end{equation}
One obtains the gradient (G) and curl (C) responses and cross-responses summing the spin-weight responses $_{ab}\mathcal R$ in the following way $(a, b = \pm 1)$:
\begin{equation}
 \begin{split}
 \mathcal R^{GG} &= \frac 12 ab\:_{ab}\mathcal R,\quad \mathcal R^{CC} = \frac 12\:_{ab} \mathcal R  \\
 \mathcal R^{GC} &= i\frac a 2 \:_{ab} \mathcal R, \quad \mathcal R^{CG} = -i\frac b 2  \:_{ab}\mathcal R
 \end{split}
\end{equation}
In order to make progress, it is natural to split the $F$ matrices into a diagonal and isotropic ($m$-independent), reference part $\Fiso_\ell$ , and a perturbation $\delta F$. There are then two sorts of terms contributing to the change in response, proportional to $\Fiso \delta F$ (and $\delta F \Fiso$), and $\delta F \delta F$, where for small enough perturbations the former dominates. In that regime, the case of the cross-spectrum (Eq.~\eqref{eq:cross2input}) is particularly simple and a fairly general result can be given which we now describe.

 From the addition rules of the Gaunt integrals, whenever two of the three legs of the pairs of matrices in Eq.~\eqref{eq:carronmatrices} are contracted (magnetic numbers summed over for the same multipole moments), the third leg numbers must then also match and the entire sum results in a simple one-dimensional integral of Wigner small-$d$ correlation functions (for example, the isotropic case contracts on the two CMB legs resulting in the well-known isotropic lensing response $\mathcal R^{MM'}_{LL'} \propto \delta_{MM'} \delta_{LL'} \mathcal R_L$). The $\Fiso F$ terms in the case of the cross-spectrum (Eq.~\eqref{eq:cross2input}) also have two contracting legs (one CMB and the lensing leg), so that these terms are just as simple to evaluate, the difference to an isotropic response calculation being the replacement on the non-contracting CMB leg of $\Fiso_\ell$ by the diagonal of the perturbation,
 \begin{equation}\label{eq:respFdF}
\Fiso_\ell \rightarrow \frac{1}{(2\ell + 1)} \sum_m \delta F^{mm}_{\ell\ell} \quad(\Fiso \delta F \textrm{ terms})
 \end{equation}
 For example when just masking very small holes (in general positions), the perturbation to the $F$ matrix in the absence of optimal filtering (as in Eq.~\eqref{eq:apofilt}, but allowing for a non-azimuthally symmetric mask, and identifying the sky and fiducial transfer functions)  may be written
 \begin{equation}
 	\delta F^{mm'}_{\ell \ell'} = -\Fiso_\ell \frac{b_\ell'}{b_\ell}\sum_{\textrm{sources }i} \sigma_i Y^\dagger_{\ell m}(\hat n_i)Y_{\ell' m'}(\hat n_i),
 \end{equation}
 where $\sigma_i$ is the excised area for each small source. Using the rule in Eq.~\eqref{eq:respFdF}, this results in the replacement
 \begin{equation}
 	\Fiso_\ell \rightarrow - \Fiso_\ell \sum_{\textrm{sources }i} \left(\frac{\sigma_i}{4\pi} \right) \quad (\textrm{no optimal filtering})
 \end{equation}
 The prefactor is just the total excised sky fraction. The total cross-spectrum to the input lensing map is then reduced exactly by twice the excised sky fraction (since there is both a $\Fiso \delta F$ and a $\delta F \Fiso$) term). On the units of Fig.~\ref{fig:PS_GR}, this is identically -2.

With optimal filtering, the CMB is reconstructed within the holes and the bias is much reduced. For the same population of sources, the matrix inverse in the optimal filtering $F$ will introduce coupling between the sources contributions, but this can be neglected in the limit of small\footnote{This can be made more precise with the requirement that the maximal resolved multipole and hole radius are related through $\theta_{\textrm{mask}}  \cdot \ell_{\textrm{max,noisefree}} \ll 2 $}, (not necessarily well separated) holes. The corresponding result becomes
 \begin{equation}
 \begin{split}
 	 	\Fiso_\ell \rightarrow -\Fiso_\ell &\left( 1 - \frac{C^{\rm fid}_\ell}{C^{\rm fid}_\ell + N_r / b_\ell^2}\right) \sum_{\textrm{sources }i} \left(\frac{\sigma_i}{4\pi} \right) \\
 	&(\textrm{optimal filtering})
  \end{split}
 \end{equation}
As expected, the bias to the cross-spectrum depends on the reconstruction signal to noise via the Wiener-filtering factor of the CMB reconstruction on the scales where the QE takes its most relevant contribution.

For larger holes, we cannot neglect the term quadratic in $\delta F$. One way to proceed is through a SVD decomposition of $\delta F$. Instead of contracting, the relevant leg separates in the large sum, and this also results in simpler one-dimensional integrals. This remains more complex, as one needs to calculate one set of integrals per singular value, per $L$ and $M$. Hence, this approach is only practical when the perturbation to the filtering matrix can be described by a reasonable number of singular value. For example, a cluster mask of $5^\prime$ radius only has $\sim 10$ relevant modes and is still perfectly doable. On the other hand, a galactic mask like that of the $\Planck$-lensing analysis has many thousands of these and we have not attempted this calculation.

\subsection{Comparison with Gaussian simulations}\label{sec:GAUSS}

To validate the pipeline described in Sec.~\ref{sec:recon} and to estimate the error bars in Figs.~\ref{fig:lens-rec_auto}, ~\ref{fig:lens-rec_cross}, ~\ref{fig:lens-rec_auto_and_cross_1p7} and ~\ref{fig:cross_pb}, we computed both the estimators in Eq.~\eqref{eq:lensing-power-spectra-esimator} and Eq.~\eqref{eq:lensing-cross-spectra-esimator} without $\fMC$ correction, averaging over 20 G simulations that are uncorrelated to the fixed \websky\ mask. As expected $W_X$ and $W^{\rm rot}_X$ masks give consistent results in the absence of mask correlations.

\begin{figure}[!]
\includegraphics[width = \columnwidth]{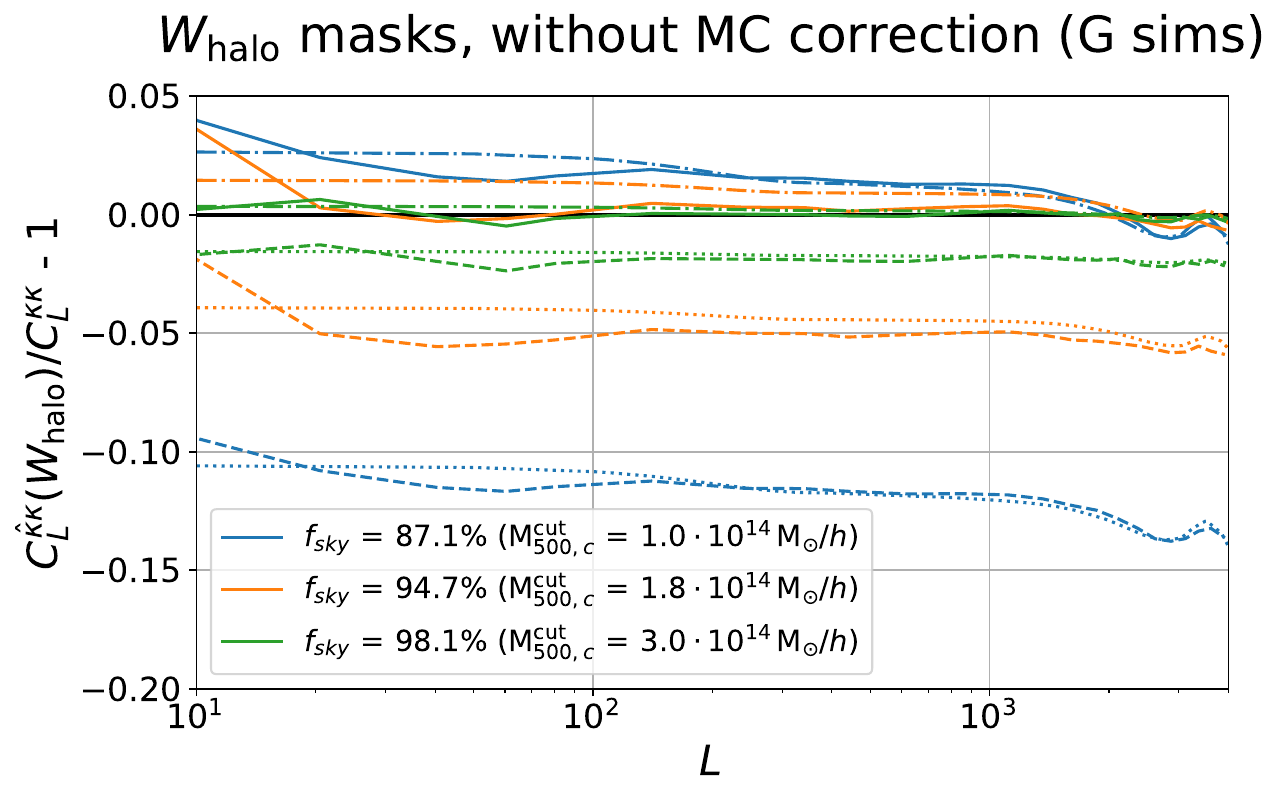}
\caption{
Relative differences between the reconstructed cross $\hat \kappa  \kappa$-spectrum and the true \websky $C_{L}^{\kappa\kappa}$ for three different uncorrelated $W_{\rm halo}$ masks normalizing with $\fsky^{-1}$ to account for the mask area but no
multiplicative Monte Carlo correction. Dashed lines show the same relative differences without the $f_{\rm sky}^{-1}$ normalization.
Dash-dotted and dotted lines show the theoretical expectations for the model of Eq.~\eqref{eq:cross2input}  with and without the $f_{\rm sky}^{-1}$ normalization respectively. In the theoretical calculation, we assumed that all the disks have the same radius equal the mean of the halo sizes (in terms of $2\theta_{500,c}$) in each mass-limited sample.
For small holes, the $f_{\rm sky}^{-1}$ normalization over-corrects for the loss of power as discussed in Appendix~\ref{sec:holes}.}

\label{fig:MCcorr_GAUSS}
\end{figure}

We also used the uncorrelated-mask simulation to compare with the analytic models for populations of small masked holes.
We compare the relative differences between the reconstructed $\hat \kappa  \kappa$-spectrum and the true \websky $\kappa\kappa$-spectrum that was used to simulate the uncorrelated lensing fields, for three different $W_{\rm halo}$ masks, without the multiplicative Monte Carlo correction defined in Eq.~\eqref{eq:fMC_correction}. Figure~\ref{fig:MCcorr_GAUSS} shows that, as predicted by the analytic model, a $f_{\rm sky}^{-1}$ normalization would be a substantial overcorrection for masks made of many tiny holes that are partially re-filled by the optimal filtering. Similar results have been found for all the other foreground masks.

\section{Effects of masking non-Gaussian fields}\label{sec:threshold_masks}

\begin{figure*} [t!]
\centering
\begin{tabular}{cccc}
\includegraphics[width=0.33\textwidth]{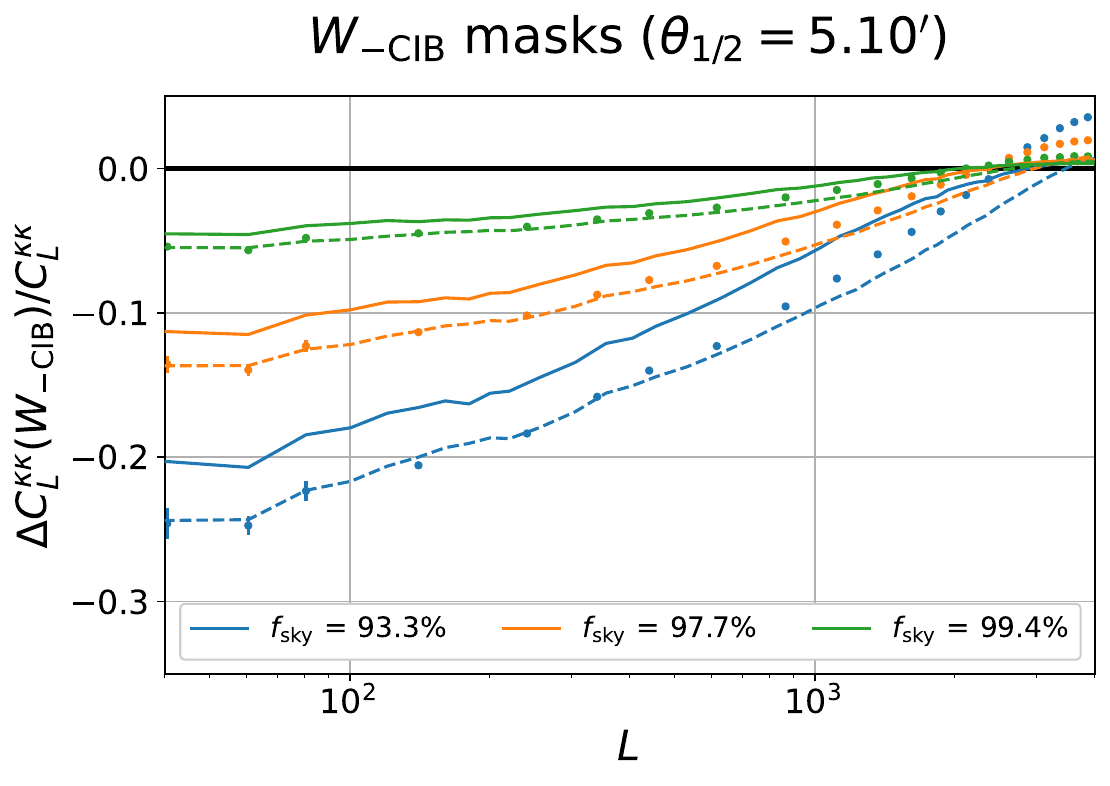} &
\includegraphics[width=0.338\textwidth]{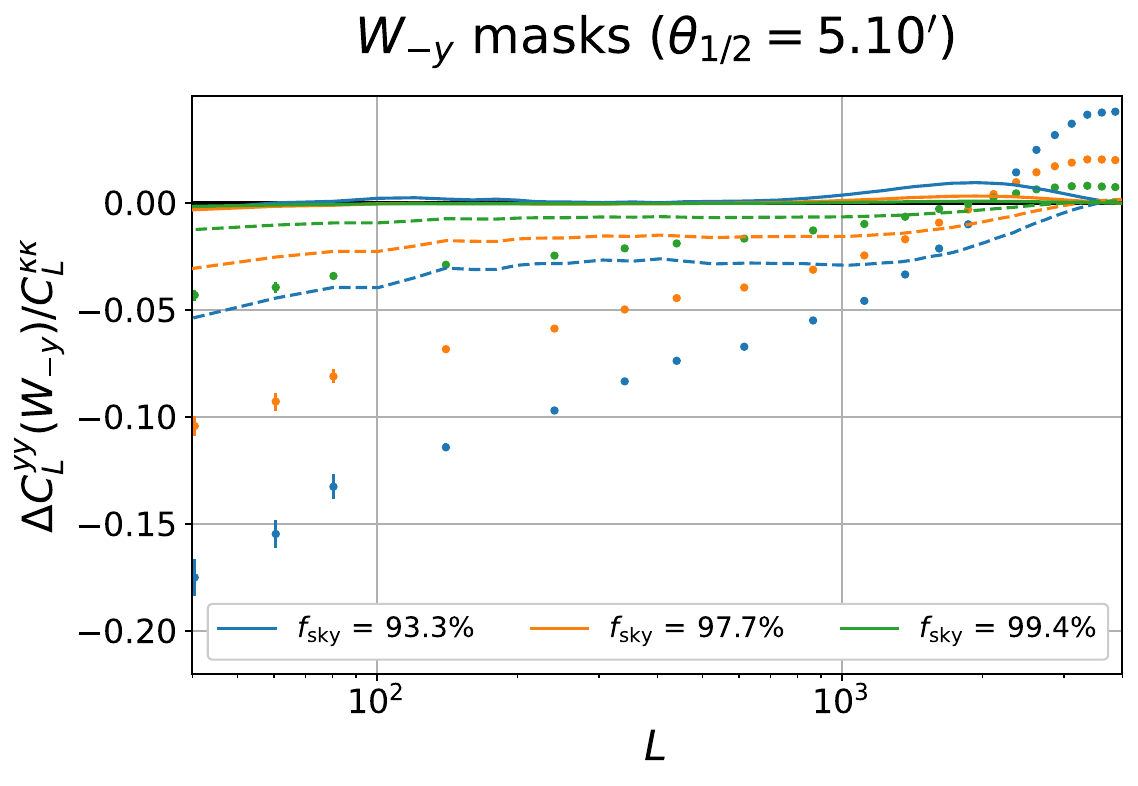} &
\includegraphics[width=0.33\textwidth]{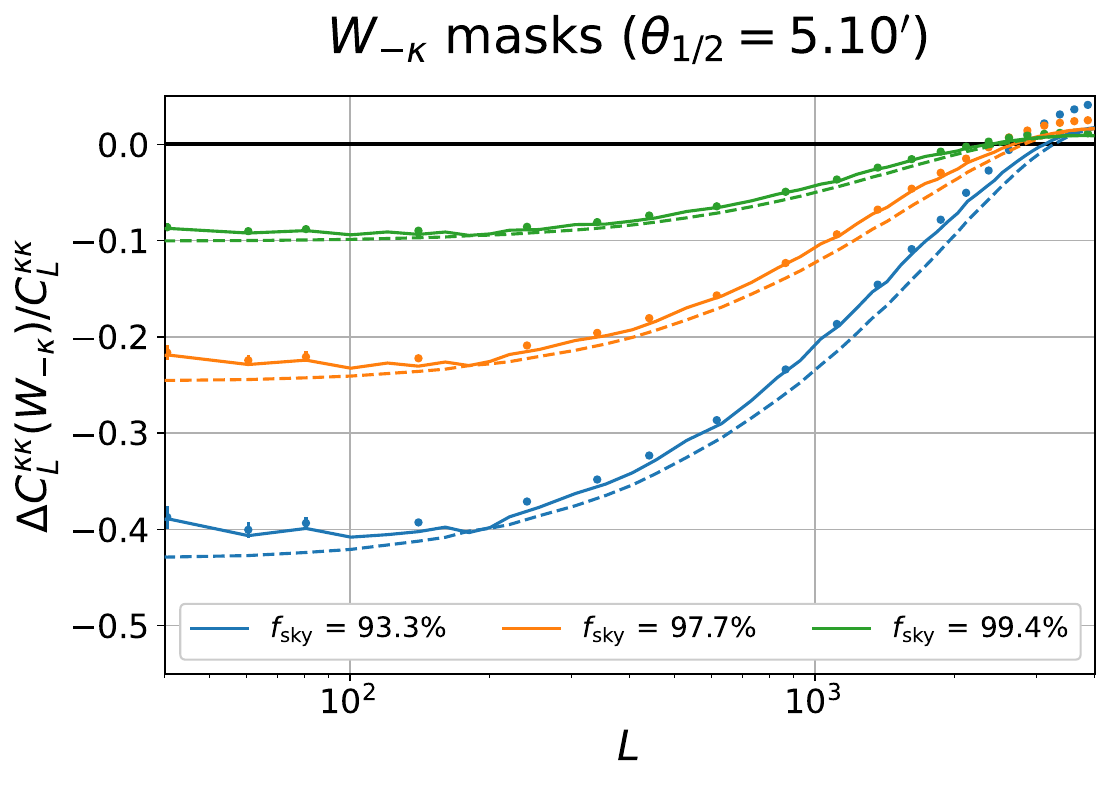}\\
\end{tabular}
\begin{tabular}{cccc}
\includegraphics[width=0.33\textwidth]{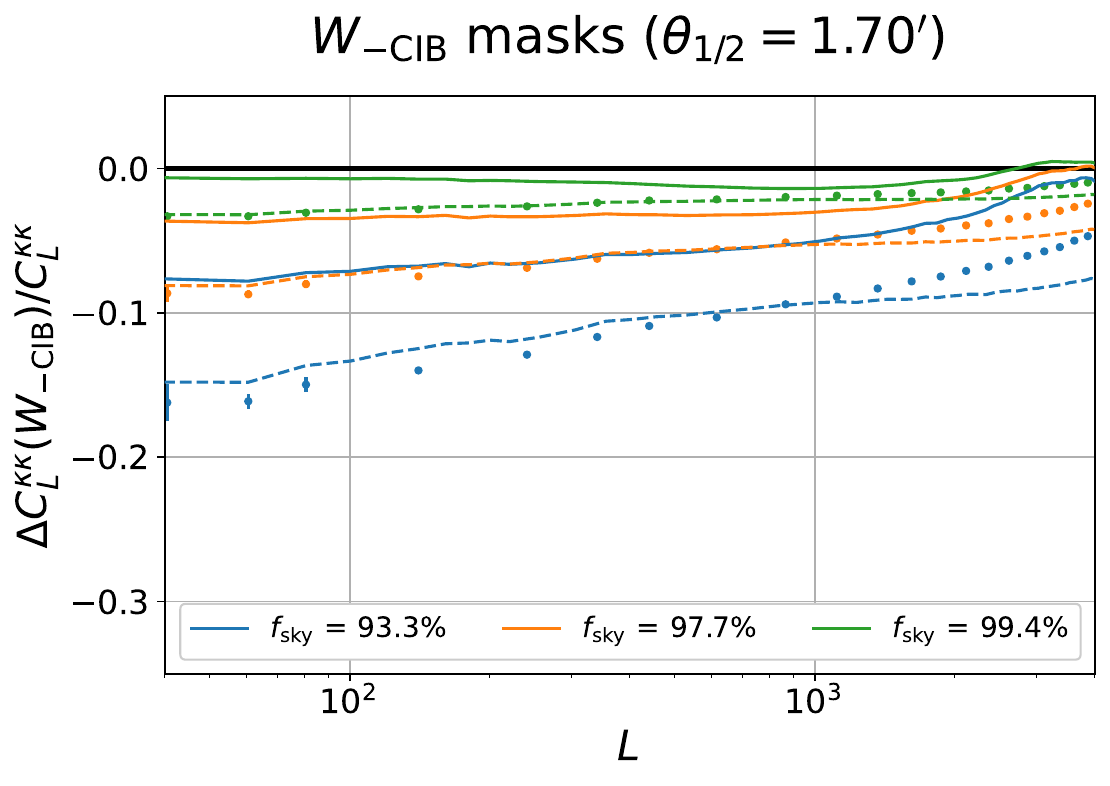}&
\includegraphics[width=0.338\textwidth]{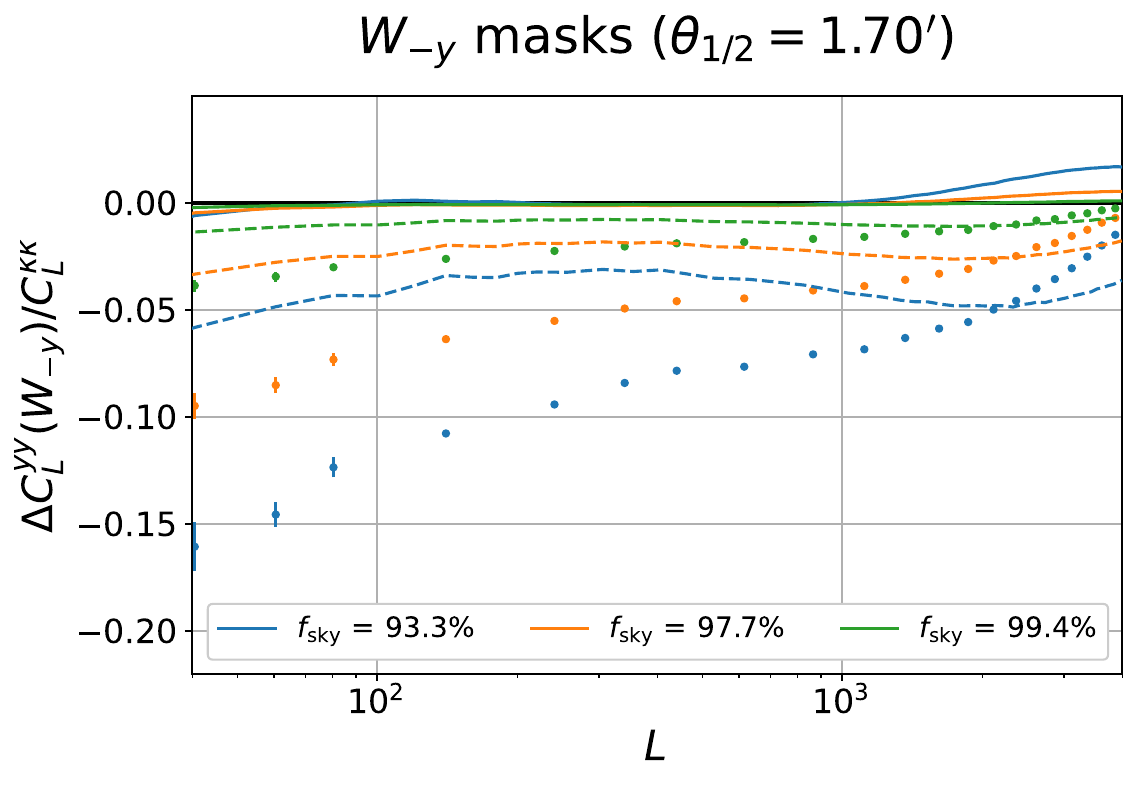} &
\includegraphics[width=0.33\textwidth]{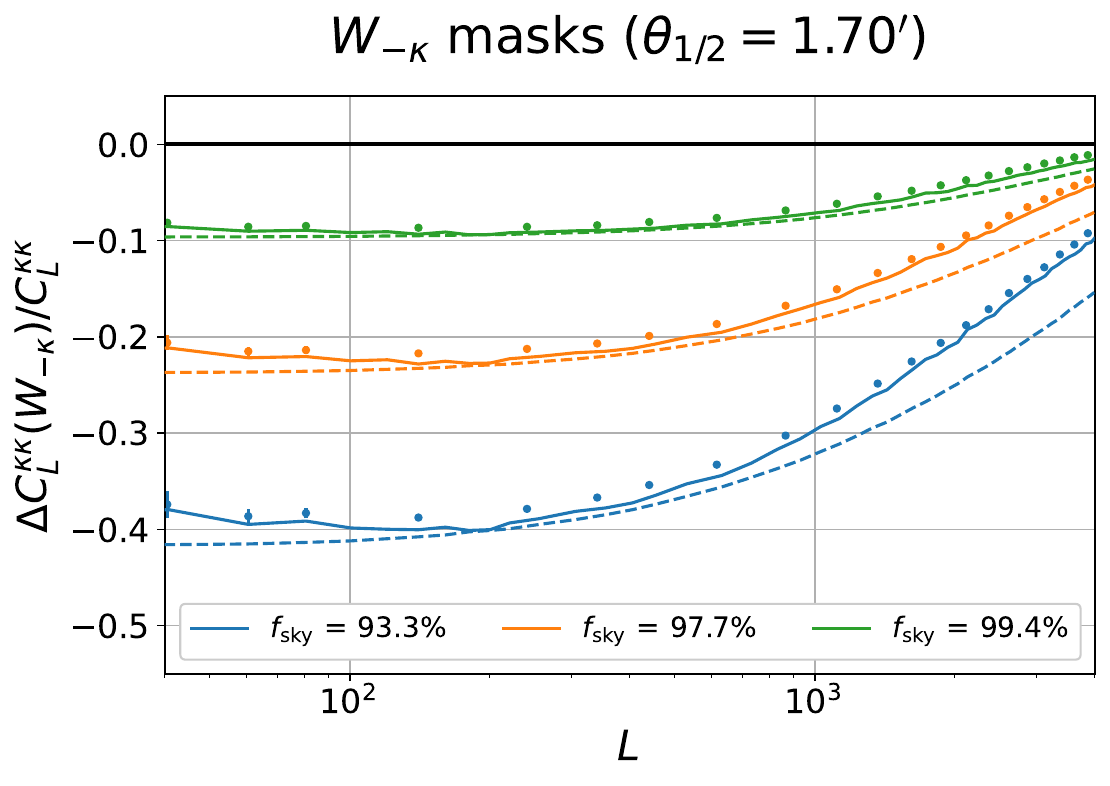}\\
\end{tabular}
\caption{
Same as Fig.~\ref{fig:masking_kappa_noapo} for the sign-flipped masks $W_{- \rm CIB}$, $W_{-y}$, and $W_{-\kappa}$ that remove minima of the foreground fields. Simulation measurements are shown as data points, the semi-analytic Gaussian theory predictions in solid and the fully-analytic Gaussian model in dashed (evaluating the expectation values in Eq.~\eqref{eq:ABf_bias} empirically from our simulations and analytically respectively).
The top and bottom rows show the masks obtained by smoothing the foreground maps with a Gaussian beam of $\theta_{1/2} = 5.1^\prime, 1.7^\prime$ respectively.}
\label{fig:masking_kappa_noapo_flip}
\end{figure*}

In this Appendix, we explore the disagreement between the theory predictions based on simple Gaussian models and the simulation results for the threshold masks, $W_\kappa$, $W_{\rm CIB}$ and $W_y$,  as shown in  Fig.~\ref{fig:masking_kappa_noapo}.

\begin{figure}[!]
\centering
\includegraphics[width =0.45 \textwidth]{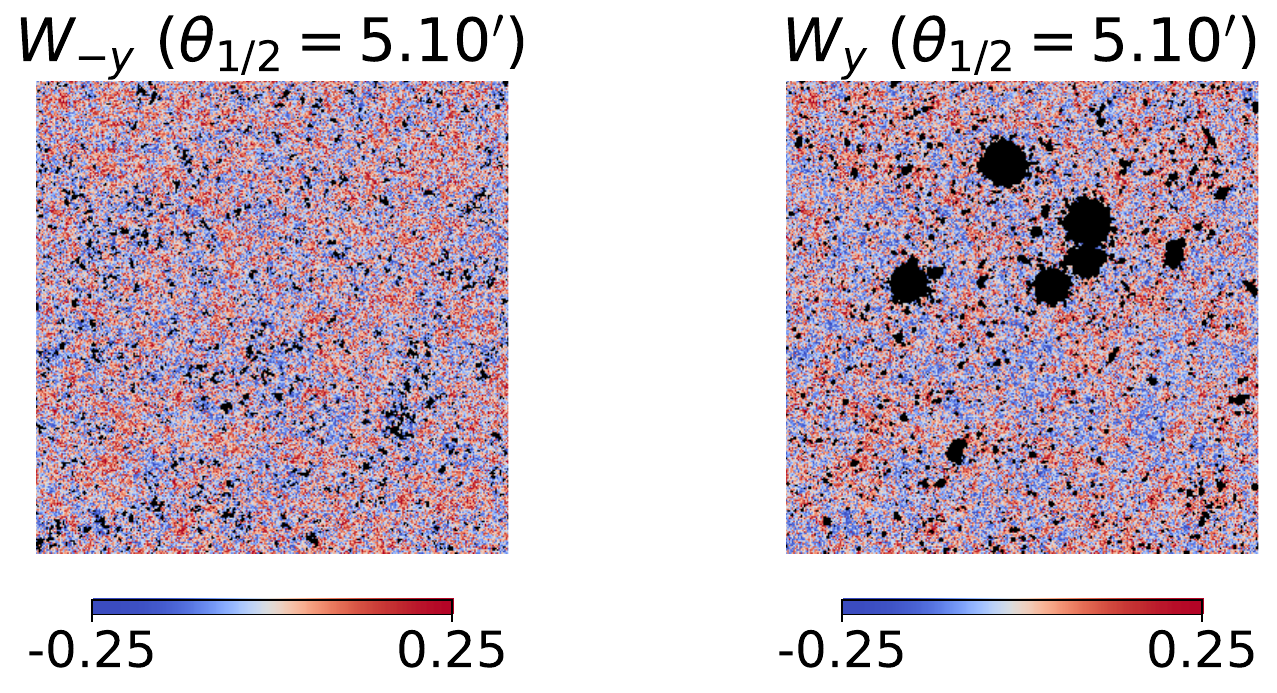}
\caption{ \websky $\kappa$ field masked with $W_y$ and $W_{- y}$ masks obtained smoothing the foreground field with a $\theta_{1/2}=5.1^\prime$ Gaussian beam prior to thresholding. The masked fraction of the sky is the same for both masks ($f_{\rm sky} =93.3\%$). The tSZ signal is highly non-Gaussian and highly skewed, so that the mask covers large clusters in the standard case, and a much larger number of small underdensities in the $-y$ case (for Gaussian fields the two masks would be statistically identical).
}
\label{fig:y_vs_flipy}
\end{figure}

To test higher-order non-Gaussian effects, as in  \PaperI\ , we constructed another set of foreground masks from the $y$, ${\rm CIB}$ and $\kappa$ fields by inverting their sign prior to the thresholding operation, $W_{- \rm CIB}$, $W_{-y}$ and $W_{-\kappa}$. This is equivalent to masking the minima instead of the peaks of the foreground fields. This helps isolating higher-order effects that involve a higher even power of the perturbations compared to the expansion considered in our analytical model (where $X$ and $-X$ masks give identical results). The results are shown in Fig.~\ref{fig:masking_kappa_noapo_flip}. We found a good agreement only for $W_{-\kappa}$ while we recover at least a good qualitative trend for $W_{-\rm CIB}$.
However for $W_y$, the predictions instead lie away from the analytic curves, consistent with strongly non-Gaussian higher-order effects (clearly visible in the mask shape difference shown in Fig.~\ref{fig:y_vs_flipy}) not being captured in our Gaussian model. Here the semi-empirical results suggest only a very small bias when masking $W_{-y}$.

\begin{figure*} [t!]
\centering
\begin{tabular}{cccc}
\includegraphics[width=0.33\textwidth]{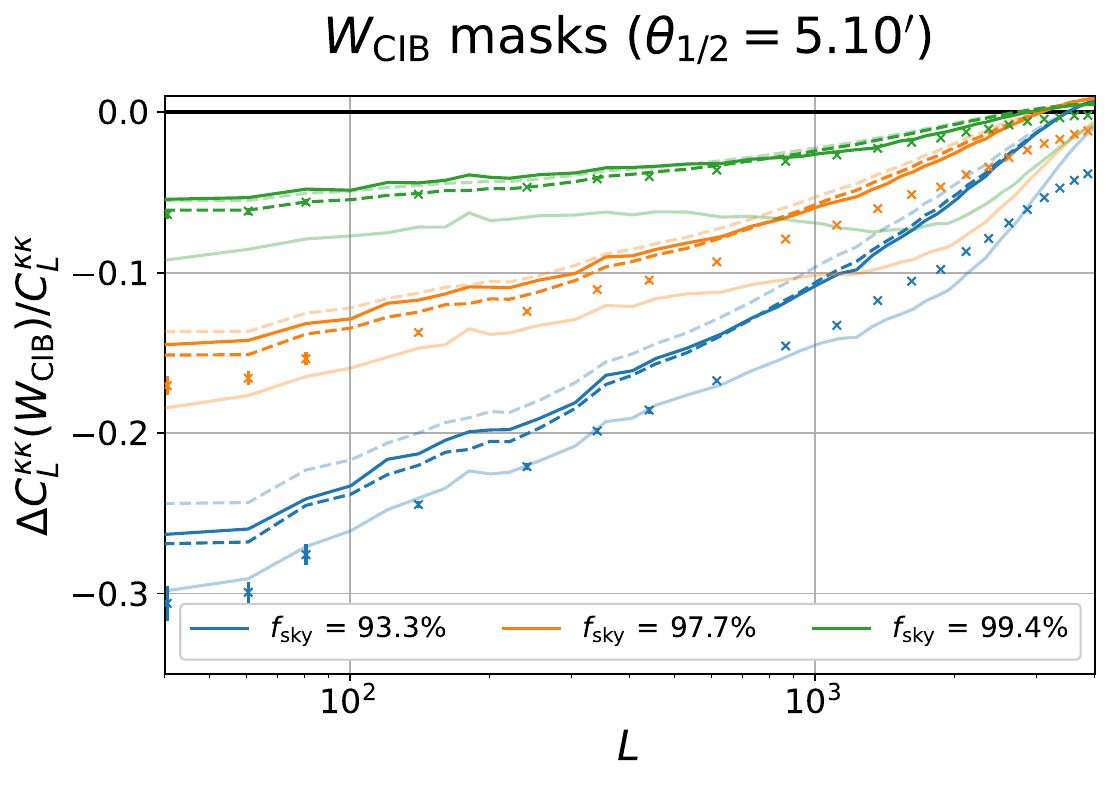} &
\includegraphics[width=0.338\textwidth]{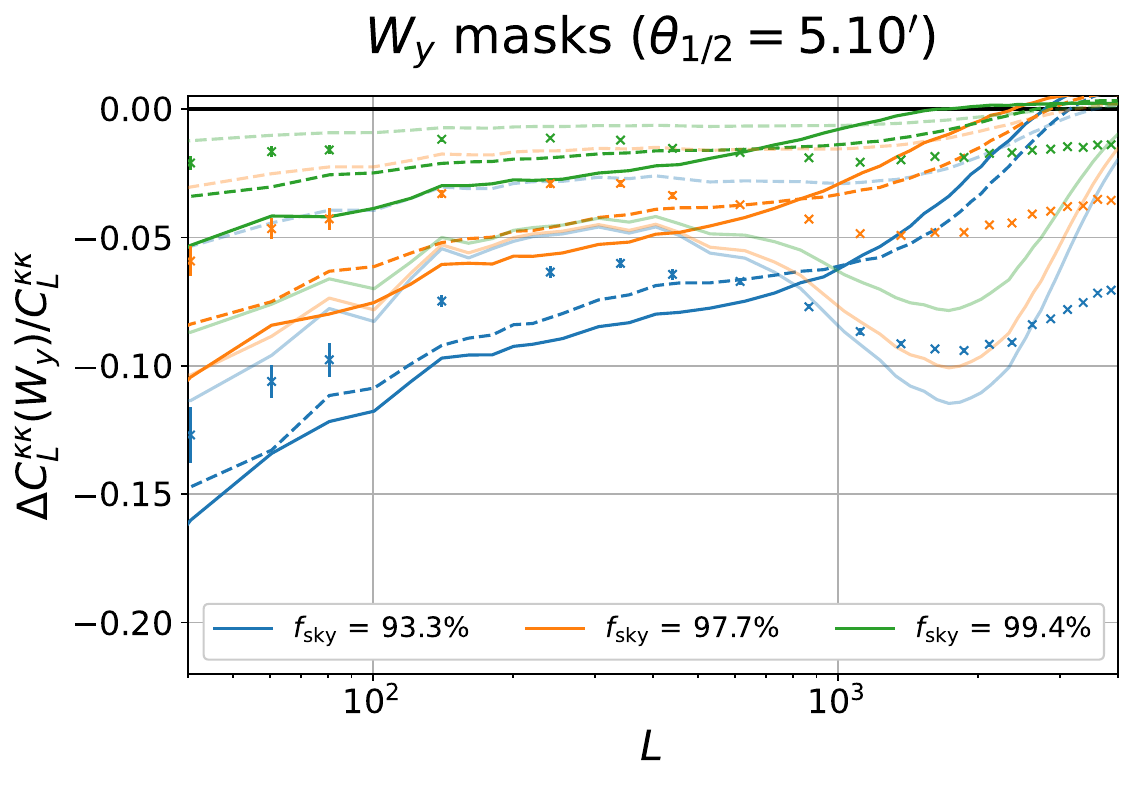} \\
\end{tabular}
\begin{tabular}{cccc}
\includegraphics[width=0.33\textwidth]{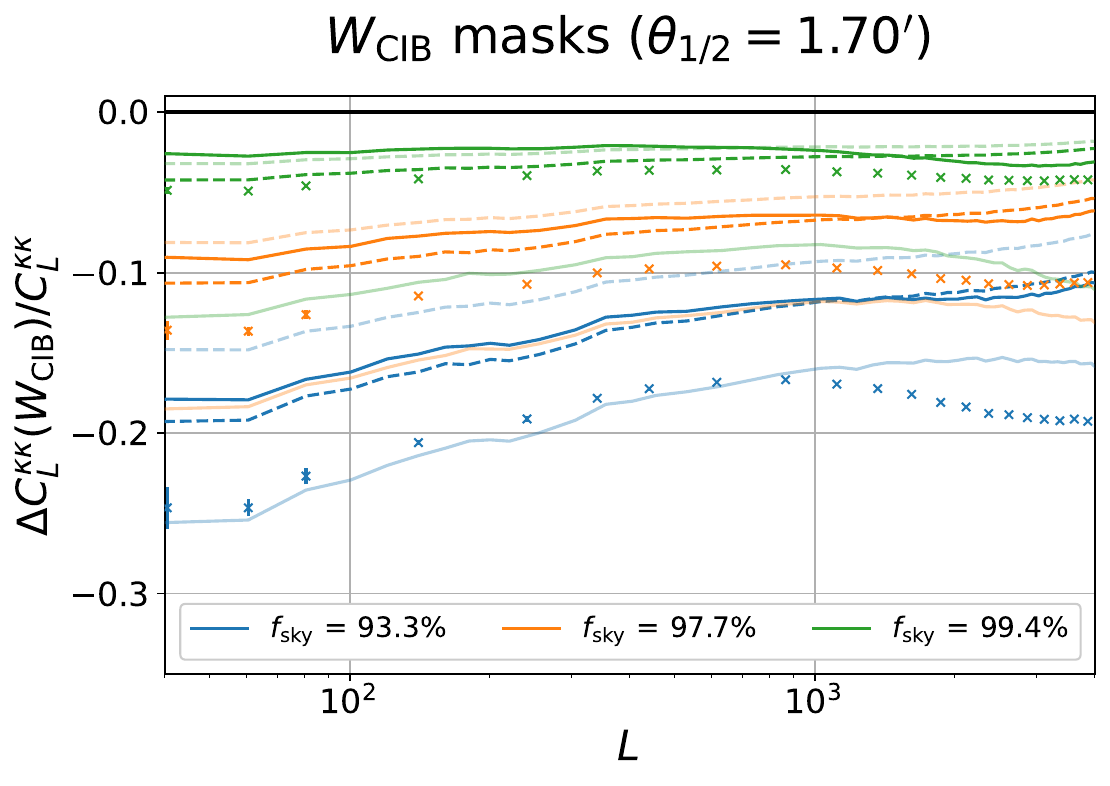}&
\includegraphics[width=0.338\textwidth]{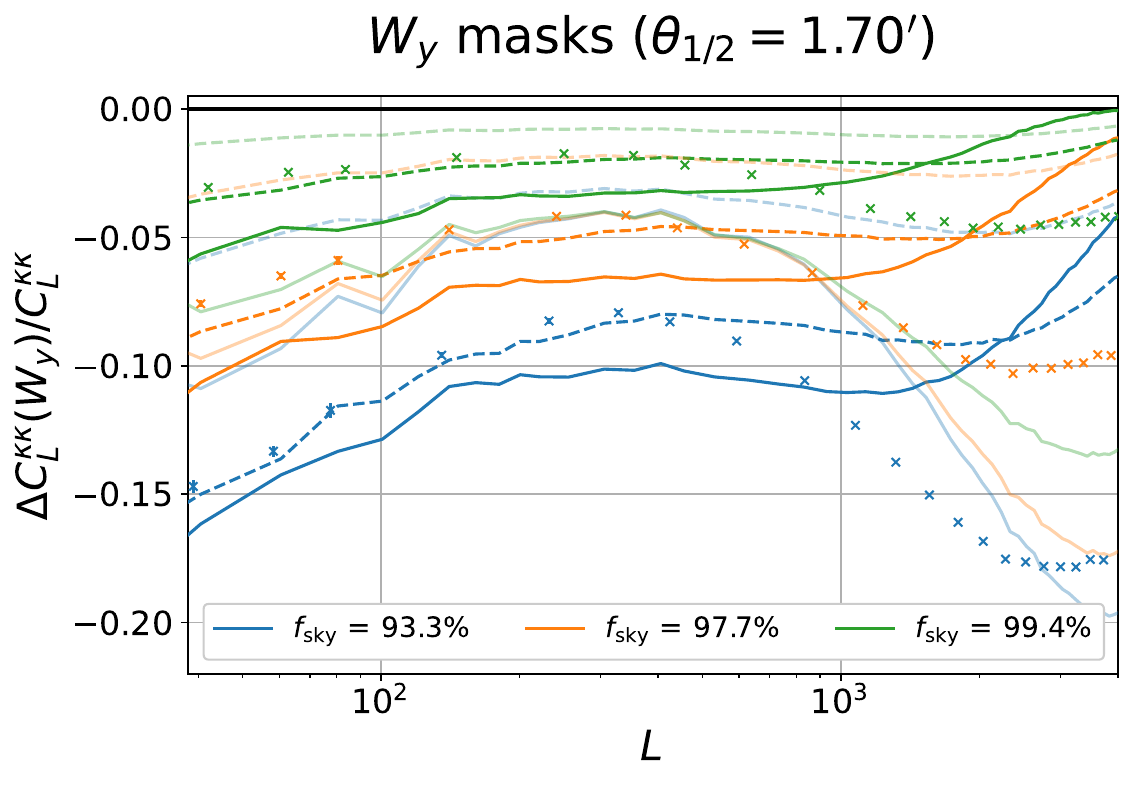} \\
\end{tabular}
\caption{Effect of LSS-correlated masking on the CMB lensing convergence power spectrum $C_L^{\kappa \kappa}$ as fractional difference between the $C_L^{\kappa \kappa}$ computed on the masked sky and its value computed on the full sky. We compare a subset of the simulation results shown in Fig.~\ref{fig:masking_kappa_noapo} with alternative analytical predictions. The semi-analytical and analytical predictions of the Gaussian model obtained using the full-sky power spectra of Gaussianized foreground maps are shown in solid and in dashed, respectively. The baseline theory predictions (both semi-analytical and fully analytical) based on non-Gaussian foreground maps and shown in Fig.~\ref{fig:masking_kappa_noapo} appear as faded lines. Top and bottom panels show results for the $W_{\rm CIB}$ and $W_{y}$ masks constructed on foregrounds field smoothed with different Gaussian beams.
}
\label{fig:masking_kappa_noapo_gaussianized}
\end{figure*}

The theory predictions are based on full-sky spectra of non-Gaussian foreground maps, and hence are sensitive to the peaks of the field inside the sky areas that get masked. Such features are enhanced for non-Gaussian fields compared to a purely Gaussian field.  To test the impact of these non-Gaussian features, and to understand if they are responsible for the disagreement with the simulation results, we tried creating a Gaussianized version of the non-Gaussian foreground fields. 
To do this, we remapped the values in each pixel  to the value that they would have in same the corresponding percentile in a Gaussian distribution having the same variance. Ideally, the masked foreground spectra after this procedure should be similar to the one obtained from non-Gaussian maps, but the full-sky ones used in Eq.~\eqref{eq:ABf_bias} may be rather different, as now all the very bright extreme pixels have had their values scaled down. In Fig.~\ref{fig:masking_kappa_noapo_gaussianized} we show the analytic prediction based on these Gaussianized maps. We still do not find a perfect agreement between simulations and theory predictions, but at least we are now able to reproduce the trends seen in the simulations: in particular, the $W_{y}$ mask predictions now recover, at intermediate scales, distinctly different curves for masks retaining different $f_{\rm sky}$s. Moreover, as expected, the semi-analytic theory predictions and the pure Gaussian model obtained using the full-sky power spectra from Gaussianized maps are closer together (see dashed and solid lines).

\begin{figure*} [t!]
\centering
\begin{tabular}{cccc}
\includegraphics[width=0.33\textwidth]{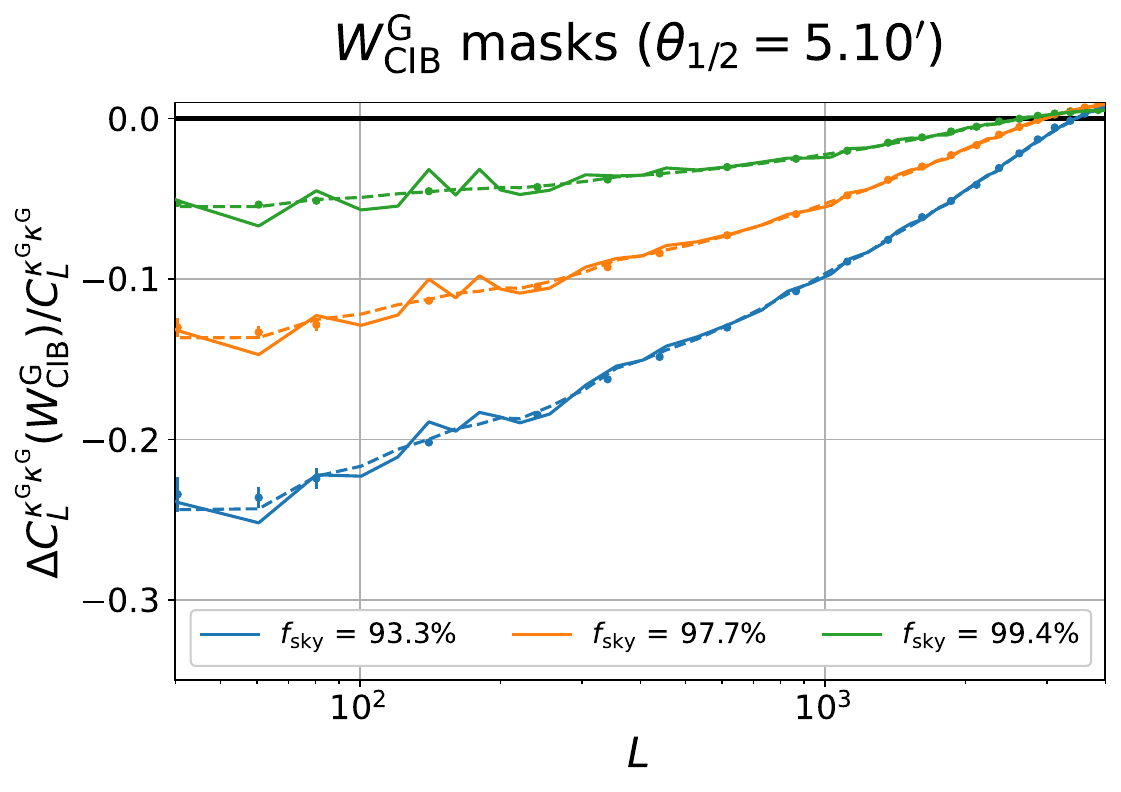} &
\includegraphics[width=0.338\textwidth]{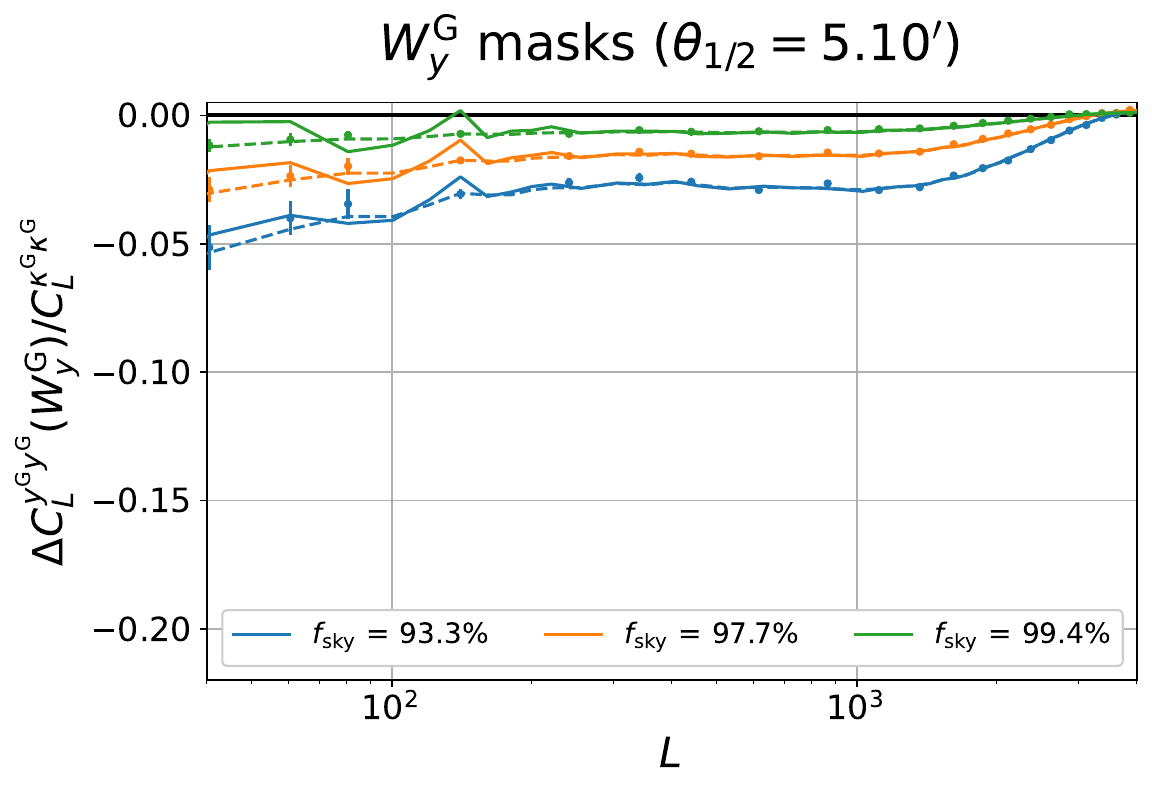} &
\includegraphics[width=0.33\textwidth]{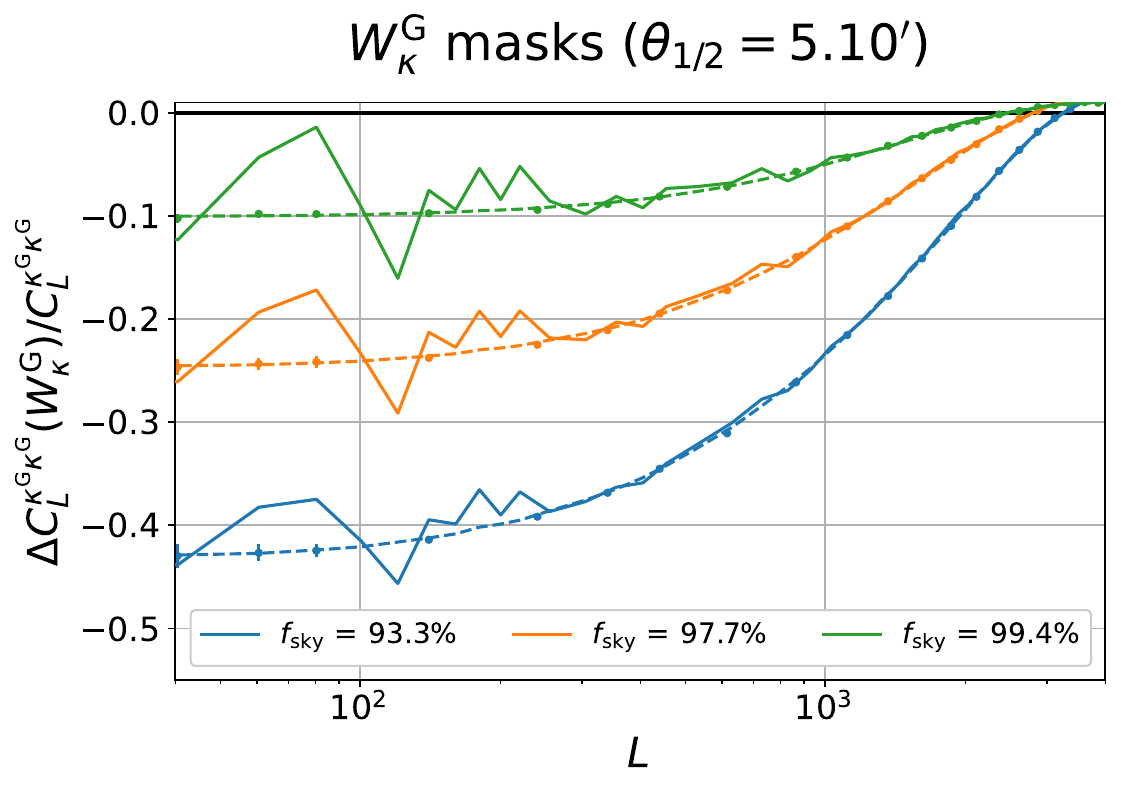}\\
\end{tabular}
\begin{tabular}{cccc}
\includegraphics[width=0.33\textwidth]{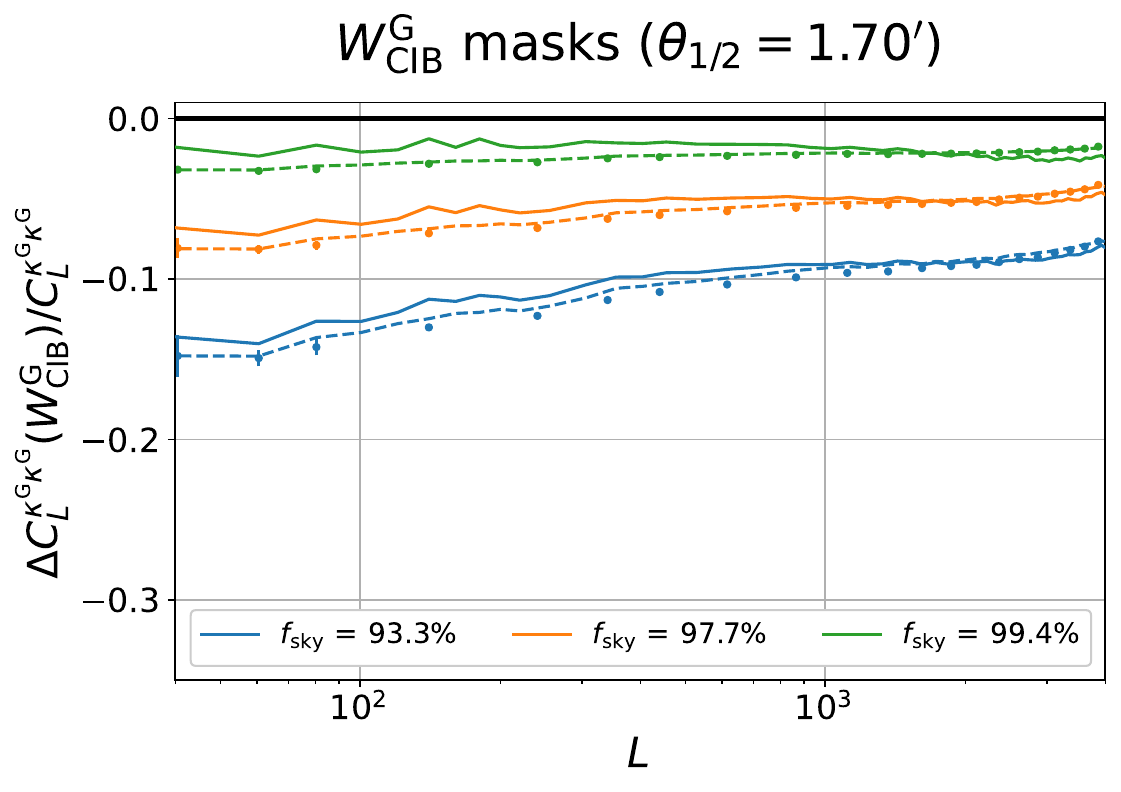}&
\includegraphics[width=0.338\textwidth]{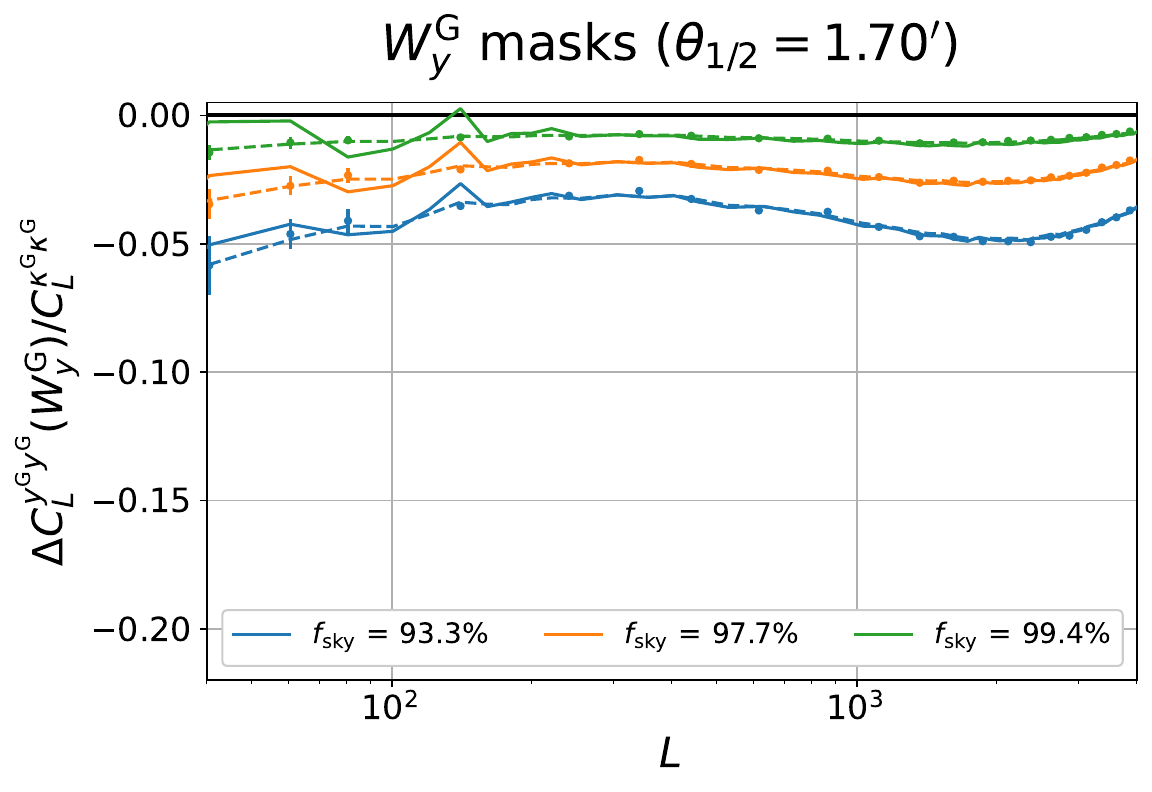} &
\includegraphics[width=0.33\textwidth]{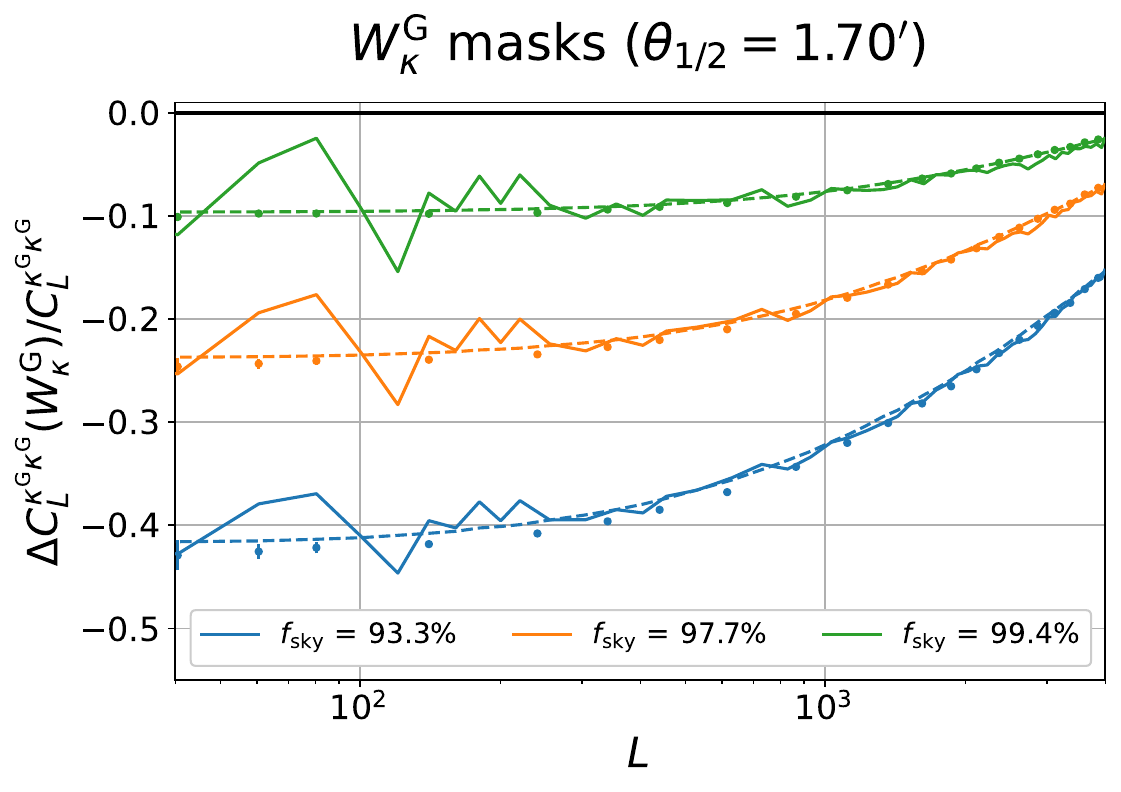}\\
\end{tabular}
\caption{Same as Fig.~\ref{fig:masking_kappa_noapo_flip} for the $W^{\rm G}_y$, $W^{\rm G}_{\rm CIB}$ and $W^{\rm G}_\kappa$ masks derived thresholding a Gaussian realization of the foreground fields and of CMB lensing keeping the same correlation between the components observed in the \websky simualations.}
\label{fig:masking_kappa_noapo_gauss}
\end{figure*}

Finally, to test the accuracy of the analytical model and to convince ourselves that discrepancies we observe between the analytical models and the simulations are simply due to non-Gaussianity of the foreground fields, we built correlated Gaussian map of the foreground fields, $y^{\rm G}$,  $\mathrm{CIB}^{\rm G}$ and $\kappa^{\rm G}$, that preserve the cross-correlation between $y$, CIB and $\kappa$ as measured from the full-sky \websky maps. We used these Gaussian realizations of the foreground fields to construct the  Gaussian foreground $W^{\rm G}_y$, $W^{\rm G}_{\rm CIB}$ and $W^{\rm G}_\kappa$ masks that we later applied on the corresponding Gaussian $\kappa^{\rm G}$ map. The results are shown in Fig.~\ref{fig:masking_kappa_noapo_gauss}  and are in excellent agreement with the analytic model described in Sec.~\ref{sec:gaussianmodel}, confirming that the differences we see in the more realistic simulations are due to the non-Gaussian statistics.

\bibliography{antony,cosmomc,julbib,maskbias}

\end{document}